\documentclass[a4paper,11pt]{article}

\usepackage{classeJBArxive}

\usepackage{arydshln}

\newcommand*\unit[1]{\bigl[\, \mathsf{#1} \,\bigr]}

\newcommand{\Bi}{\mathrm{Bi}}

\newcommand\deltaT{\Delta_{\,T}^{\,\rf}}
\newcommand\deltaP{\Delta_{\,P_{\,1}}^{\,\rf}}
\newcommand{\Fo}{\mathrm{Fo}}
\newcommand{\fosw}{\mathrm{f}_{\,\mathrm{osw}}}
\newcommand{\Psat}{P_{\,\mathrm{sat}}}
\newcommand\tf{t_{\,\mathrm{f}}}

\newcommand{\rf}{\mathrm{ref}}

\title{Dimensionless formulation and similarity to assess the main phenomena of heat and mass transfer in building porous material
\vspace{2pt}
}

\author{Julien Berger \textsuperscript{a}$^{\ast}$,  Clemence Legros\textsuperscript{b,c}, Madina Abdykarim\textsuperscript{b}\\
\date{\today\vspace{-0.5cm}}}

\begin{document}

\maketitle

\begin{center}
\small
\textsuperscript{a}  Laboratoire des Sciences de l’Ingénieur pour l’Environnement (LaSIE), UMR 7356 CNRS, La Rochelle Université, CNRS, 17000, La Rochelle, France \\
\textsuperscript{b} Univ. Savoie Mont Blanc, CNRS, LOCIE, 73000 Chambéry, France \\
\textsuperscript{c} Univ. Grenoble Alpes, CEA, LITEN, DTS, INES, 38000 Grenoble, France\\
$^{\ast}$corresponding author, e-mail address : julien.berger@univ-lr.fr\\
\end{center}

\begin{abstract}

Within the environmental context, several tools based on simulations have been proposed to analyze the physical phenomena of heat and mass transfer in porous materials. However, it is still an open challenge to propose tools that do not require to perform computations to catch the dominant processes. Thus, this article proposes to explore advantages of using a dimensionless analysis by scaling the governing equations of heat and mass transfer. Proposed methodology introduces dimensionless numbers and their nonlinear distortions. The relevant investigation enables to enhance the preponderant phenomena to \emph{(i)} compare different categories of materials, \emph{(ii)} evaluate the competition between heat and mass transfer for each material or \emph{(iii)} describe the transfer in multi-layered wall configurations. It also permits to define hygrothermal kinetic, geometric and dynamic similarities among different physical materials. Equivalent systems can be characterized in the framework of experimental or wall designs. Three cases are presented for similarity studies in terms of \emph{(i)} equivalent material length, \emph{(ii)} time of heat and mass transfer and \emph{(iii)} experimental configurations. All these advantages are illustrated in the given article considering $49$ building materials separated in $7$ categories.



\end{abstract}

\section{Introduction}

Within the environmental issues, modeling the building energy efficiency is an important challenge for designers and engineers. The governing equations of heat and mass transfer in building porous materials have been proposed in \cite{Luikov_1966}. Since this early work of \textsc{Luikov}, several numerical models have been developed and recently reported in \cite{Mendes_2017}. These tools enable to compute accurately the prediction of the physical phenomena as illustrated in \cite{Chikhi_2016}. However, without computational tools, it is an open challenge to catch the dominant physical phenomena. More precisely, it is still impossible to answer the following questions: which of heat or mass transfer is preponderant in the material? which material is more exposed to sensible or latent heat transfer? which material is better to stop the penetration of heat and mass transfer? 

The numerical models permit to answer these questions as illustrated, for instance, in \cite{Xu_2019,Simotagne_2019}. However, the numerical models can have an important computational cost. Thus, the main issue is to propose tools that enable to assess easily, \emph{i.e.} without computations, the main features of the phenomena. For heat transfer exclusively, the so-called $U$-value is commonly used to describe the efficiency of a building wall material such as insulation \cite{Iso_9869,Rodler_2019}. For mass transfer, there is an equivalent indicator with the vapor permeability coefficient $\mu\,$. However, no similar indicators exist to describe the material behavior regarding both heat and mass transfer despite the proven impact of latent transfer on building energy efficiency \cite{Antretter_2012,Mendes_2003,Berger_2015}. 

To tackle this issue, the dimensionless analysis can be carried out. It is based on the scaling of equations to get a nondimensional formulation of the mathematical model describing the physical phenomena. It introduces dimensionless numbers and their nonlinear distortions. As successfully presented in \cite{Ruzicka_20088}, those dimensionless numbers have several important advantages. The most important one is that the scaling parameters, such as the \textsc{Biot} and \textsc{Fourier} numbers, permit a clear physical understanding of the main phenomena of heat and mass transfer. An analysis of their order of magnitude produces estimation of the dominant physical processes. Thus, it can eventually lead to a simplification of the problem \citep{Nayfeh_2000}. Another important advantage of those numbers is to define similitude laws useful for the design of walls or even experimental set-ups. 

The dimensionless analysis is broadly used in other domains such as fluid dynamics. Among several studies, \emph{i.e.} \cite{Li_2016}, a sensitivity analysis is carried out for the dimensionless numbers of the problem. Asymptotic expansion can be performed as in \cite{Chhay_2017}. It can be remarked that the dimensionless equations have been used recently in \cite{Trabelsi_2018} for the analysis of heat and mass transfer in a Medium Density Fiberboard. However, this study focused only on one material. Thus, the present article proposes to extend the methodology to seven classes of building materials, for a total of nearly $50$ materials. The purpose of this study is twofold. First, it proposes a methodology to enhance and compare the dominant processes of a transfer among the materials. In this way, it is possible to identify materials which are more subject to heat or mass transfer or have more inertia. Thereby, the potential behavior of a material can be estimated without any further numerical or experimental studies. The second purpose is to use the dimensional analysis to determine kinetic, geometric and dynamic similarities. Such similarities are very useful for scaling up/down experimental designs and thus reduce time, space, etc.

The article is organized as follows. First, the physical model of heat and mass transfer is presented together with its dimensionless representation. In Section~\ref{sec:results_discussion}, the advantages of the method are presented for almost $50$ materials. In Section~\ref{sec:mapping_phenomena}, the results demonstrate how the dimensionless analysis is used to analyze the dominant processes of heat and mass transfer. Then, in Section~\ref{sec:similarity_analysis}, the use of similarity analysis is also shown to determine equal physical systems regarding the phenomena. The investigations are carried considering the nonlinear behavior of the material properties and multi-layered walls. Last, similitude laws are established to define an equivalent experimental configuration with a reduced time of investigations. 

\section{Physical model}

\subsection{Physical quantities in porous material}

The porous media is composed of a solid porous matrix and two species: water vapor and liquid water, denoted with index $i \egal 1$ and $i \egal 2\,$, respectively. The porous matrix is indexed by $i \egal 0\,$. The spatial domain is defined as $\boldsymbol{x} \, \in \, \Omega_{\,x}\,$.
The phenomena are investigated over the time domain $t \, \in \, \bigl[\,0\,,\,\tf \,\bigr] \,$. The important quantities are the temperature $T \ \unit{K}$ and the vapor pressure $P_{\,1} \ \unit{Pa}\,$. The relative humidity $\phi \ \unit{-}$ is related to the vapor pressure through the saturation pressure: 
\begin{align*}
\phi \egal \frac{P_{\,1}}{\Psat\,(\,T\,)}
\end{align*}
where the saturation pressure $\Psat  \ \unit{Pa}$ is given by a temperature $T \ \unit{K}$  dependent function according to \textsc{Antoine}'s law:
\begin{align*}
\Psat \bigl(\, T \,\bigr) \egal \Psat^{\,\circ} \cdot \Biggl(\, \frac{T \moins T_{\,a}}{T_{\,b}} \,\Biggr)^{\,\alpha} \,, && T \, \geqslant \, T_{\,a} \,, 
\end{align*}
with $\Psat^{\,\circ} \egal 997.3 \ \text{Pa} \,$, $T_{\,a} \egal 159.5 \ \text{K} \,$, $T_{\,b} \egal 120.6 \ \text{K}$ and $\alpha \egal 8.275 \,$.  The capillary pressure in the material is denoted as $P_{\,2}$ and the relation with the relative humidity is given by the \textsc{Kevin} equation:
\begin{align}
\label{eq:capillary_pressure}
P_{\,2} \egal \rho_{\,2} \cdot R_{\,1} \cdot T \cdot \log \bigl(\, \phi \,\bigr) \,,
\end{align}
where $\rho_{\,2} \ \unit{kg\,.\,m^{\,-3}}$ is the liquid water density, $R_{\,1} \ \unit{ J \,.\, kg^{\,-1} \,.\,K^{\,-1}}$ the gas constant of water vapor. Last, the moisture content is denoted by $\omega \ \unit{kg\,.\,m^{\,-3}}\,$.

\subsection{Heat and mass transfer in porous material}

Based on the original work of \textsc{Luikov} \cite{Luikov_1966}, the model of heat and mass transfer is described by a system of two coupled partial differential equations. From this work, several numerical models have been elaborated and implemented in software such as EnergyPlus \citep{Crawley_2001}, ESP-r \citep{Clarke_2013}, BSim \citep{Rode_2003} or Wufi \cite{Kunzel_2005}. An extended overview of its use and development for research and engineering purposes is given in \cite{Mendes_2017}. Here, the thermal equilibrium is presumed. In addition, the gradient of total pressure is assumed as negligible \cite{Busser_2017,Busser_2017c}. With this assumption, the heat and mass flows are driven only by diffusion processes. The advection processes, induced by air velocity through the porous matrix (\textsc{Darcy} law), are thus omitted. Note that this hypothesis required to be revised for innovative hygroscopic materials as detailed in \cite{Berger_2019d,Perre_2019}. The mass transfer is driven by liquid and vapor flows, while thermo-diffusion process is neglected:
\begin{align*}
\pd{\omega}{t} & \egal \div \biggl(\, k_{\,1} \cdot \grad P_{\,1} \plus k_{\,2} \cdot \grad P_{\,2} \,\biggr) \,.
\end{align*}
The chosen potential to write the equation is the vapor pressure. Thus, the capillary pressure can be decomposed as:
\begin{align*}
\partial P_{\,2} \egal \pd{P_{\,2}}{P_{\,1}} \cdot \partial P_{\,1} \plus \pd{P_{\,2}}{T} \cdot \partial T \,.
\end{align*}
Using the \textsc{Kelvin} equation~\eqref{eq:capillary_pressure} and the \textsc{Clausius}-\textsc{Clapeyron}'law, one obtains:
\begin{align*}
\pd{P_{\,2}}{P_{\,1}} \egal \frac{\rho_{\,2} \cdot R_{\,1} \cdot T}{P_{\,1}} \,, &&
\pd{P_{\,2}}{T} \egal \rho_{\,2} \cdot R_{\,1} \cdot \log \bigl(\, \phi \,\bigr) \moins \rho_{\,2} \cdot \frac{r_{\,12}\,(\,T\,)}{T} \,.
\end{align*}
Moreover, the differential of the moisture content can be extended as:
\begin{align*}
\partial \omega \egal \pd{\omega}{T} \cdot \partial T \plus \pd{\omega}{\phi} \cdot \partial \phi \,.
\end{align*}
It is assumed that the moisture content does not vary with temperature. Thus,
\begin{align*}
\partial \omega \egal \pd{\omega}{\phi} \cdot \biggl(\, 
\frac{1}{\Psat\,(\,T\,)} \cdot \partial P_{\,1} 
\moins \frac{P_{\,1}}{\Psat\,(\,T\,)^{\,2}} \cdot \pd{\Psat\,(\,T\,)}{T} \cdot \partial T \,\biggr) \,.
\end{align*}
Using the \textsc{Clausius}-\textsc{Clapeyron}'law the following coefficients can be defined:
\begin{align*}
c_{\,m} \, \eqdef \, \pd{\omega}{\phi} \cdot \frac{1}{\Psat\,(\,T\,)} \,, && 
c_{\,mq} \, \eqdef \, c_{\,m} \cdot \frac{P_{\,1} \cdot r_{\,12}\,(\,T\,)}{R_{\,1} \cdot T} \,.
\end{align*} 
Thus, one gets:
\begin{align*}
\partial \omega \egal c_{\,m} \cdot \partial P_{\,1} \moins c_{\,mq} \cdot \partial T \,.
\end{align*}
In the end, the mass balance equation is written as:
\begin{align}
\label{eq:mass_cons_equation}
c_{\,m} \cdot \pd{P_{\,1}}{t} 
\egal \div \, \bigl(\, k_{\,m} \cdot \grad \, P_{\,1} 
\plus k_{\,mq} \cdot \grad \, T \,\bigr) 
\plus c_{\,mq} \cdot \pd{T}{t} \,,
\end{align}
where the coefficients $k_{\,m}$ and $k_{\,mq}$ are given by:
\begin{align*}
k_{\,m} \, \eqdef \, k_{\,1} \plus k_{\,2} \cdot \pd{P_{\,2}}{P_{\,1}}  \,, &&
k_{\,mq} \, \eqdef \, k_{\,2} \cdot \pd{P_{\,2}}{T} \,.
\end{align*}

The second partial differential is obtained from the energy conservation law:
\begin{align}
\label{eq:heat_cons_equation}
c_{\,q} \cdot \pd{T}{t}
\egal \, \div \, \bigl(\, k_{\,q} \cdot \grad T \,\bigr)
\plus r_{\,12} \cdot \div \, \bigl(\, k_{\,qm} \cdot \grad P_{\,1} \,\bigr)\,.
\end{align}
where the heat flux is driven by sensible and latent phenomena. The coefficient $k_{\,qm} \ \unit{s}$ corresponds to the vapor permeability of the material $k_{\,qm} \, \eqdef \, k_{\,1} \,$, while $k_{\,q} \ \unit{W\,.\,m^{\,-1}\,.\,K^{\,-1}}$ to the thermal conductivity. The latent heat $r_{\,12} \ \unit{J \,.\, kg^{\,-1}}$ is temperature dependent according to:
\begin{align*}
r_{\,12}  \bigl(\, T \,\bigr) \, \eqdef \, r_{\,12\,,\,0}
\plus \Bigl(\, c_{\,1} \moins c_{\,2} \,\Bigr) \cdot \Bigl(\, T \moins T_{\,c} \, \Bigr) \,,
\end{align*}
where  $c_{\,1} \ \unit{J \,.\, kg^{\,-1} \,.\, K^{\,-1}}$ and $c_{\,2} \ \unit{J \,.\, kg^{\,-1} \,.\, K^{\,-1}}$ are the specific heat of vapor and liquid water, respectively. The numerical values for the physical constants are $ r_{\,12\,,\,0} \egal 2.5\e{6} \ \mathsf{J\,.\,kg^{\,-1}}\,$, $c_{\,1} \egal 1870 \ \mathsf{J \,.\, kg^{\,-1} \,.\, K^{\,-1}}\,$, $c_{\,2} \egal 4180 \ \mathsf{J \,.\, kg^{\,-1} \,.\, K^{\,-1}}$ and $T_{\,c} \egal 273.15 \ \mathsf{K}\,$. The heat storage coefficient $c_{\,q}$ is defined as follows:
\begin{align*}
c_{\,q} \, \eqdef \, \rho_{\,0} \cdot c_{\,0} \plus c_{\,2} \cdot \omega \,,
\end{align*}
where
$\rho_{\,0} \ \unit{kg \,.\, m^{\,-3}}$ is the dry heat density and $c_{\,0} \ \unit{J \,.\, kg^{\,-1} \,.\, K^{\,-1}}$ is the specific heat of the material.

In terms of boundary conditions, at the interface between the porous material and the ambient air, the equality of the moisture and heat flows are stated:
\begin{align*}
k_{\,m} \cdot \pd{P_{\,1}}{n} 
\plus k_{\,mq} \cdot \pd{T}{n}  & 
\egal h_{\,m} \cdot \bigl(\, P_{\,1} \moins P_{\,1}^{\,\infty}\,\bigr) \,, \\[4pt]
k_{\,q} \cdot \pd{T}{n} \plus r_{\,12} \cdot k_{\,qm} \cdot \pd{P_{\,1}}{n} 
& \egal   h_{\,q} \cdot \bigl(\, T \moins T^{\,\infty}\,\bigr)
\plus r_{\,12} \cdot h_{\,m} \cdot \bigl(\, P_{\,1} \moins P_{\,1}^{\,\infty}\,\bigr) \,,
\end{align*}
where $h_{\,m} \ \unit{s\,.m^{\,-1}}$ and $h_{\,q} \ \unit{W\,.\,m^{\,-1}\,.\,K^{\,-1}}$ are the surface transfer coefficients. The ambient air fields are denoted by $ P_{\,1}^{\,\infty}$ and $T^{\,\infty}\,$. It can be remarked that the liquid flows due to wind driven rains or the heat flux due to radiation transfer are not taken into account in this model. The operator $\pd{y}{n}$ represents the projection $\grad \, y \cdot \boldsymbol{n}$ where $\boldsymbol{n}$ is the surface normal. For multi-layer walls, at the interface between two materials, the continuity of the heat and moisture flows is assumed. Furthermore, the temperature and vapor pressure are continuous at the interface. As initial conditions, the temperature and vapor pressure are given by:
\begin{align*}
T \egal T^{\,0} \,, \qquad P \egal P^{\,0} \,, \qquad t \egal 0 \,, 
\end{align*}
where $T^{\,0}$ and $P^{\,0}$ are given functions that can depend on space.

\subsection{Porous material properties}
\label{sec:porous_mat_properties}

Several material properties are introduced. First, the so-called sorption curve gives the relation between the total amount of water $\omega \ \unit{kg\,.\,m^{\,-3}}$ as a function of the relative humidity $\phi \ \unit{-}\,$. In this work, the \textsc{Oswin} model is adopted \cite{Oswin_1946} for the sorption curve:
\begin{align}
\label{eq:sorption_model}
\omega \egal \fosw \,\bigl(\,\phi\,\bigr) \egal \omega_{\,1} \cdot \Biggl(\, \frac{\phi}{1 \moins \phi}\,\Biggr)^{\,\alpha_{\,1}} \,,
\end{align} 
where $\omega_{\,1}  \ \unit{kg\,.\,m^{\,-3}} $ and $\alpha_{\,1}  \ \unit{-}$ are defined parameters depending on the material. The derivative of the sorption curve according to the relative humidity is given by:
\begin{align*}
\pd{\omega}{\phi} \egal \pd{\fosw}{\phi} \egal \frac{\alpha_{\,1}}{\bigl(\, 1 \moins \phi \,\bigr) \cdot \phi} \cdot \fosw \,\bigl(\,\phi\,\bigr)\,.
\end{align*}

The vapor resistance factor $\mu \ \unit{-}$ is another important property of the material. It is defined as the ratio between the air permeability and the vapor permeability $k_{\,1} \ \unit{s}$  of the material \cite{Casnedi_2020}. According to \cite{Kunzel_1996}, the vapor permeability can be related to $\mu$ using the following relation:
\begin{align*}
k_{\,1} \egal \frac{2 \cdot 10^{\,-7} \cdot T^{\,0.81}}{\mu \cdot P_{\,3}} \,,
\end{align*}
where $P_{\,3} \egal 10^{\,5} \ \mathsf{Pa}$ is the air pressure assumed as constant. The liquid permeability $k_{\,2} \ \unit{s}$ of the material is given by:
\begin{align*}
k_{\,2} \egal D_{\,2} \cdot \pd{\omega}{P_{\,2}} \,,
\end{align*}
where $D_{\,2} \ \unit{m^{\,2}\,.\,s^{\,-1}}$ is the liquid transport coefficient given by the following expression:
\begin{align*}
D_{\,2} \egal 3.8 \cdot \biggl(\, \frac{A}{\omega_{\,f}} \biggr)^{\,2} \cdot 
\exp \Biggl(\, \frac{3}{\omega_{\,f}} \cdot \ln (\, 10 \,) \cdot \biggl(\, \omega \moins \omega_{\,f} \,\biggr) \,\Biggr) \,,
\end{align*}
where $\omega_{\,f} \ \unit{kg\,.\,m^{\,-3}}$ is the free water content and $A \ \unit{kg\,.\,m^{\,-2}\,.\,s^{\,-0.5}}$ is the water adsorption coefficient. Using the equation~\eqref{eq:capillary_pressure} of the capillary pressure $P_{\,2}$ and the sorption equation~\eqref{eq:sorption_model}, the liquid permeability can be expressed as:
\begin{align*}
k_{\,2} \egal D_{\,2} \cdot \pd{\phi}{P_{\,2}} \cdot \pd{\omega}{\phi} 
\egal D_{\,2}  \cdot \frac{\phi}{\rho_{\,2} \cdot R_{\,1} \cdot \Psat\,(\,T\,)} \cdot \pd{\omega}{\phi} \,.
\end{align*}

Last, the dry density and specific heat of the material are denoted by $c_{\,0} \ \unit{J\,.\,kg^{\,-1}\,.\,K^{\,-1}}$ and $\rho_{\,0} \ \unit{kg\,.\,m^{\,-3}}\,$, respectively. These parameters are independent of vapor pressure or temperature. The thermal conductivity $k_{\,q} \ \unit{W\,.\,m^{\,-1}\,.\,K^{\,-1}}$ depends on the water content:
\begin{align}
\label{eq:thermal_conductivity_model}
k_{\,q} \egal k_{\,q\,,\,0} \plus \beta \cdot \omega \,,
\end{align}
where $k_{\,q\,,\,0}$ is the  thermal conductivity at the dry state. As a synthesis, $9$ material properties are required for the model: the dry density $\rho_{\,0}\,$, the specific heat $c_{\,0}\,$, the vapor resistant factor $\mu\,$, the two coefficients of the \textsc{Oswin} sorption curve $\alpha_{\,1}$ and $\omega_{\,1} \,$, the two coefficients of liquid transport $\omega_{\,f}$ and $A\,$, the dry thermal conductivity $k_{\,q\,,\,0}$ and its dependency with moisture content through the coefficient $\beta\,$.

\subsection{Relation with the coefficients of heat and mass conservation equations}

The relation between the coefficients involved in the heat and mass conservation equations~\eqref{eq:mass_cons_equation} and \eqref{eq:heat_cons_equation} with the material properties is now introduced. The mathematical formalism is used to underline the dependency of the coefficients with the field of interests, \emph{i.e.} temperature and vapor pressure. The heat and mass capacity coefficients are defined by:
\begin{align*}
c_{\,q} \,:\, \bigl(\,T\,,\,P_{\,1}\,\bigr) & \longmapsto \ \rho_{\,0} \cdot c_{\,0} \plus c_{\,2} \cdot \fosw \,\bigl(\,T\,,\,P_{\,1}\,\bigr) \,,\\[4pt]
c_{\,m} \,:\, \bigl(\,T\,,\,P_{\,1}\,\bigr) & \longmapsto \ \pd{\fosw\,\bigl(\,T\,,\,P_{\,1}\,\bigr)}{\phi} \cdot \frac{1}{\Psat\,(\,T\,)} \,.
\end{align*}
The coupling capacity coefficient is given by:
\begin{align*}
c_{\,mq} \,:\, \bigl(\,T\,,\,P_{\,1}\,\bigr) \ \longmapsto \ c_{\,m}\,\bigl(\,T\,,\,P_{\,1}\,\bigr) \cdot \frac{r_{\,12}\,(\,T\,)}{R_{\,1} \cdot T^{\,2}} \cdot P_{\,1} \,.
\end{align*}
The heat and mass permeability coefficients are expressed as:
\begin{align*}
k_{\,m} \,:\, \bigl(\,T\,,\,P_{\,1}\,\bigr) & \longmapsto \ k_{\,1}\,\big(\,T\,\bigr) \plus k_{\,2}\,\bigl(\,T\,,\,P_{\,1}\,\bigr) \cdot \frac{\rho_{\,2} \cdot R_{\,1} \cdot T}{P_{\,1}} \,, \\[4pt]
k_{\,q} \,:\, \bigl(\,T\,,\,P_{\,1}\,\bigr) & \longmapsto \ k_{\,q\,,\,0} \plus \beta \cdot \fosw \,\bigl(\,T\,,\,P_{\,1}\,\bigr) \,.
\end{align*}
The coupling coefficient $k_{\,qm}$ considering the influence of the latent heat transfer on the heat conservation equation is:
\begin{align*}
k_{\,qm}  \,:\, T & \longmapsto \ k_{\,1}\,\big(\,T\,\bigr) \,.
\end{align*}
The other coupling coefficient $k_{\,mq}$ representing the moisture migration under temperature gradient  is written as:
\begin{align*}
k_{\,mq} \,:\, \bigl(\,T\,,\,P_{\,1}\,\bigr) & \longmapsto \ 
k_{\,2} \,\bigl(\,T\,,\,P_{\,1}\,\bigr) \cdot \Biggl(\, 
\rho_{\,2} \cdot R_{\,1} \cdot \log \biggl(\, \frac{P_{\,1}}{\Psat\,(\,T\,)} \,\biggr) 
\moins \rho_{\,2} \cdot \frac{r_{\,12}\,(\,T\,)}{T} \,\Biggr) \,.
\end{align*}
In the end, it can be remarked that all coefficients are depending at least on one field and the system of partial differential equations~\eqref{eq:mass_cons_equation} and \eqref{eq:heat_cons_equation} is nonlinear.

\subsection{Dimensionless representation}

To propose a material properties classification, the dimensionless representation of the heat and mass transfer equations is presented according to \cite{Luikov_1966,Mikhailov_1984}. For this, the vapor pressure and temperature are scaled according to:
\begin{align*}
u \egal \frac{P_{\,1} \moins P_{\,1}^{\,\rf}}{\deltaP} \,,
&& v \egal \frac{T \moins T^{\,\rf}}{\deltaT} \,,
\end{align*}
where $T^{\,\rf}$ and $P_{\,1}^{\,\rf}$ are reference temperature and vapor pressure values, respectively. The quantities $\deltaT$ and $\deltaP$ are denoted for a temperature and vapor pressure differences. The choice of those parameters will be discussed in the next sections. In a similar way, the space and time domains are transformed into a dimensionless representation:
\begin{align*}
\boldsymbol{\chi} \egal \frac{\boldsymbol{x}}{L^{\,\rf}} \,,
&& \tau \egal \frac{t}{t^{\,\rf}} \,.
\end{align*}
The heat and mass transfer coefficients are changed:
\begin{align*}
c_{\,q}^{\,\star} & \egal \frac{c_{\,q}}{c_{\,q}^{\,\rf}} \,, &&
c_{\,m}^{\,\star} \egal \frac{c_{\,m}}{c_{\,m}^{\,\rf}} \,, &&
c_{\,mq}^{\,\star} \egal \frac{c_{\,mq}}{c_{\,mq}^{\,\rf}} \,, &&
r_{\,12}^{\,\star} \egal \frac{r_{\,12}}{r_{\,12}^{\,\rf}} \\
k_{\,q}^{\,\star} & \egal \frac{k_{\,q}}{k_{\,q}^{\,\rf}} \,, &&
k_{\,m}^{\,\star} \egal \frac{k_{\,m}}{k_{\,m}^{\,\rf}} \,, && 
k_{\,mq}^{\,\star} \egal \frac{k_{\,mq}}{k_{\,mq}^{\,\rf}} \,, &&
k_{\,qm}^{\,\star} \egal \frac{k_{\,qm}}{k_{\,qm}^{\,\rf}} \,.
\end{align*}
In this way, several dimensionless numbers are introduced. First, the \textsc{Fourier} numbers represent the ease of heat or mass transfer inside the material:
\begin{align*}
\Fo^{\,q} & \  \eqdef \ \frac{k_{\,q}^{\,\rf} \cdot t^{\,\rf}}{c_{\,q}^{\,\rf} \cdot (\,L^{\,\rf}\,)^{\,2}}  \,,
&& \Fo^{\,m}\ \eqdef \ \frac{k_{\,m}^{\,\rf} \cdot t^{\,\rf}}{c_{\,m}^{\,\rf} \cdot (\,L^{\,\rf}\,)^{\,2}} \,.
\end{align*}
The \textsc{Fourier} numbers represent the part of the heat/mass flux that is transmitted through the material in relation to the heat/mass stored. In other words, it is the ratio of the diffusive transport to the rate of storage. The greater $\Fo$ is, the easier the heat/mass flows inside the material during a given time. Thus, the overall dynamics of heat and mass transfer is faster. Proportionally, more heat/mass is stored in the material for low $\Fo \,$. Note that the \textsc{Darcy} number does not appear here since the transfer by convection has been neglected in the mathematical model \cite{ArticleC,ArticleD,ArticleE}.

Three other numbers are defined to describe the influence of the heat or mass transfer on each other:
\begin{align*}
\eta & \ \eqdef \ \frac{c_{\,mq}^{\,\rf} \cdot \deltaT}{c_{\,m}^{\,\rf} \cdot \deltaP} \,, &&
\gamma \  \eqdef \  \frac{k_{\,qm}^{\,\rf} \cdot \deltaP}{k_{\,q}^{\,\rf} \cdot \deltaT} \cdot r_{\,12}^{\,\rf} \,, &&
\delta \  \eqdef \ \frac{k_{\,mq}^{\,\rf} \cdot \deltaT}{k_{\,m}^{\,\rf} \cdot \deltaP} \,.
\end{align*}
As it can be remarked in Eq.~\eqref{eq:dimless_heat_and_mass} below, the numbers $\delta$ and $\gamma$ modify the \textsc{Fourier}  numbers of mass and heat transfer, respectively. Thus, $\delta$ describes the influence of heat transfer on mass transfer. If $\delta \, < \, 1\,$, the impact of heat transfer on mass transfer is minor compared to the mass transfer driven by vapor pressure. Similar analysis can be carried with $\gamma$ and the heat transfer equation. The number $\eta$ is a coupling parameter for the dynamics of temperature over mass transfer. It arises directly from the perfect gas law and the choice of vapor pressure as potential. 

For the boundary conditions, the \textsc{Biot} number is introduced for the heat and mass transfer at the interface between the air and the porous material:
\begin{align*}
\Bi^{\,q} \ \eqdef \ \frac{h_{\,q}^{\,\rf} \cdot L^{\,\rf}}{k_{\,q}^{\,\rf}} \,, &&
\Bi^{\,qm} \ \eqdef \ \frac{h_{\,m}^{\,\rf} \cdot L^{\,\rf}}{k_{\,qm}^{\,\rf}} \,, &&
\Bi^{\,m} \  \eqdef \ \frac{h_{\,m}^{\,\rf} \cdot L^{\,\rf}}{k_{\,m}^{\,\rf}} \,.
\end{align*}
The \textsc{Biot} numbers translate the ratio of heat/mass transfer resistance inside of a material and at the interface with the air. It quantifies whether the field vary significantly when a drop of temperature or vapor pressure occurs in the ambient air. For instance, when $\Bi \, > \, 1\,$, the diffusion inside the material is slower than the heat  arising at the interface. As a consequence, important gradient may occur if combined with a small \textsc{Fourier} number. The greater the \textsc{Biot} number, the more exchanges happen at the interface between the air and the material. Note that this number can be referenced as the \textsc{Nusselt} number in fluid mechanics.

In the end, the dimensionless representation of the model for heat and mass transfer is:
\begin{subequations}
\label{eq:dimless_heat_and_mass}
\begin{align}
c_{\,m}^{\,\star} \cdot \pd{u}{\tau} 
& \egal \Fo^{\,m} \cdot \div \, \bigl(\, k_{\,m}^{\,\star} \cdot \grad \, u \,\bigr)
\plus \Fo^{\,m} \cdot \delta \cdot \div \, \bigl(\, k_{\,mq}^{\,\star} \cdot \grad \, v \,\bigr)
\moins \eta \cdot c_{\,mq}^{\,\star} \cdot \pd{v}{\tau} \,, \\[4pt]
c_{\,q}^{\,\star} \cdot \pd{v}{\tau}
& \egal \Fo^{\,q} \cdot \div \, \bigl(\, k_{\,q}^{\,\star} \cdot  \grad v \,\bigr)
\plus \Fo^{\,q} \cdot \gamma \cdot r_{\,12}^{\,\star} \cdot \div \, \bigl(\, k_{\,qm}^{\,\star} \cdot  \grad u \,\bigr) \,.
\end{align}
\end{subequations}
The boundary conditions are:
\begin{subequations}
\label{eq:dimless_heat_and_mass_BC}
\begin{align}
k_{\,m}^{\,\star} \cdot \pd{u}{n} \plus \delta \cdot k_{\,mq}^{\,\star} \cdot \pd{v}{n} 
& \egal \Bi^{\,m} \cdot \bigl(\, u \moins u^{\,\infty} \,\bigr) \,, \\[4pt]
k_{\,q}^{\,\star} \cdot \pd{v}{n} \plus \gamma \cdot r_{\,12}^{\,\star} \cdot k_{\,qm}^{\,\star} \cdot \pd{u}{n} 
& \egal \Bi^{\,q} \cdot \bigl(\, v \moins v^{\,\infty} \,\bigr) 
\plus \gamma \cdot r_{\,12}^{\,\star} \cdot \Bi^{\,qm} \cdot \bigl(\, u \moins u^{\,\infty} \,\bigr) \,.
\end{align}
\end{subequations}
It is important to remark that in Eqs.~\eqref{eq:dimless_heat_and_mass} and \eqref{eq:dimless_heat_and_mass_BC}, the space operators, such as divergence and gradient, are expressed according to the dimensionless space variable $\chi\,$.

\section{Results and discussion}
\label{sec:results_discussion}

First of all, the material properties and the reference conditions are described. Then, the two advantages of the dimensionless approach are presented: \emph{(i)} the physical understanding of the main phenomena of heat and mass transfer and \emph{(ii)} the use of similitude laws for practical applications.

\subsection{Materials and reference conditions}

To compute the dimensionless numbers and their nonlinear distortion, it is required to set the reference parameters. This choice has to be suitable to get the dimensionless variables scaling with the unity $\mathcal{O}(\,1\,)\,$. The usual time scale in building physics is the hour, so the reference time is set as $t_{\,0} \egal 1 \ \mathsf{h}\,$. The reference length of materials varies according to their type. Since different analysis will be carried out in the next sections, the reference length will be specified thereafter. In addition, the presence of multiple scales and directional scaling are disregarded for the sake of simplicity. Here, the material reference length corresponds to its thickness used in a wall configuration. In buildings, the admitted temperature range is $\bigl[\,5 \,,\, 35 \, \bigr] \ ^{\circ}\mathsf{C}\,$ \cite{Trabelsi_2018}. The relative humidity range is defined as $\bigl[\, 0.05 \,,\, 0.90 \, \bigr]\,$. By choosing $P_{\,1}^{\,\rf} \egal 43.572 \ \mathsf{Pa}\,$, $\deltaP \egal 5021.5 \ \mathsf{Pa}\,$, $T^{\,\rf} \egal 278.15 \ \mathsf{K}$ and $\deltaT \egal 30 \ \mathsf{K}\,$, the dimensionless variables $u$ and $v$ are both in the range $\bigl[\, 0 \,,\, 1\,\bigr]\,$. 

The material properties are taken from the MASEA database in \texttt{WUFI} \cite{Wufi_database}. The dimensionless analysis is carried out for $49$ materials, classified in seven categories as illustrated in Figure~\ref{fig:Lref}. An identification number is also defined for each material. Since, discrete data are given for the sorption curve $\omega$ as a function of the relative humidity and for the thermal conductivity $k_{\,q}$ as a function of the moisture content, a fitting procedure is realized to estimate the coefficients of the \textsc{Oswin} model, Eq.~\eqref{eq:sorption_model}, and the thermal conductivity function, Eq.~\eqref{eq:thermal_conductivity_model}. All coefficients used in the definition of the material properties are synthesized in Tables~\ref{tab:mat_properties_1} and \ref{tab:mat_properties_2} in Appendix~\ref{sec:mat_properties}. Then, the reference coefficient values are defined as follows:
\begin{align*}
c_{\,m}^{\,\rf} & = \, c_{\,m}\bigl(\,T^{\,\circ} \,,\, P_{\,1}^{\,\circ}\,\bigr)  \,, &&
k_{\,m}^{\,\rf} \egal k_{\,m}\bigl(\,T^{\,\circ}\,,\,P_{\,1}^{\,\circ}\,\bigr)  \,, &&
k_{\,mq}^{\,\rf} \egal k_{\,mq}\bigl(\,T^{\,\circ}\,,\,P_{\,1}^{\,\circ}\,\bigr)  \,, && 
c_{\,mq}^{\,\rf} \egal c_{\,mq}\bigl(\,T^{\,\circ}\,,\,P_{\,1}^{\,\circ}\,\bigr)  \,, && \\[4pt]
c_{\,q}^{\,\rf} & = \, c_{\,q}\bigl(\,T^{\,\circ}\,,\,P_{\,1}^{\,\circ}\,\bigr)  \,, &&
k_{\,q}^{\,\rf} \egal k_{\,q}\bigl(\,T^{\,\circ}\,,\,P_{\,1}^{\,\circ}\,\bigr)  \,, &&
k_{\,qm}^{\,\rf} \egal k_{\,qm}\bigl(\,T^{\,\circ}\,\bigr)  \,, &&
r_{\,12}^{\,\rf} \egal r_{\,12\,,\,0} \,, && 
\end{align*}
where $T^{\,\circ} \egal 293.15 \ \mathsf{K}$ and $P_{\,1}^{\,\circ} \egal 1166.9 \ \mathsf{Pa}\,$, which corresponds to a relative humidity $\phi \egal 0.5\,$. This choice permits to compute the dimensionless number for standard conditions in buildings and analyze the behavior of the material in such references. For the boundary conditions, two reference surface transfer coefficients are chosen. The first one stands for an interface between the material and the outside air: 
\begin{align*}
h_{\,m}^{\,\rf\,,\,o} \egal 10^{\,-7} \ \mathsf{s\,.m^{\,-1}} \,, && h_{\,q}^{\,\rf\,,\,o} \egal 15 \ \mathsf{W\,.\,m^{\,-2}\,.\,K^{\,-1}}\,.
\end{align*}
The second set of reference values is defined for an interface between the material and the inside air:
\begin{align*}
h_{\,m}^{\,\rf\,,\,i} \egal 5 \e{-9} \ \mathsf{s\,.m^{\,-1}}  \,, && h_{\,q}^{\,\rf\,,\,i} \egal 5 \ \mathsf{W\,.\,m^{\,-2}\,.\,K^{\,-1}}\,.
\end{align*}

\subsection{Mapping the main phenomena}
\label{sec:mapping_phenomena}

\subsubsection{Comparison between material categories}
\label{sec:range_dimensionless_numbers}

\begin{figure}
\centering
\subfigure[\label{fig:FoM}]{\includegraphics[width=.45\textwidth]{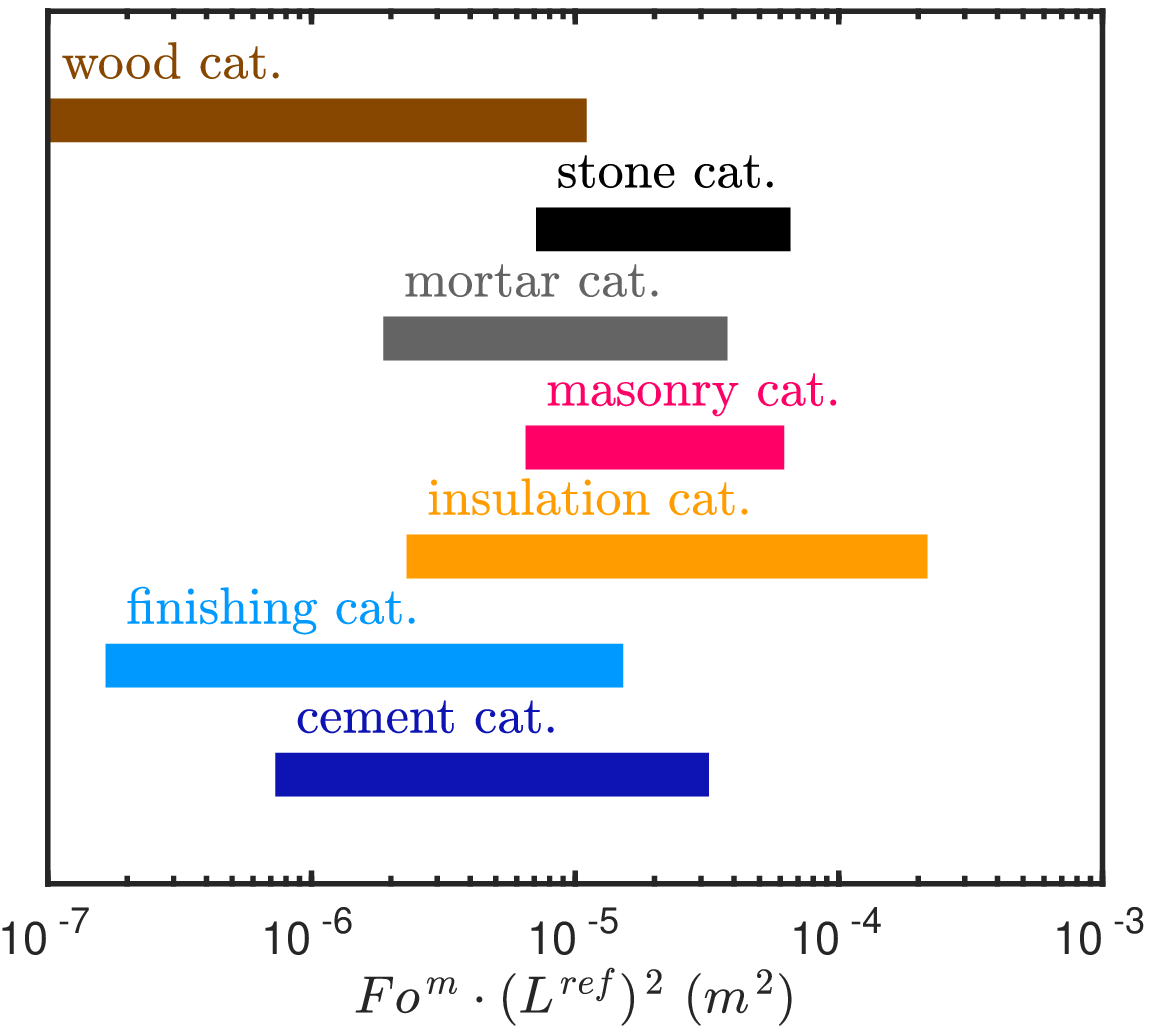}} \hspace{0.2cm}
\subfigure[\label{fig:FoQ}]{\includegraphics[width=.45\textwidth]{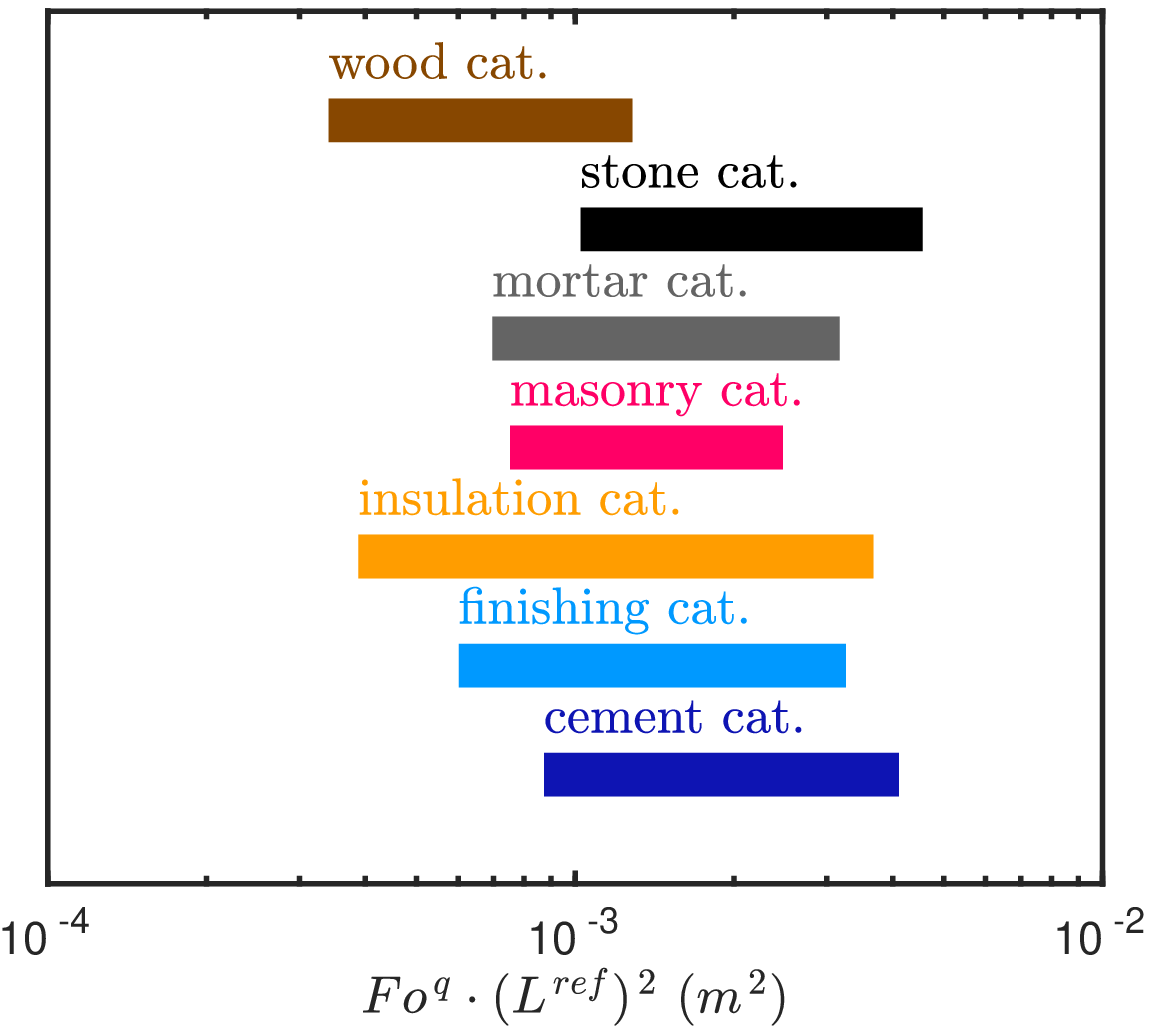}} \\
\subfigure[\label{fig:delta}]{\includegraphics[width=.45\textwidth]{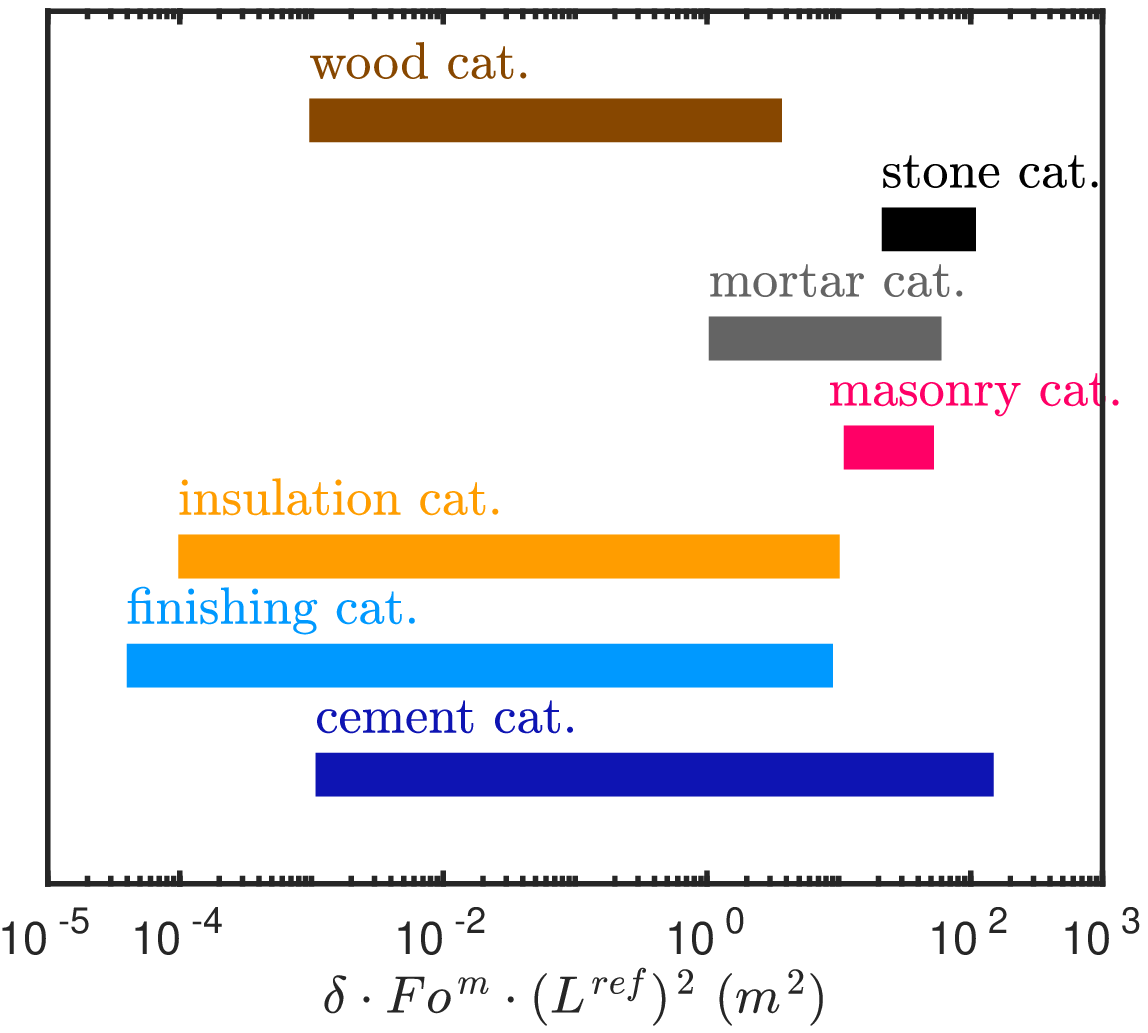}} \hspace{0.2cm}
\subfigure[\label{fig:gamma}]{\includegraphics[width=.45\textwidth]{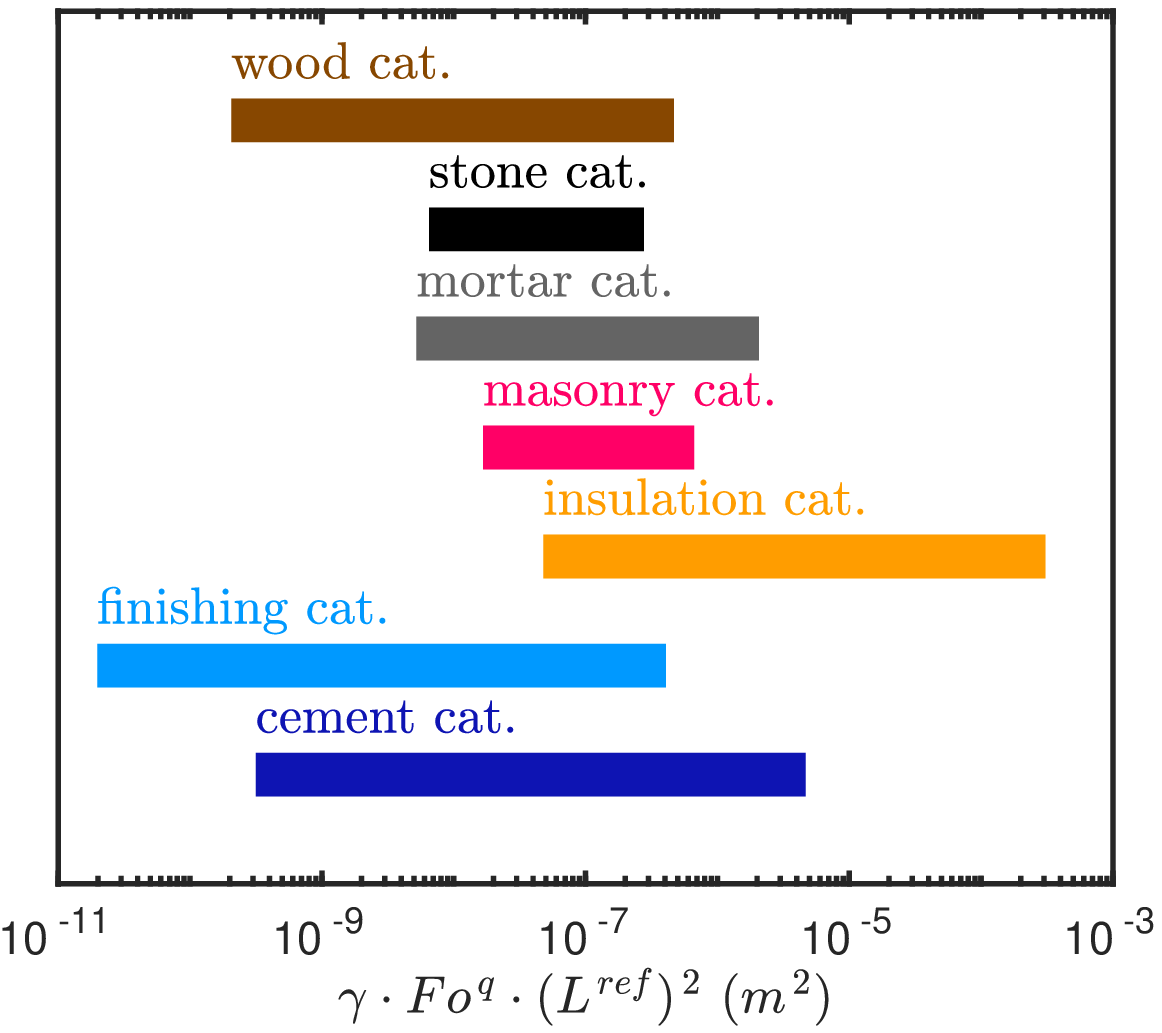}} \\
\subfigure[\label{fig:BiM}]{\includegraphics[width=.45\textwidth]{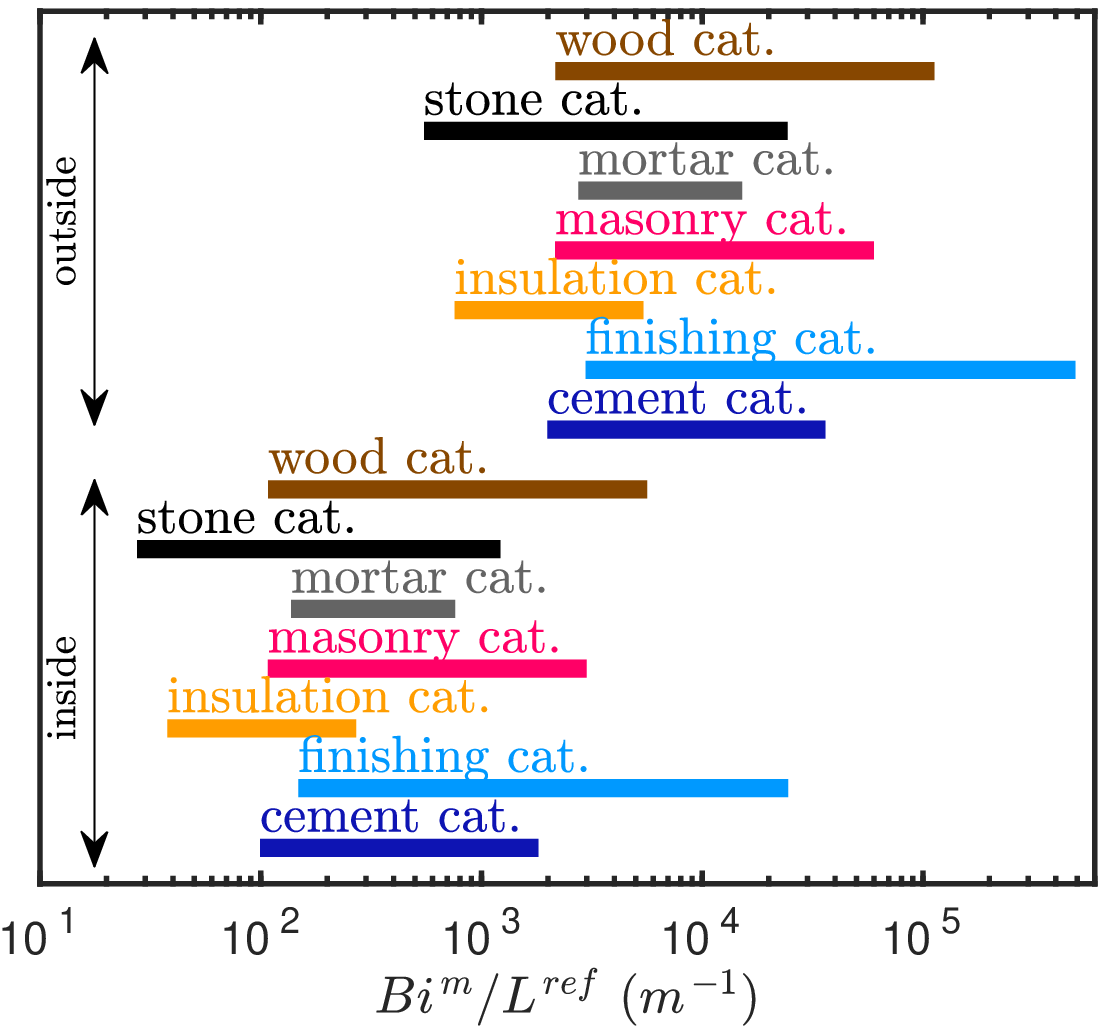}} \hspace{0.2cm}
\subfigure[\label{fig:BiQ}]{\includegraphics[width=.45\textwidth]{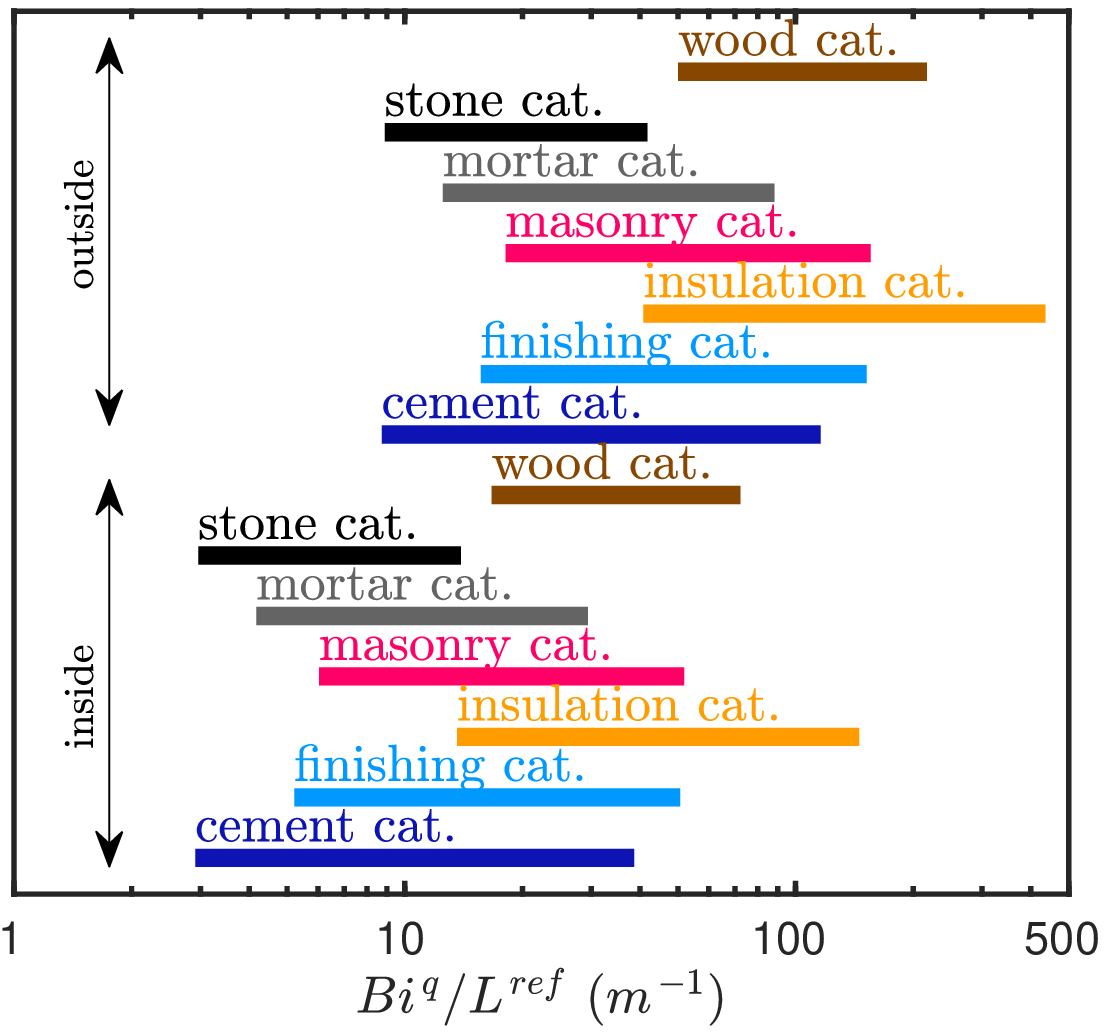}} 
\caption{Variation of the dimensionless numbers according to materials categories.}
\label{fig:Dim_number_Lfixed}
\end{figure}

The reference length of the materials is set as a constant equal for all materials. A first analysis is carried to evaluate the order of magnitudes of the dimensionless numbers for the different categories of materials, as presented in Figure~\ref{fig:Dim_number_Lfixed}. For the moment, it is not required to set a value to the reference length since the figures are presented depending on this value. 

As a first approximation, it can be stated that the heat and mass transfer are roughly driven by $\Fo^{\,q}$ and $\Fo^{\,m}\,$, respectively. The mass \textsc{Fourier} number, presented in Figure~\ref{fig:FoM}, varies strongly according to the different categories of a material. It means that each material category has its own behavior. For instance, as the $\Fo^{\,m}$ is greater for masonry and stone categories than for wood or finishing ones, it implies that the dynamics of mass transfer is faster in stones and masonry elements. In addition, almost all the bars are spread out. Thus, some materials belonging to the same category can have very different behaviors from each other such as wood or insulation for instance. It demonstrates the important variability between materials regarding mass transfer. 

Looking at Figure~\ref{fig:FoQ}, it can be assumed that all the materials have similar behavior. Nevertheless, it can be remarked that the heat diffusion is the slowest for the wood and insulation categories. The heat \textsc{Fourier} numbers are less scattered than the mass ones. It shows that the materials are less different from each other regarding heat transfer. In Figure~\ref{fig:FoM} and \ref{fig:FoQ}, the mass and heat transfer is easier for the materials located on the right side of the plots. On the contrary, the mass or heat transfer is the most difficult, with slower dynamics, for materials on the left side of the plot. Finally, by comparing Figures~\ref{fig:FoM} and ~\ref{fig:FoQ}, for the given reference conditions, the heat transfer is around thousand times faster than mass one for most categories excepting the masonry and stone. This time scale information can be relevant when assessing the contribution of separate phenomena on a general behavior.

One one hand, the he variations of the dimensionless number $\delta$ are illustrated in Figure~\ref{fig:delta}. It investigates the coupling effects and more precisely the impact of heat transfer on the mass one. There are some important differences within the finishing, insulation or cement categories. The coupling parameter $\delta$ varies around six orders of magnitude. It indicates that the behavior of the materials regarding this phenomena is very fluctuating for these categories. For the stone category, the parameter scales with $\mathcal{O}(\,10^{\,2}\,)\,$, revealing an important impact of heat transfer on the mass one. The analysis is similar for the masonry, but the impact is lower since the parameter magnitude is around $\mathcal{O}(\,10^{\,1}\,)\,$. On the other hand, Figure~\ref{fig:gamma} shows the dimensionless parameter $\gamma\,$, transcribing the impact of mass transfer on the heat one. Here, important variations are noted for the finishing and cement categories. The insulation category is the one where mass transfer have the greatest impact on the heat one. Moreover, for this category, the speed of transfer is similar for latent and sensible heat. These results highlight that dimensionless numbers provides analysis of the dominant phenomena for both heat and mass transfer. It is particularly interesting for insulation materials where mass transfer has a non-negligible impact on heat transfer through latent phenomena. The number $\gamma$ can be a useful indicator to assess the impact of mass transfer in insulation material and thus be a guarantee of their thermal efficiency.

Last, the penetration of heat and mass transfer at the interface between the material and the outside or inside air can be analyzed in Figures~\ref{fig:BiM} and \ref{fig:BiQ} through the \textsc{Biot} numbers. The materials located on the left side of the plots are characterized by the lowest \textsc{Biot} numbers, describing the slowest penetration of heat/mass transfer. On the opposite, the materials located on the most right part of the plots are characterized by the fastest penetration of heat/mass transfer. If those materials have a low \textsc{Fourier} then high gradients can occur. As shown in Figure~\ref{fig:BiM}, the finishing materials have one of the highest \textsc{Biot} numbers. They also have a relatively small mass \textsc{Fourier} number. It explains how those materials can be used to protect from intense moisture content other material placed behind them.
Moreover, the \textsc{Biot} number of all studied materials is higher than one, confirming the hypothesis of a non-negligible gradient of temperature/vapour pressure within it for the given reference conditions. As expected, the penetration of heat and mass transfer is more important on the outside interface than on the inside one. The penetration of mass transfer is the lowest and highest for the insulation and finishing categories, respectively. The stone category has the slowest heat \textsc{Biot} numbers $\Bi^{\,q}\,$, revealing a slow penetration of the heat transfer. Conversely, important heat passes through the interface for the insulation category. In other words, for such insulation materials, the thermal resistance is higher in the material than at its edges.

\subsubsection{Comparison between heat and mass transfer for each material}
\label{sec:comparison_materials}

\begin{figure}
\centering
\includegraphics[width=.95\textwidth]{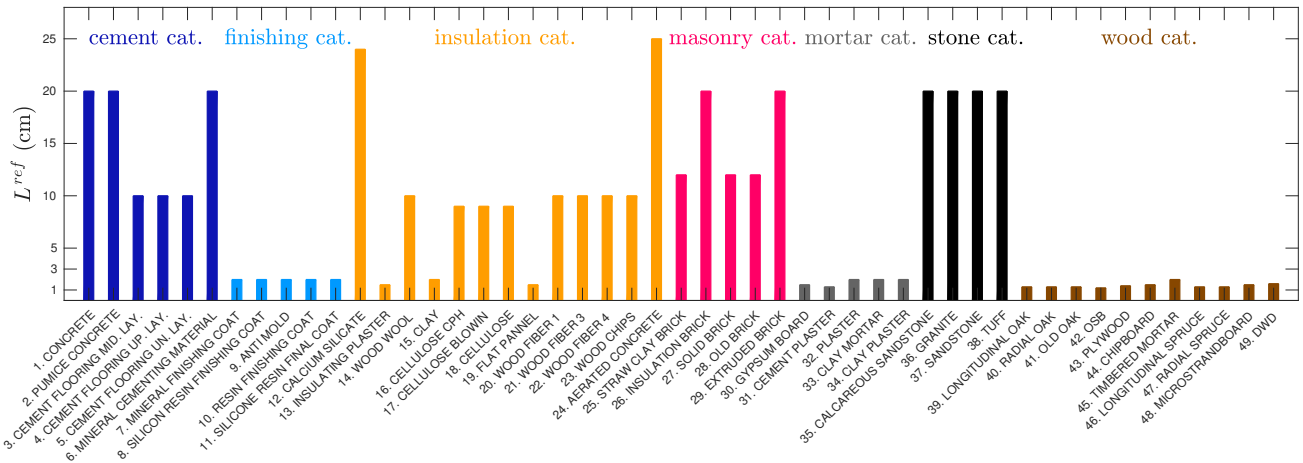}
\caption{Reference length for each material.}
\label{fig:Lref}
\end{figure}

\begin{figure}
\centering
\includegraphics[width=.95\textwidth]{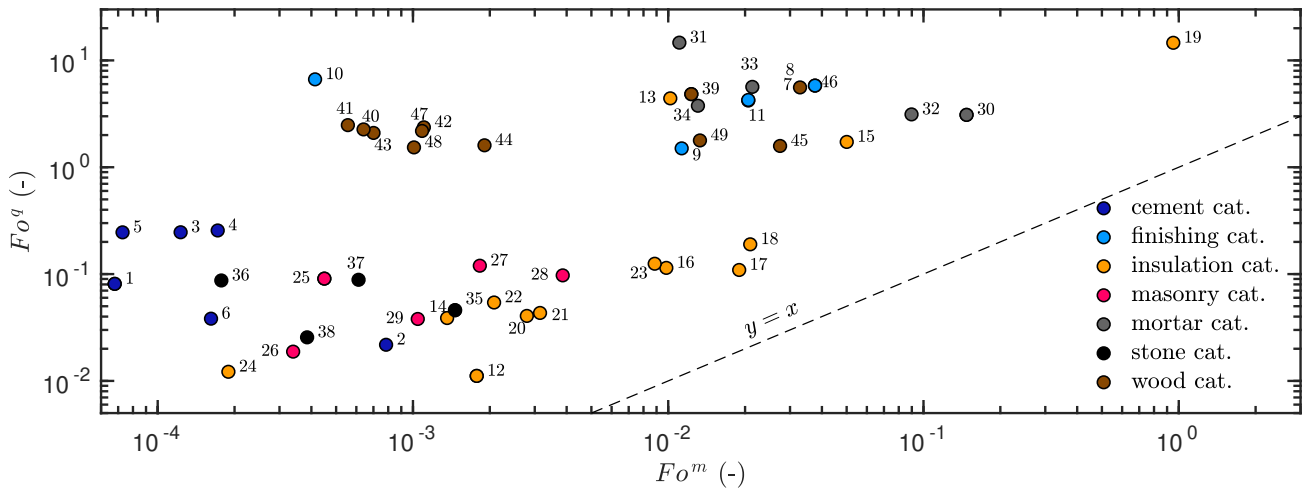}
\caption{Comparison of the competition between heat and mass transfer by analyzing their \textsc{Fourier} numbers.}
\label{fig:FoQvsFoM}
\end{figure}

To carry out a comparison of the heat and mass transfer properties between the materials, a specified reference length $L_{\,0}$ is defined for each material according to Figure~\ref{fig:Lref}. Figure~\ref{fig:FoQvsFoM} gives the heat \textsc{Fourier} number according to the mass one. It analyzes the competition between the heat and mass transfer within the materials. The dotted line in Figure~\ref{fig:FoQvsFoM} represents the case where heat transfer is as important as mass one. Thus, as all the materials are above this line, it can be noted that the heat transfer is always more important than the mass one. The wood materials have a similar behavior regarding heat transfer with a heat \textsc{Fourier} number $\Fo^{\,q}$ scaling with $\mathcal{O}(\,1\,)\,$. However, they have strongly different behaviors for mass transfer since the mass \textsc{Fourier} number $\Fo^{\,m}$ varies by two orders in the range $\bigl[\, 5 \cdot 10^{\,-4} \,,\, 5 \cdot 10^{\,-2} \,\bigr]\,$. Similar conclusions can be observed for most of the insulation materials. Their heat \textsc{Fourier} number is of the order $\mathcal{O}(\,10^{\,-1}\,)$, indicating a slow heat transfer process in the material. However, very different nature arises for the mass transfer. Both heat and mass transfers are the fastest for the mortar category for the given reference length. On the contrary, for concrete and mineral cementing materials, the transfers are the slowest since they are located in the bottom left region. 

\begin{figure}
\centering
\includegraphics[width=.95\textwidth]{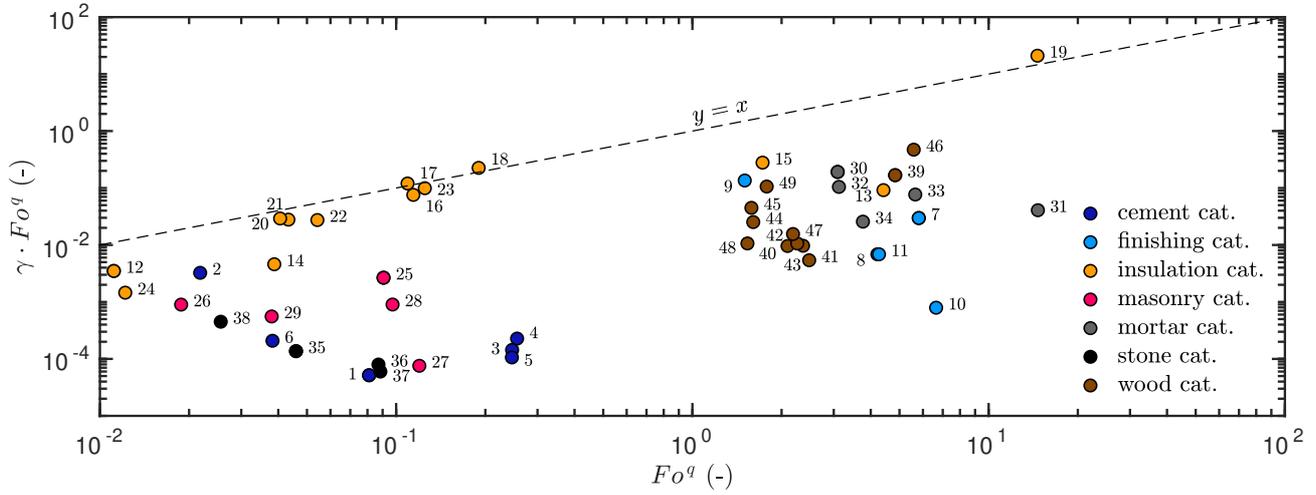}
\caption{Comparison of the competition between latent and sensible heat transfer by analyzing the heat \textsc{Fourier} number $\Fo^{\,q}$ and the coupling parameters $\gamma\,$.}
\label{fig:FoQvsGammaFoQ}
\end{figure}

Figure~\ref{fig:FoQvsGammaFoQ} contrasts the competition between sensible and latent heat transfer. Almost all materials are under the identity function $y \, = \, x\,$. Thus, since the ratio $\displaystyle \frac{\Fo^{\,q}}{\gamma \cdot \Fo^{\,q}}$ scales with $\mathcal{O}(\,10^{\,2}\,)\,$, the sensible heat transfer is around hundred times higher than latent one. An interesting remark is that a group of insulation materials has an equivalent competition between both transfers. The wood category is grouped in the same region, where sensible heat transfer is preponderant to latent one. The flat panel is the most vulnerable to both processes. 

\begin{figure}
\centering
\includegraphics[width=.95\textwidth]{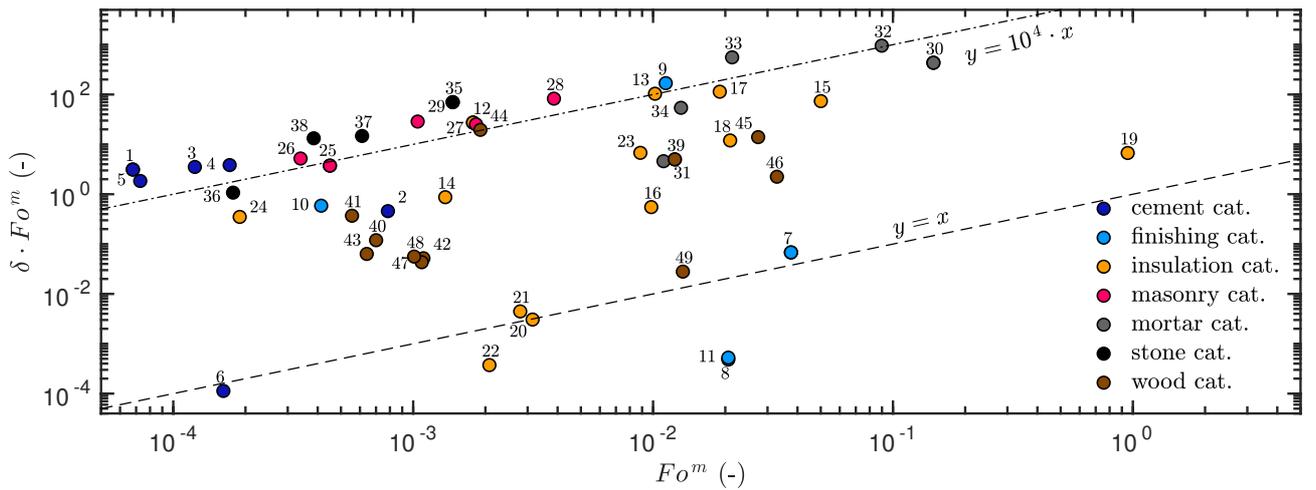}
\caption{Comparison of the competition between mass transfer under vapor pressure and temperature gradient by analyzing the mass \textsc{Fourier} number $\Fo^{\,m}$ and the coupling parameter $\delta \,$.}
\label{fig:FoMvsdeltaFoM}
\end{figure}

Figure~\ref{fig:FoMvsdeltaFoM} shows the mass transfer \textsc{Fourier} number $\Fo^{\,m}$ according to the product $\Fo^{\,m} \cdot \delta \,$. This analysis enables to evaluate the competition between the two mass transfer processes driven by vapor pressure and temperature gradients, respectively. For most of the materials, the mass transfer is driven by the coupling with temperature. A few materials, as wood fiber or mineral cementing material, have equivalent impacts between vapor pressure and temperature gradients. 

\begin{figure}
\centering
\subfigure[\label{fig:BiQvsBiM_L}]{\includegraphics[width=.95\textwidth]{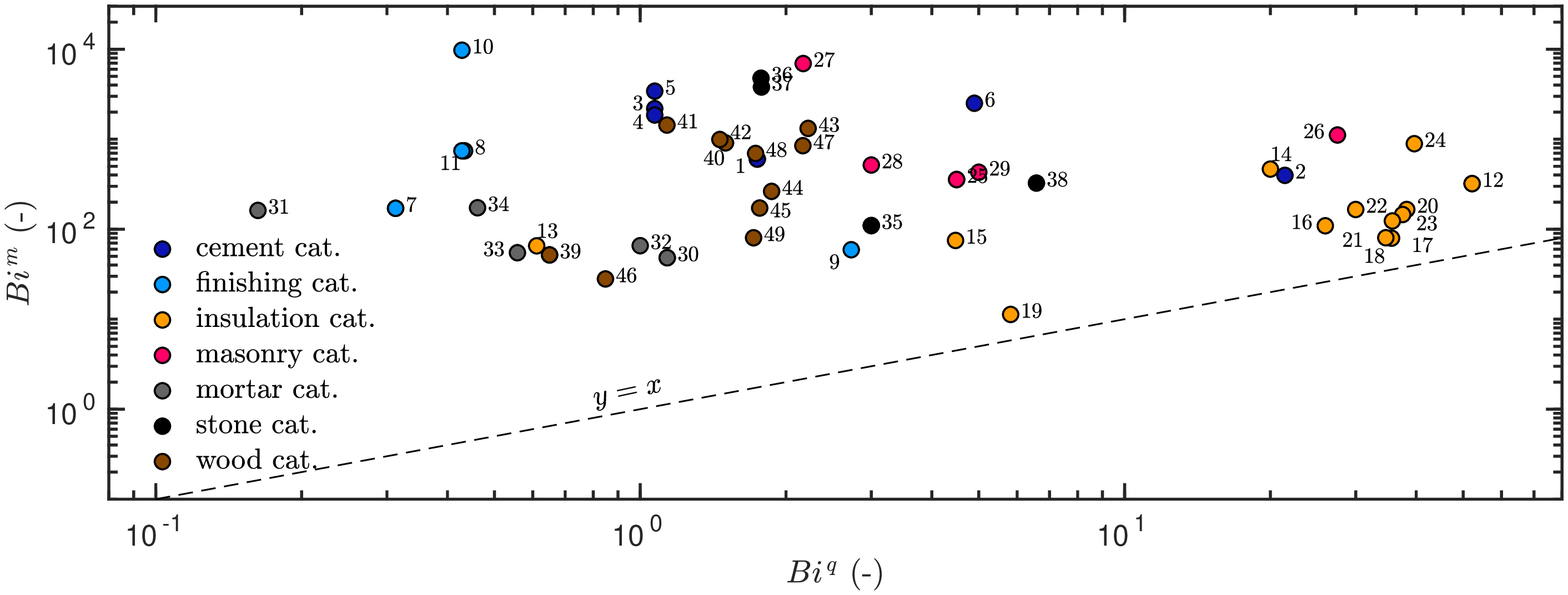}} \\
\subfigure[\label{fig:BiQvsBiM_R}]{\includegraphics[width=.95\textwidth]{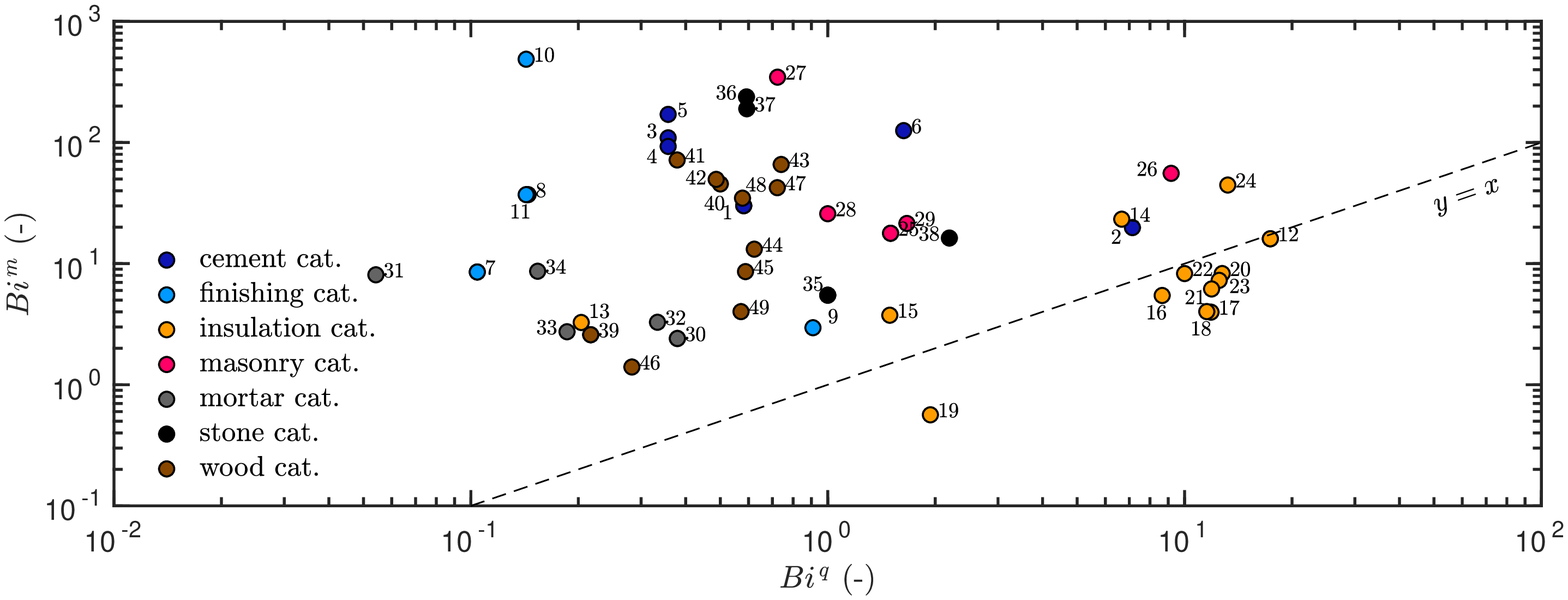}}
\caption{Comparison of the penetration of heat and mass transfer for outside \emph{(a)} and inside \emph{(b)} interfaces of the wall by analyzing the \textsc{Biot} numbers.}
\end{figure}

Figure~\ref{fig:BiQvsBiM_L} compares the quantity of heat and mass that passes through the interface between the wall and the outside air. As it can be remarked, the mass transfer through the interface is more important for the all categories. A few materials such as pumice concrete or anti mold are close to the identity function $y \egal x\,$, indicating an equivalent balance between the two phenomena. The insulating materials are grouped in one region near to the bisector. The penetration of heat and mass through the interface is the highest for this category of materials. Figure~\ref{fig:BiQvsBiM_R} analyzes the similar quantities for the interface with the inside building air. The location of the materials is very similar to those in Figure~\ref{fig:BiQvsBiM_L} since only the surface transfer coefficients are modified in the \textsc{Biot} numbers. The penetration through the inside interface is more balanced for the heat and mass transfer. Last, the mass transfer is more preponderant than heat one when looking only at the phenomena at the interface between the material and the ambient air.

It is important to remark that this analysis is valid for the defined reference lengths for each material. The question of selecting the best optimum porous media can be raised. However, it is difficult to give the sole answer since it depends on the objective and conditions of use. Considering only heat transfer, most of the time, materials are selected to reduce the heat losses of the wall. Thus, materials with low heat \textsc{Fourier} number can be chosen. However, when considering coupling with mass transfer, the question is more complex. On the one hand, materials with high mass \textsc{Fourier} numbers are interesting since they do not block the mass transfer. Thus, moisture disorders (mold, condensation) may not appear. On the other hand, the latent heat transfer is important for such materials and the wall heat losses are increased. A good compromise would be materials with a higher mass \textsc{Fourier} number but a low coupling number $\gamma$ so that the latent heat transfer is reduced. The question becomes more difficult since walls are composed of several materials.

\subsubsection{Considering the nonlinear behavior}
\label{sec:non_linear_behavior}

The previous analysis focused on the dimensionless numbers to indicate the dominant processes at the given reference conditions. With the scaled equations~\eqref{eq:dimless_heat_and_mass} and \eqref{eq:dimless_heat_and_mass_BC}, the investigations can be carried out considering the nonlinear aspect of the physical phenomena. For the sake of clarity, only two materials are compared, namely the concrete and the solid brick, with reference lengths of $L^{\,\rf} \egal 20 \ \mathsf{cm}$ and $L^{\,\rf} \egal 12 \ \mathsf{cm}\,$, respectively.

\begin{figure}
\centering
\subfigure[\label{fig:cMs}]{\includegraphics[width=.45\textwidth]{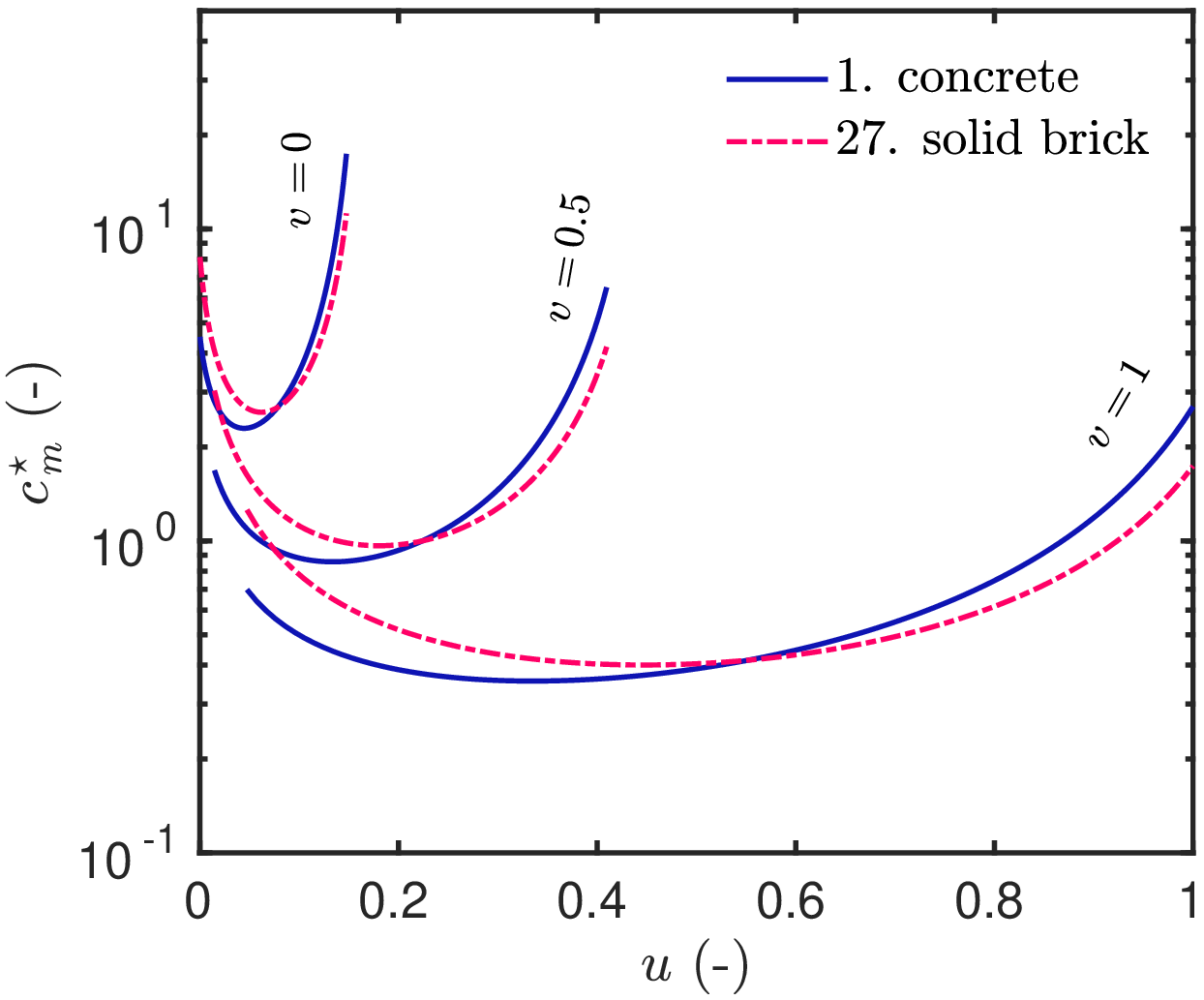}} \hspace{0.2cm}
\subfigure[\label{fig:cQs}]{\includegraphics[width=.45\textwidth]{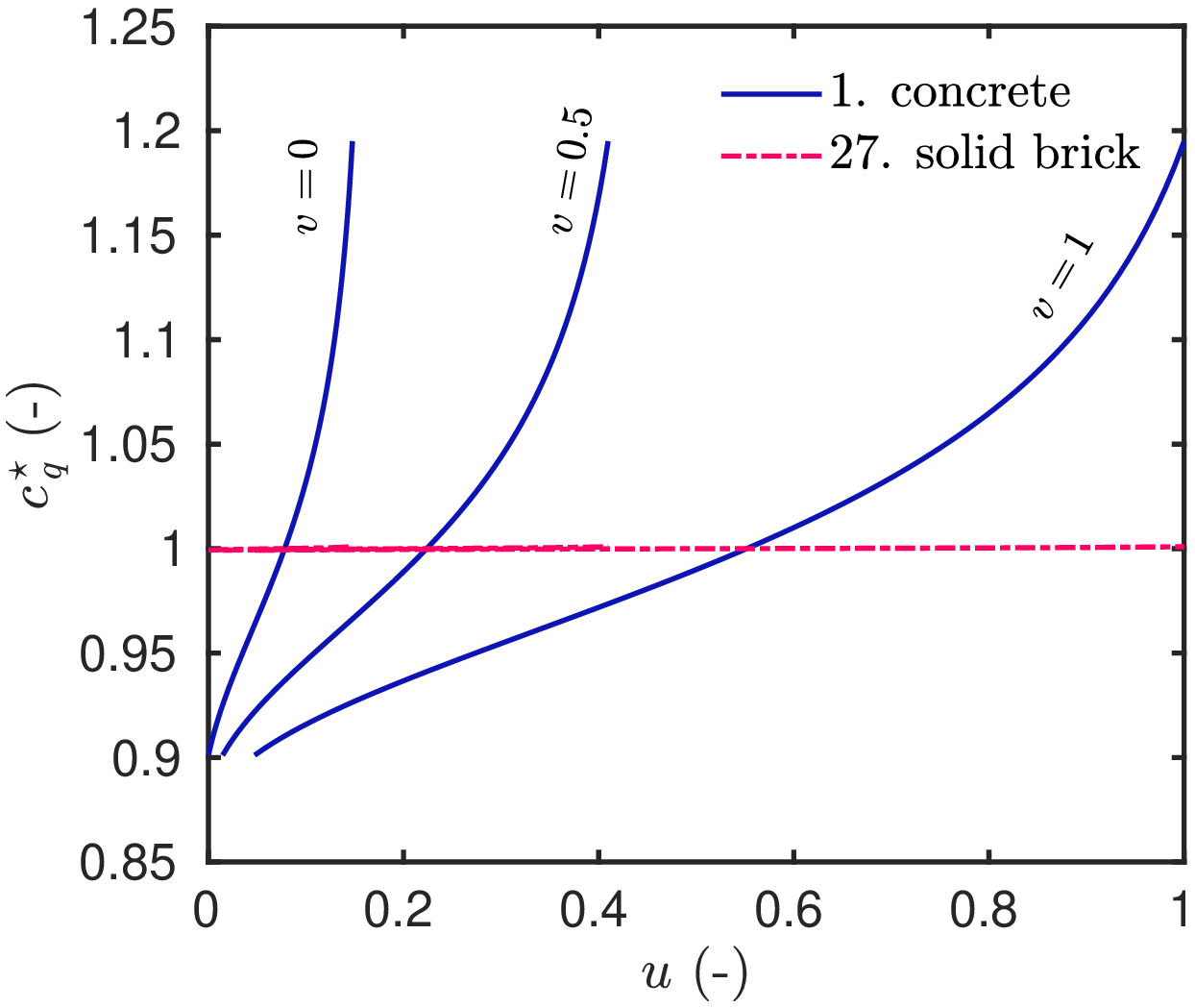}} \\
\subfigure[\label{fig:kMs}]{\includegraphics[width=.45\textwidth]{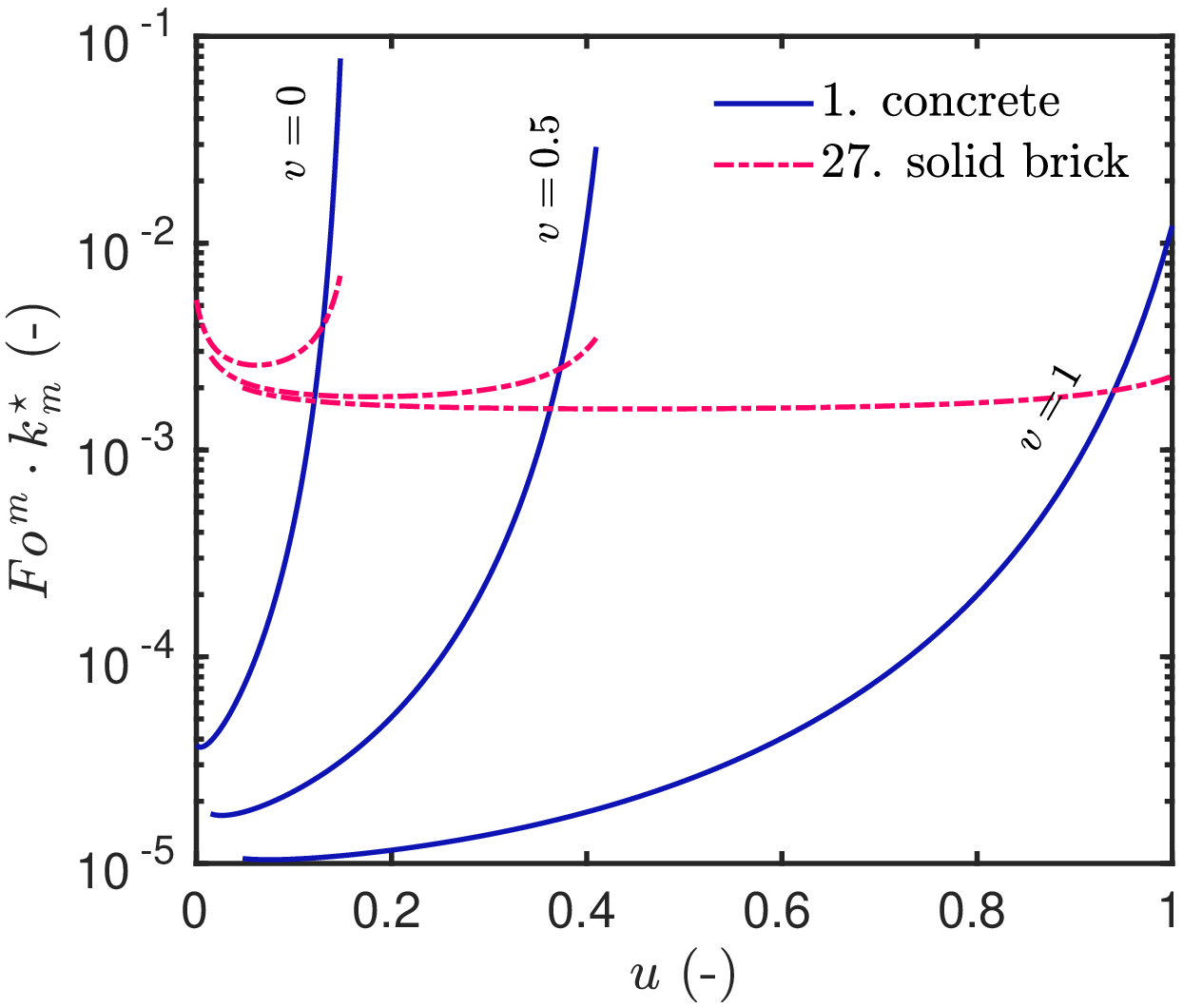}} \hspace{0.2cm}
\subfigure[\label{fig:kQs}]{\includegraphics[width=.45\textwidth]{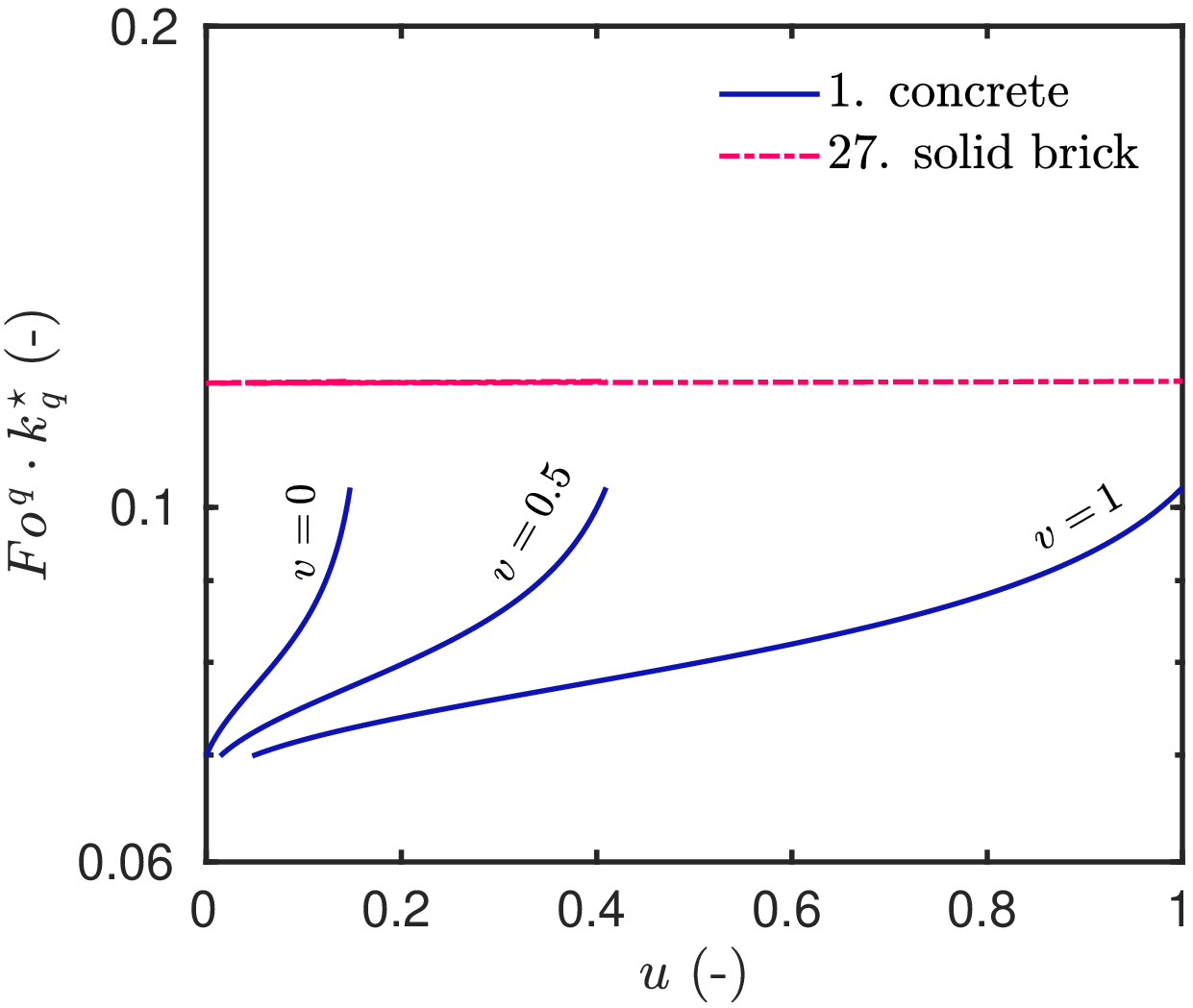}} \\
\subfigure[\label{fig:kMQs}]{\includegraphics[width=.45\textwidth]{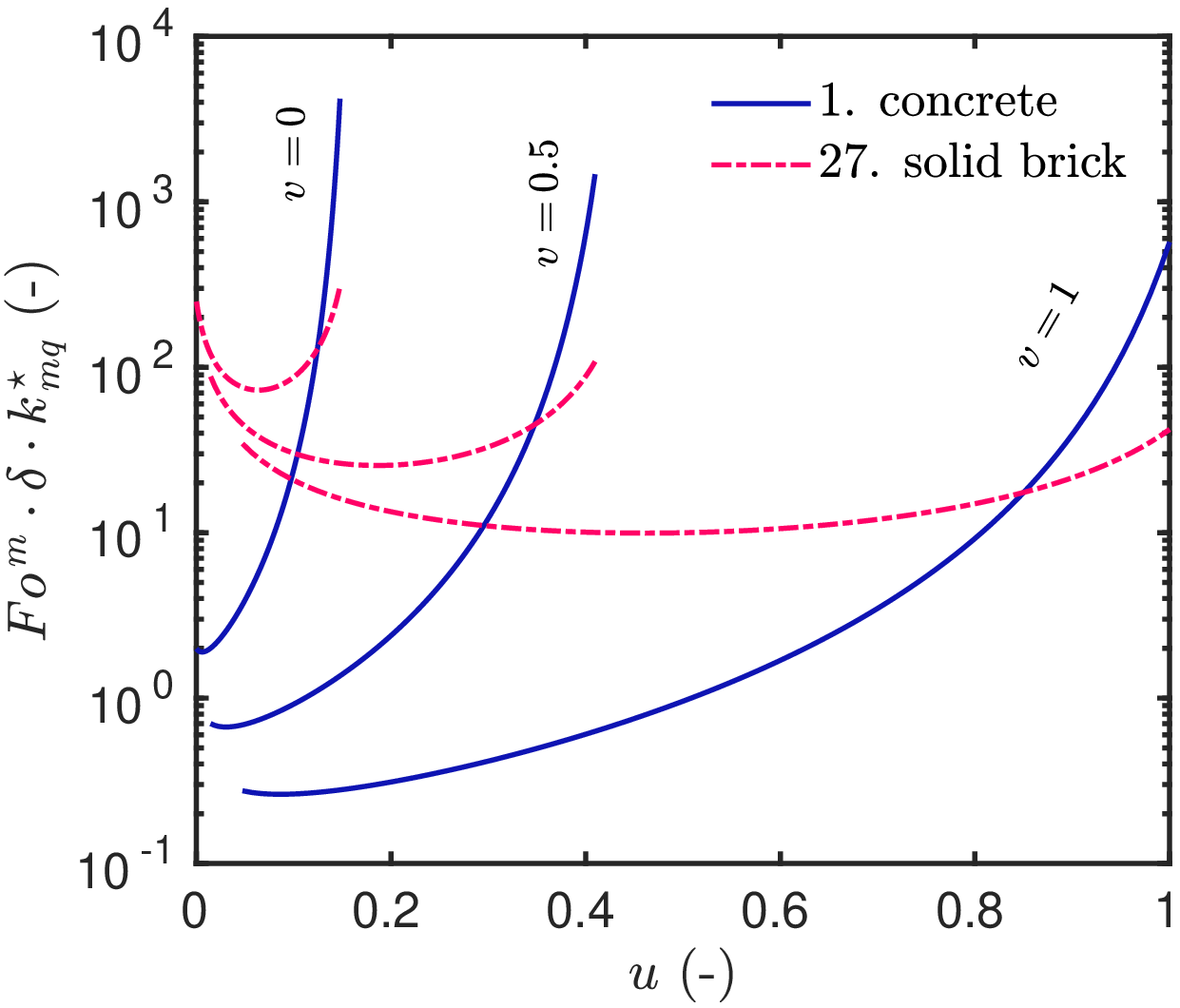}} \hspace{0.2cm}
\subfigure[\label{fig:kQMs}]{\includegraphics[width=.45\textwidth]{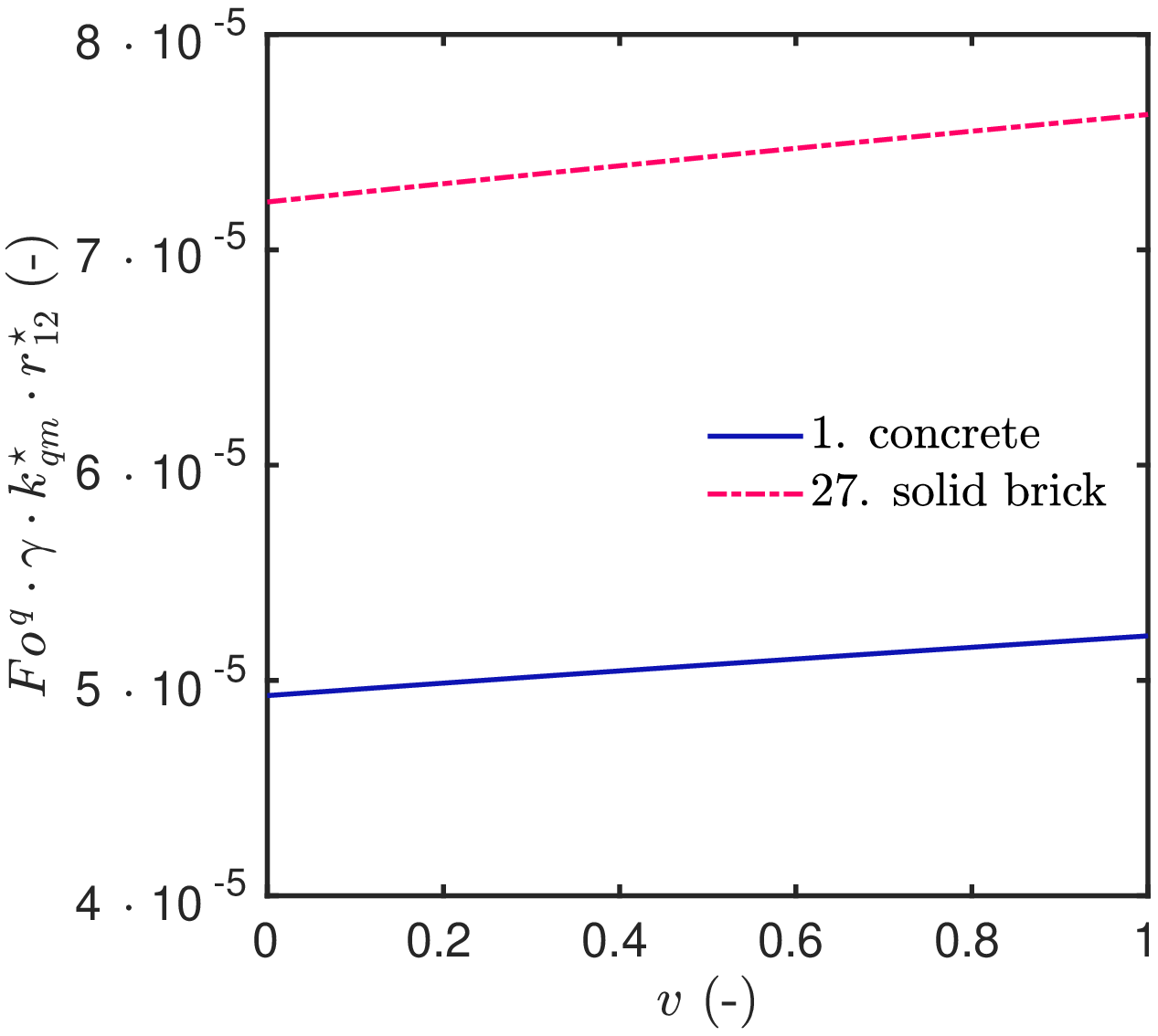}} 
\caption{Variation of the nonlinear distorsion coefficients with the fields $u$ and $v\,$: mass capacity \emph{(a)}, heat capacity \emph{(b)}, mass permeability \emph{(c)},  heat permeability \emph{(d)}, coupled mass permeability \emph{(e)}, coupled heat permeability \emph{(f)}.}
\label{fig:distorsion_NL}
\end{figure}

\begin{figure}
\centering
\includegraphics[width=.45\textwidth]{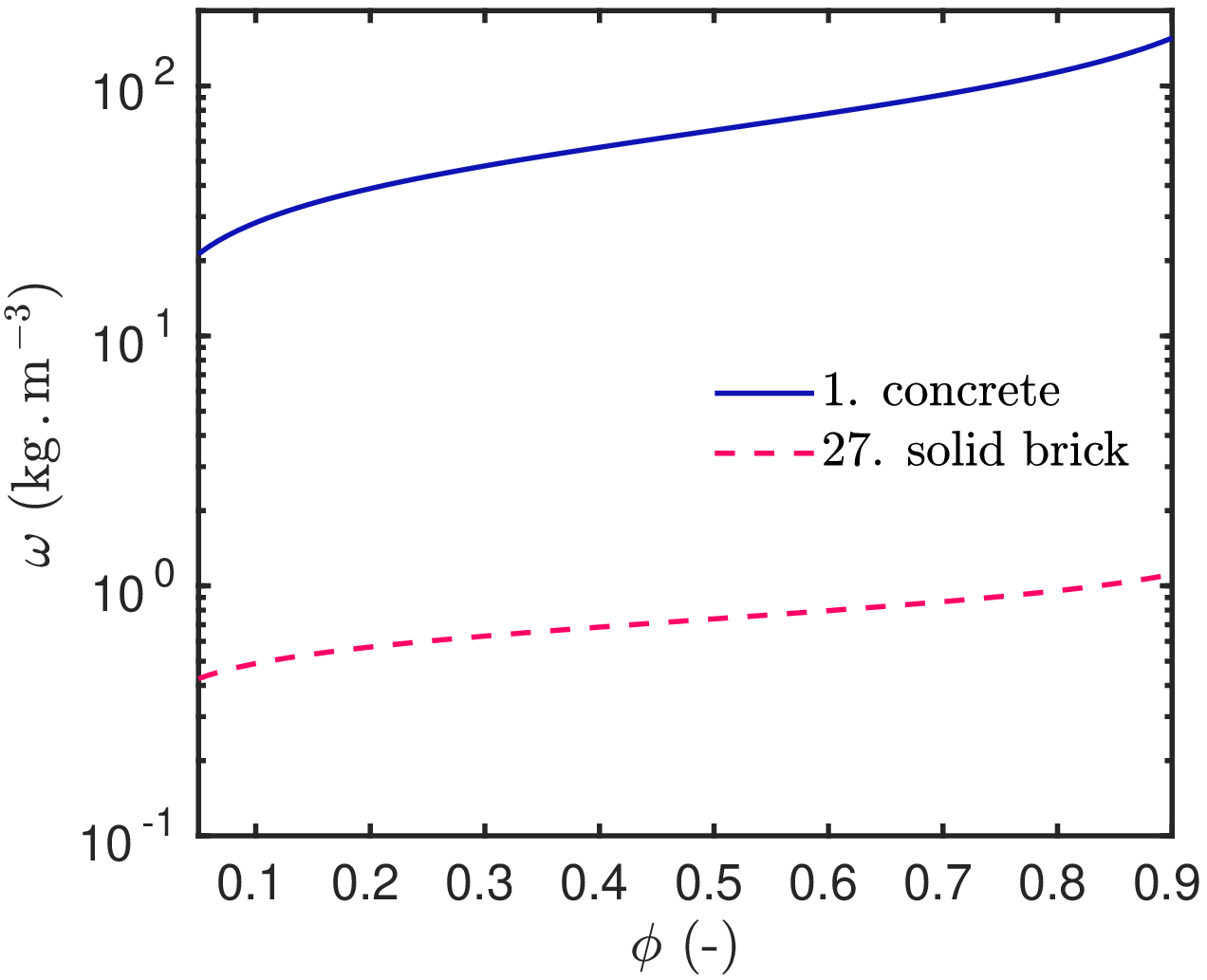}
\caption{Water adsorption curve for the concrete and the solid brick materials.}
\label{fig:omega_fphi}
\end{figure}

Figures~\ref{fig:cMs} and~\ref{fig:cQs} give the variation of the heat and mass capacities according to the field $u\,$. Those coefficients enable to analyze the (thermal and mass) inertial behavior of the material. First, the mass capacity is more important for the solid brick at low levels. In other words, the solid brick has a higher mass inertia than the concrete. Then, there is an inversion (at $u \egal 0.5$ and $v \egal 1$ for instance) where the concrete has a higher mass capacity. Therefore, for higher levels of vapor pressure, the concrete has a higher mass inertia. A similar analysis can be realized for the heat capacity. At a low level of vapor pressure, the concrete has a lower thermal inertia. It can be remarked that the heat capacity of the solid brick almost does not vary with the mass content. This result is related to the adsorption curve of both materials, illustrated in Figure~\ref{fig:omega_fphi}. The heat capacity of the material varies linearly with the moisture content in the material. Thus, higher magnitude variations in the moisture adsorption curve imply higher variations in the heat capacity coefficient.

In terms of transfer, Figures~\ref{fig:kMs} to \ref{fig:kQMs} show the variation of the different permeabilities. The mass transfer is more important in the brick than in the concrete. By comparing Figures~\ref{fig:kMs} and \ref{fig:kMQs}, for both materials, the mass transfer is driven mostly by the temperature gradient. It is consistent with the results obtained previously in Figure~\ref{fig:FoMvsdeltaFoM}. The permeability of the material to heat transfer is equivalent for both materials as remarked in Figure~\ref{fig:kQs}. It is more impacted by the water content for the concrete while it remains almost constant for the solid brick. From Figure~\ref{fig:kQMs}, it appears that latent heat transfer has a very small impact on the total heat process in the material. 

The dimensionless analysis shows how the dominant processes of heat and mass transfer can be studied without disregarding the nonlinear aspect of the phenomena. It also permits to simplify the mathematical modeling if required. As remarked in Figures~\ref{fig:cQs} and \ref{fig:kQs} for the solid brick, the coefficients $c_{\,q}^{\,\star}\,\bigl(\,u\,,\,v\,\bigr)$ and $k_{\,q}^{\,\star}\,\bigl(\,u\,,\,v\,\bigr)$ are almost invariant with $u$ and $v\,$. Thus, from a numerical point of view, the mathematical model can be simplified by disregarding the dependency of the functions $c_{\,q}^{\,\star}\,\bigl(\,u\,,\,v\,\bigr)$ and $k_{\,q}^{\,\star}\,\bigl(\,u\,,\,v\,\bigr)$ to the two fields. A similar conclusion can be drawn for both materials for the coefficient $k_{\,qm}^{\,\star} \cdot r_{\,12}$ and its variation with the field $v\,$. Thus, the dimensionless analysis can also be used to simplify the governing equations.

\subsubsection{Multi-layered walls}
\label{sec:multi_layer}

\begin{figure}
\centering
\subfigure[\label{fig:wall_configuration1} configuration 1]{\includegraphics[width=.90\textwidth]{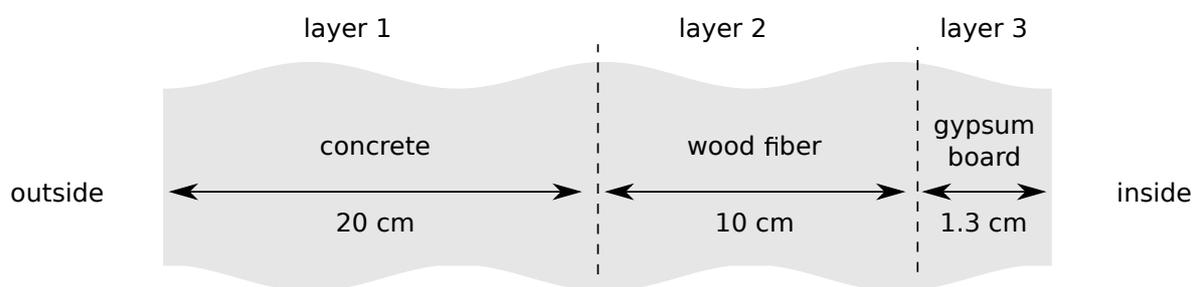}} \\
\subfigure[\label{fig:wall_configuration2} configuration 2]{\includegraphics[width=.90\textwidth]{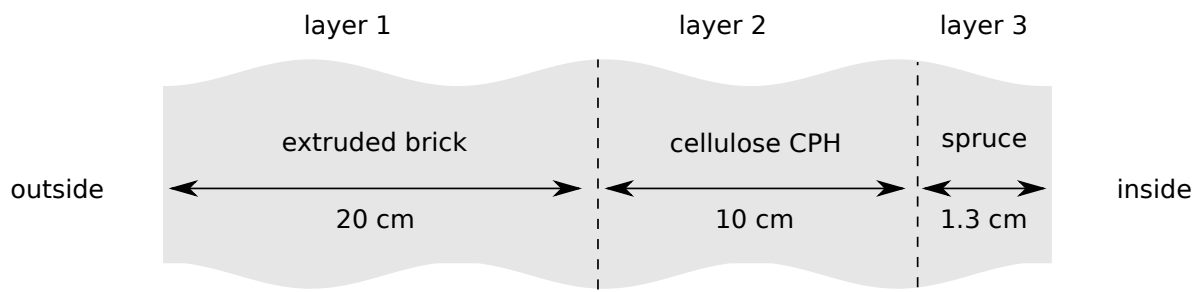}} 
\caption{Illustration of multi-layer configurations.}
\label{fig:wall_configuration}
\end{figure}

The purpose of this section is to illustrate the use of dimensionless scaling for analyzing the general tendencies of heat and mass transfer in a multi-layer wall. Two wall configurations are considered and illustrated in Figure~\ref{fig:wall_configuration}. The first one is made of concrete, wood fiber and gypsum board. The second is composed of extruded brick, cellulose and radial spruce. 

\begin{figure}
\centering
\subfigure[\label{fig:FoM_ML}]{\includegraphics[width=.45\textwidth]{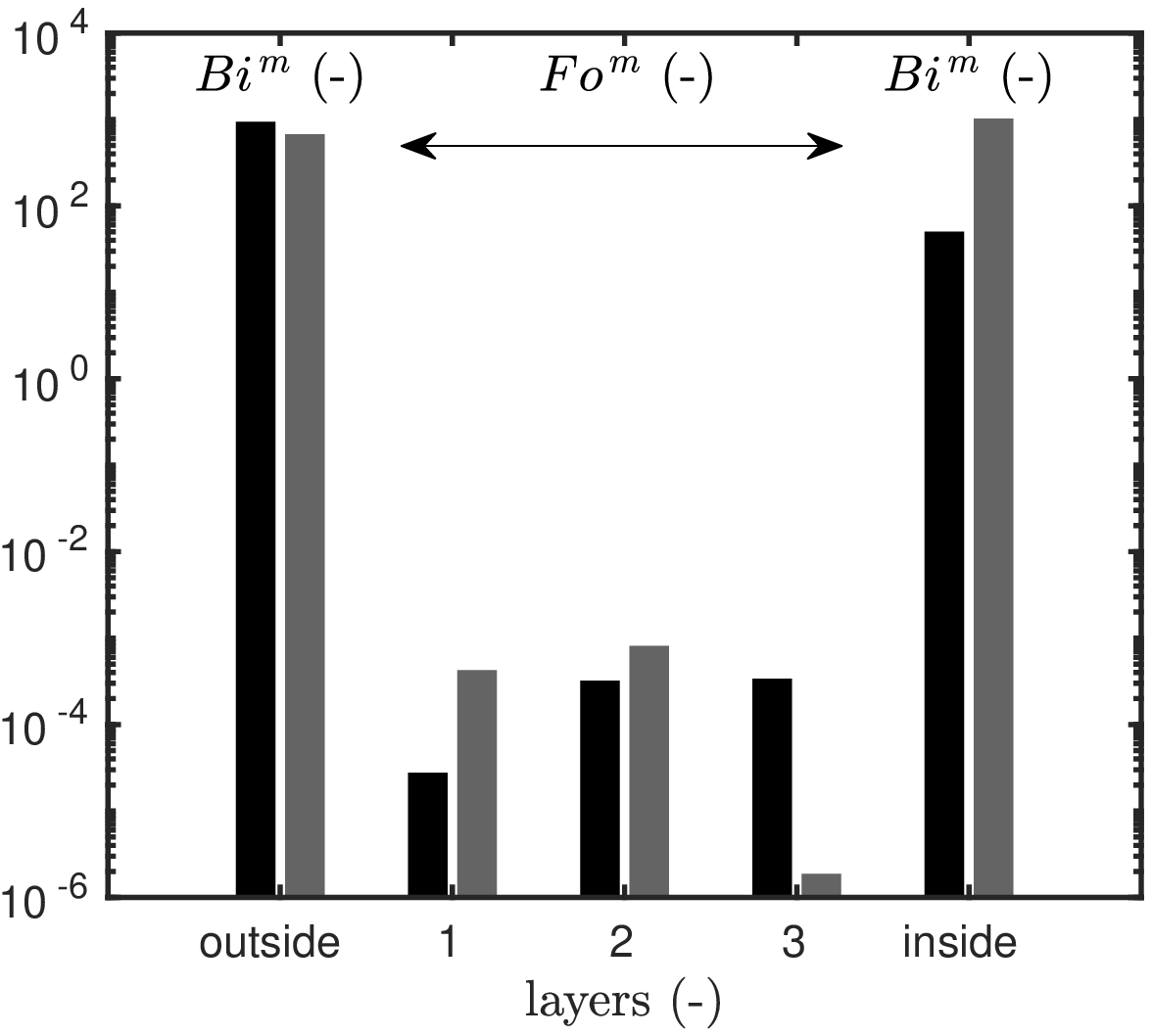}} \hspace{0.2cm}
\subfigure[\label{fig:FoQ_ML}]{\includegraphics[width=.45\textwidth]{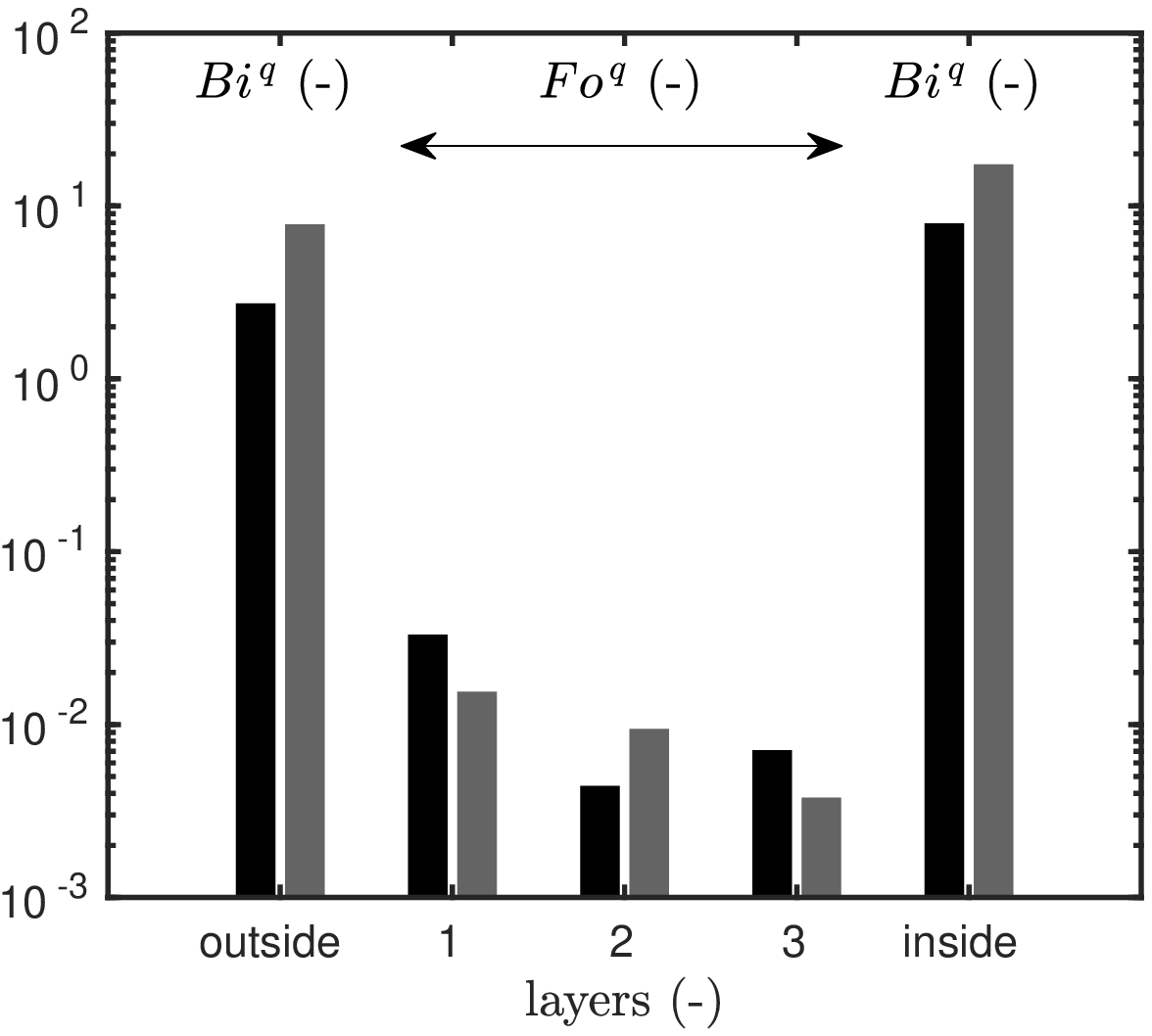}} \\
\subfigure[\label{fig:gamma_FoM_ML}]{\includegraphics[width=.45\textwidth]{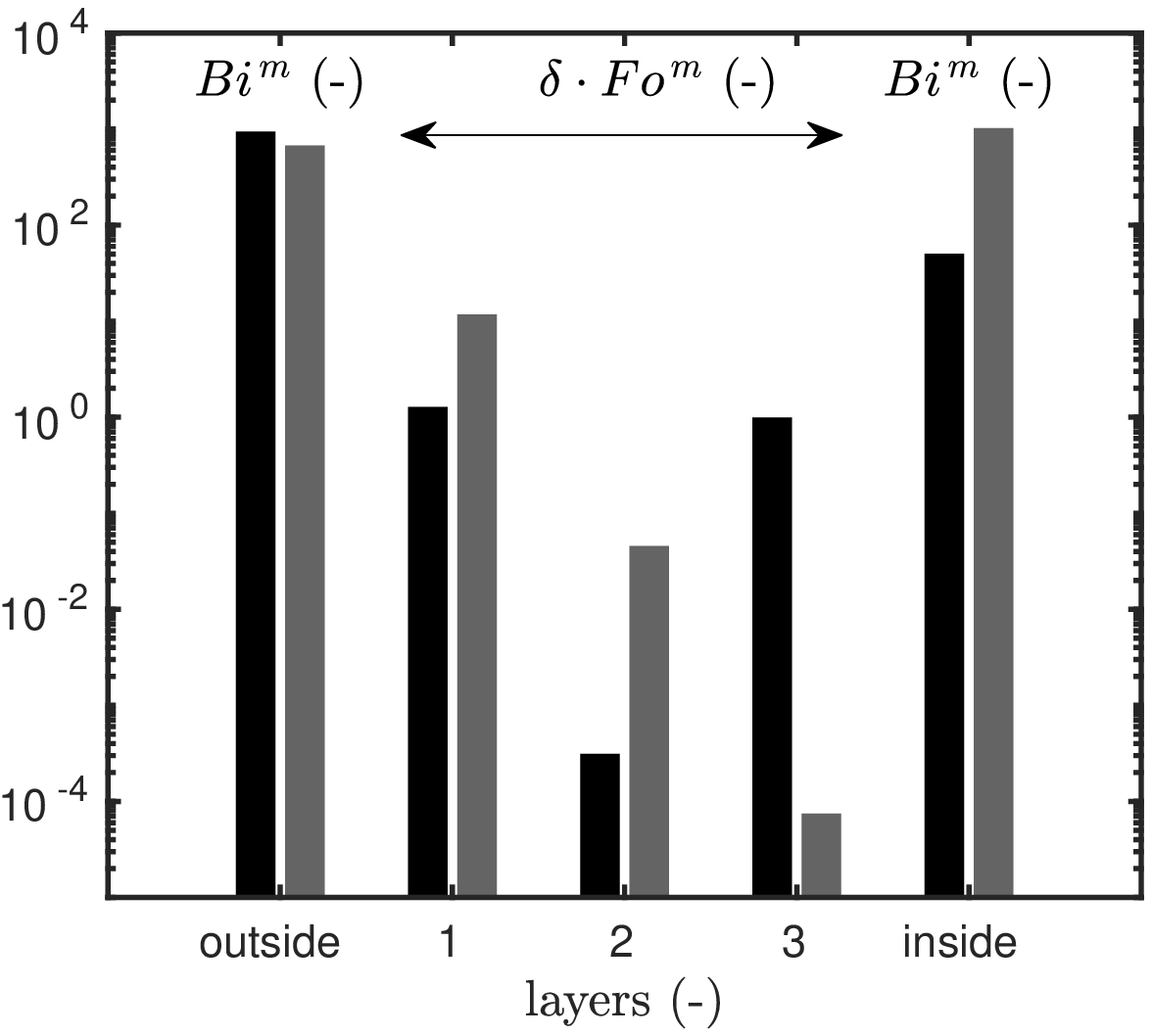}} \hspace{0.2cm}
\subfigure[\label{fig:delta_FoQ_ML}]{\includegraphics[width=.45\textwidth]{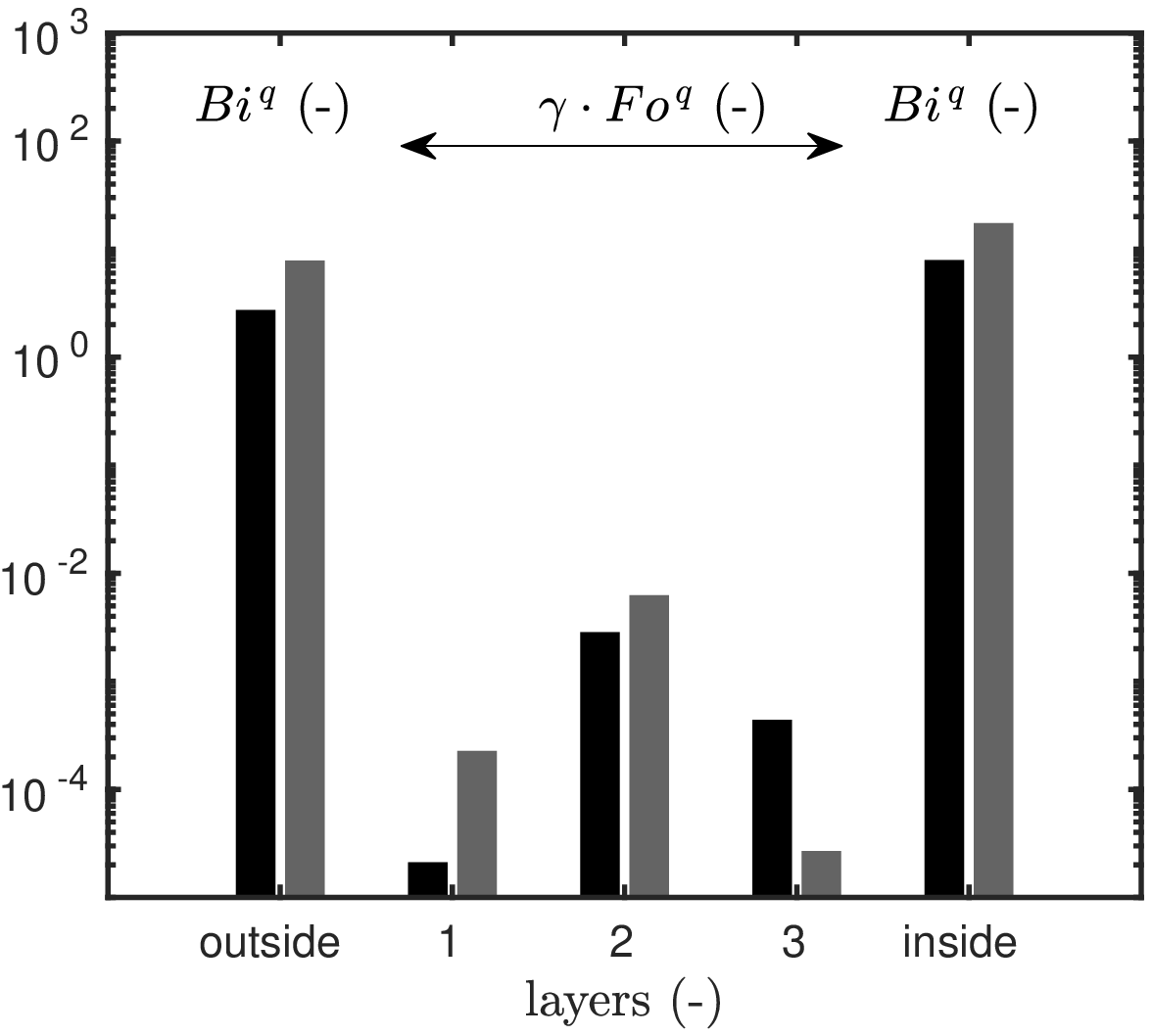}}
\caption{Variation of the dimensionless numbers for mass \emph{(a,c)} and heat \emph{(b,d)} transfer for two wall configurations. Black and grey colors correspond to the configuration $1$ (concrete / wood fiber / gipsum board) and to the configuration $2$ (extruded brick / cellulose / radial spruce), respectively.}
\label{fig:distorsion_NL}
\end{figure}

For both configurations, the dimensionless numbers are given in Figures~\ref{fig:FoM_ML} for the mass transfer and in \ref{fig:gamma_FoM_ML} for the coupled mass one. First of all, its is assumed a mass flow directed from the outside to the inside, driven by a temperature and/or mass gradient. The mass \textsc{Biot} numbers are equivalent and so the penetration of mass transfer at the outside interface for both configurations. Then, the comparison focuses on the mass \textsc{Fourier} numbers. They are higher in the layers $1$ and $2$ for the  second configuration. Thus, the mass transfer is faster in those layers of the configuration. This conclusion cannot be extended for the three layers since the quantities $\Fo^{\,m}$ and $\delta \cdot \Fo^{\,m}$ are higher for the layer $3$ in configuration $1\,$. 

One can remark that the magnitudes of $\Fo^{\,m}$ and $\delta \cdot \Fo^{\,m}$ are really lower for the layer $3$ in the configuration $2\,$. If the mass transfer reaches this layer in this configuration, they will be blocked compared to configuration $1\,$. It can be a favorable situation to the development of moisture disorders \cite{Berger_2015}. The layer $3$ in the configuration $2\,$ is combined with a high inside mass \textsc{Biot} number. If there is a mass flow from the inside part to the outside, the gradient of mass transfer will be important in this layer. The latter is more protective from inside moisture loads in configuration $2$ than $1\,$.

The analysis is now carried for the dimensionless numbers for the sensible and latent heat transfer presented in Figures~\ref{fig:FoQ_ML} and \ref{fig:delta_FoQ_ML}, respectively. Both configurations have similar \textsc{Biot} numbers. In the layer $1$ and $3\,$, the \textsc{Fourier} numbers are higher for the configuration $1\,$. Thus, the sensible heat transfer is faster in this configuration in layer $1$ or $3$ if the heat flux arises from the outside or inside, respectively. The conclusion does not hold for the whole configuration since the order in the \textsc{Fourier} numbers inverts. Further investigations, based on simulation programs are required. 

For the latent heat, the process differs depending on the direction of the flux. For outside/inside directed flux, the latent heat transfer is faster in layers $1$ and $2$ in configuration $2$ since the quantity $\gamma \cdot \Fo^{\,q}$ is higher. Inversely, for flux coming from the inside, the layer $3$ in configuration $1$ is more affected by the latent heat transfer due to its higher dimensionless number. Again, no general conclusions for the whole configuration can be addressed without further investigations. Note that generally, such constructing type includes vapor-barriers which modify the proposed analysis.

\subsection{The use of similitude to define experimental design}
\label{sec:similarity_analysis}

Several similitude laws can be defined for \emph{(i)} kinetic, \emph{(ii)}  geometric or \emph{(iii)} transfer dynamics. They conserve the characteristic time, the reference length and the phenomena of transfer, respectively.

\subsubsection{Kinetic similarity analysis for insulation length equivalency}

\begin{figure}
\centering
\subfigure[\label{fig:FoM_vL2} $\Fo_{\,m}$]{\includegraphics[width=.45\textwidth]{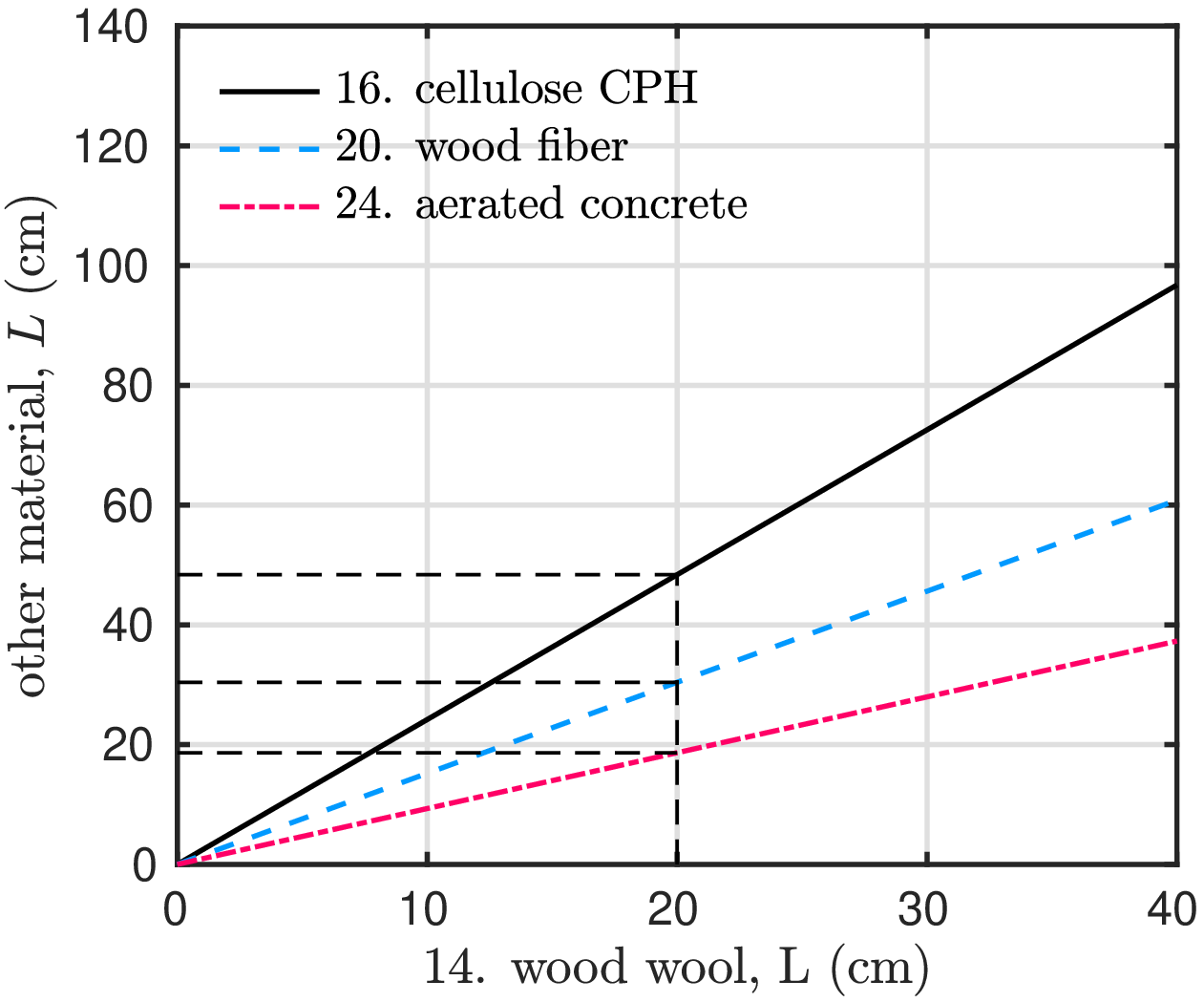}} \hspace{0.2cm}
\subfigure[\label{fig:FoQ_vL2} $\Fo_{\,q}$]{\includegraphics[width=.45\textwidth]{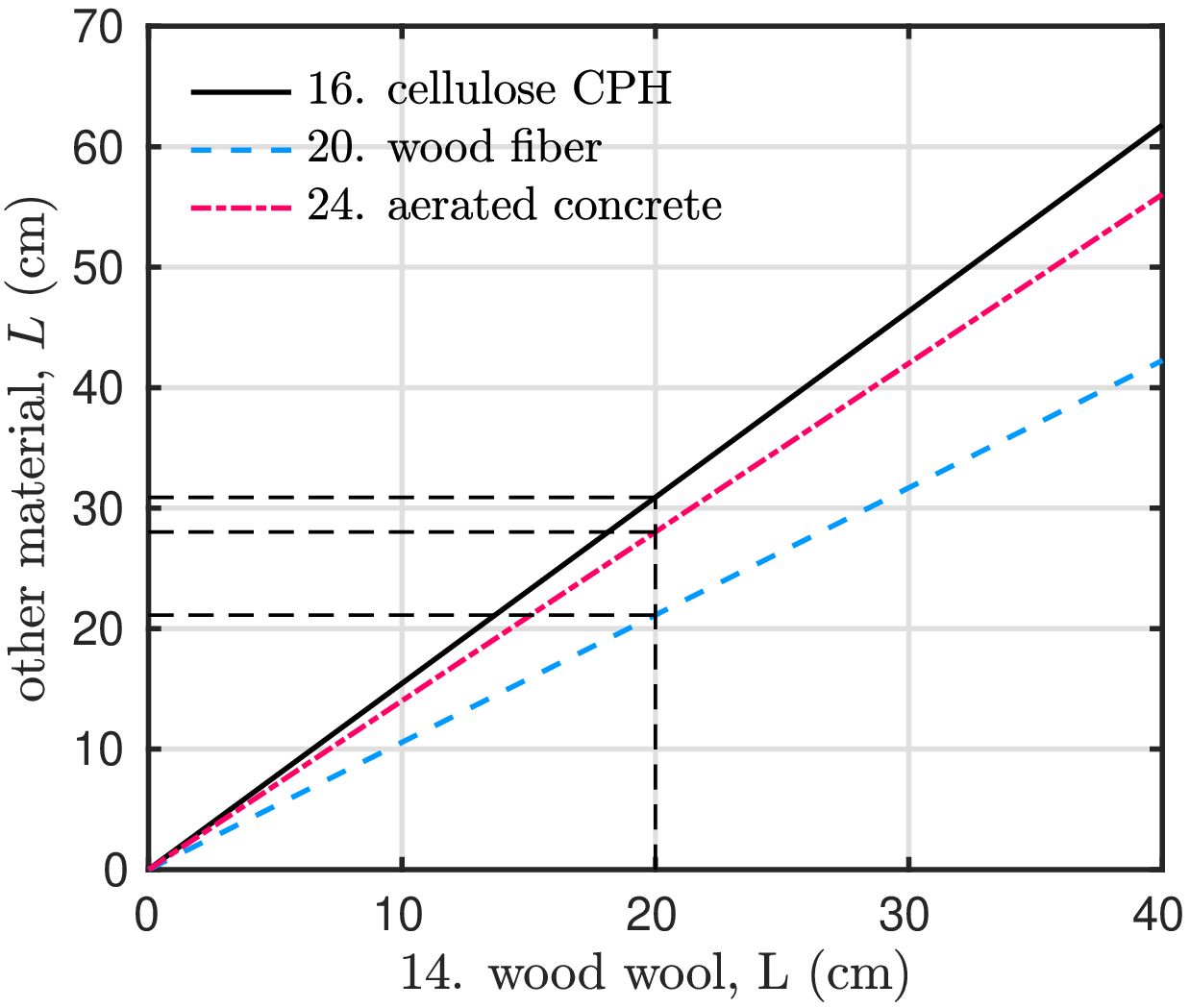}} \\
\subfigure[\label{fig:delta_FoM_vL2} $\delta \cdot \Fo_{\,m}$]{\includegraphics[width=.45\textwidth]{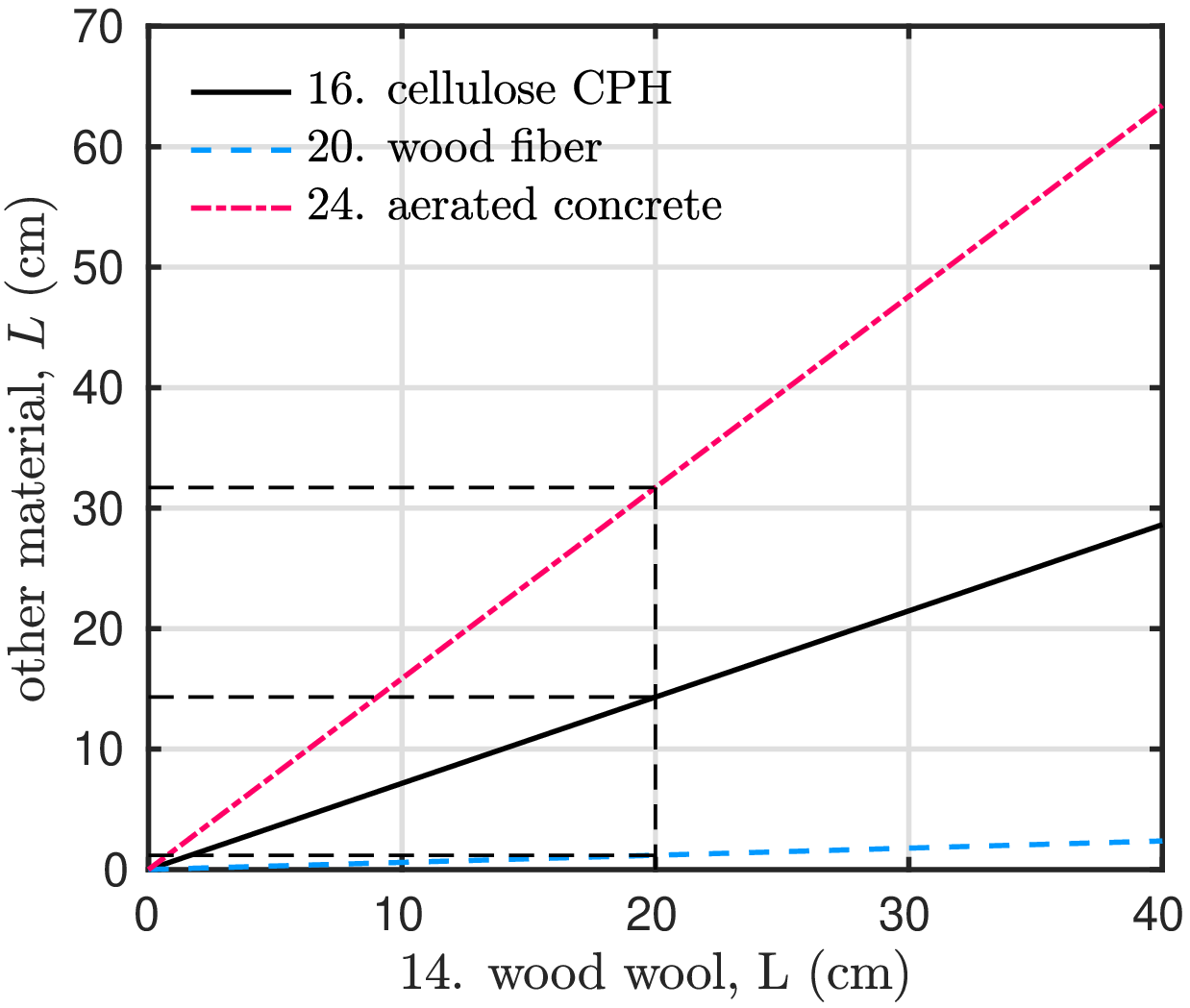}} \hspace{0.2cm}
\subfigure[\label{fig:gamma_FoQ_vL2} $\gamma \cdot \Fo_{\,q}$]{\includegraphics[width=.45\textwidth]{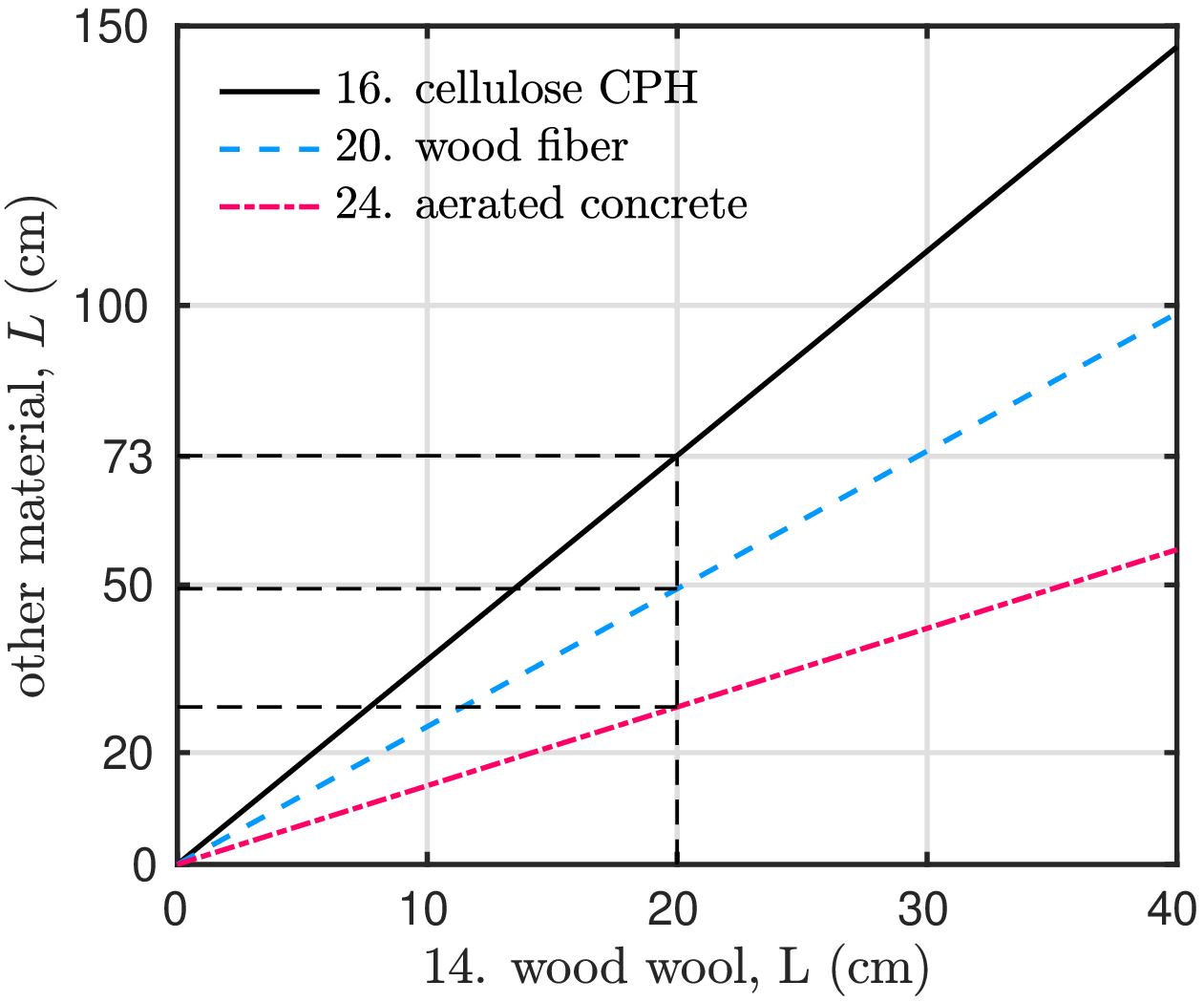}}
\caption{Kinetic similarities, in terms of equivalent length, for the mass \emph{(a,c)} and heat \emph{(b,d)} \textsc{Fourier} numbers and coupling parameters for four insulating materials.}
\label{fig:vL2}
\end{figure}

Now, one can observe an illustration of the similarity analysis thanks to the dimensionless analysis. Figures~\ref{fig:FoM_vL2} and \ref{fig:FoQ_vL2} give the equivalent length of three insulation materials to obtain the same mass and heat \textsc{Fourier} numbers as the wood wool, respectively. Figures~\ref{fig:delta_FoM_vL2} and \ref{fig:gamma_FoQ_vL2} provide similar analysis for the coupling parameters. For instance, $L \egal 20 \ \mathsf{cm}$ of wood wool has an equivalent mass \textsc{Fourier} number of $L \egal 48 \ \mathsf{cm}$ of cellulose CPH, $L \egal 30 \ \mathsf{cm}$ of wood fiber or $L \egal 19 \ \mathsf{cm}$ of aerated concrete. It can be remarked that the curves for cellulose and wood fiber are above the identity function $y \egal x\,$ for the mass \textsc{Fourier} number. A higher length is required for those two materials to get similar mass transfer as in wood wool. It also reveals that for a fixed length, the mass \textsc{Fourier} numbers are lower for the wood wool than for the cellulose and wood fiber. A similar analysis can be carried with the heat \textsc{Fourier} number where $L \egal 20 \ \mathsf{cm}$ of wood wool is equivalent to $L \egal 31 \ \mathsf{cm}$ of cellulose CPH, $L \egal 21 \ \mathsf{cm}$ of wood fiber or $L \egal 28 \ \mathsf{cm}$ of aerated concrete. In terms of coupling parameters, important differences are remarked among the materials. Indeed, for the sensible heat transfer, \emph{i.e.} the coupling parameter $\gamma \cdot \Fo_{\,q}\,$, $L \egal 20 \ \mathsf{cm}$ of wood wool is similar to $L \egal 73 \ \mathsf{cm}$ of cellulose CPH, $L \egal 49 \ \mathsf{cm}$ of wood fiber or $L \egal 28 \ \mathsf{cm}$ of aerated concrete. Such analysis enables to define physically similar systems, by choosing correspondences in geometry or dynamics of transfer.

\subsubsection{Geometric similarity analysis for characteristic time of transfer}

\begin{figure}
\centering
\subfigure[\label{fig:FoM_vt2} $\Fo_{\,m}$]{\includegraphics[width=.45\textwidth]{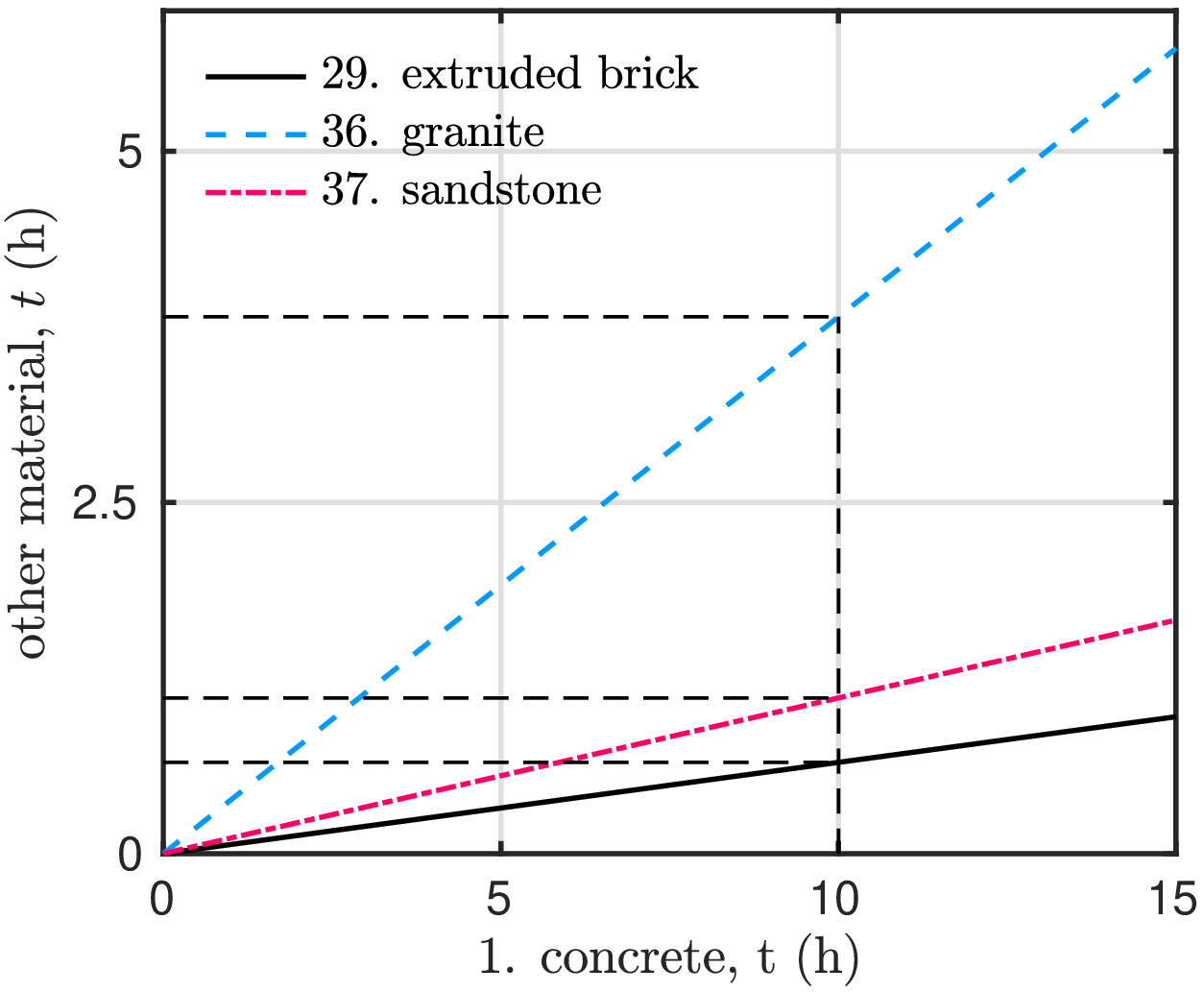}} \hspace{0.2cm}
\subfigure[\label{fig:FoQ_vt2} $\Fo_{\,q}$]{\includegraphics[width=.45\textwidth]{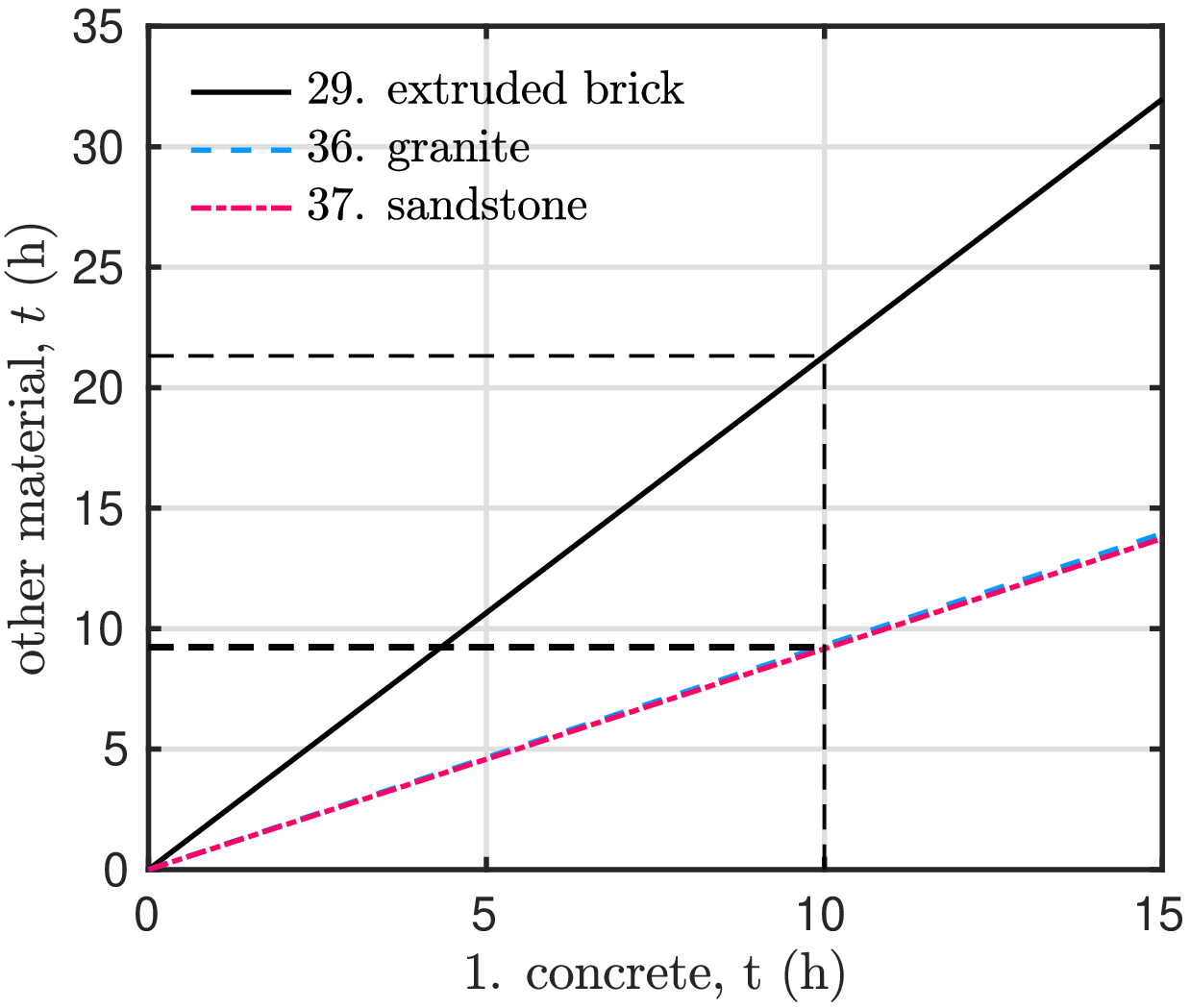}} \\
\subfigure[\label{fig:delta_FoM_vt2} $\delta \cdot \Fo_{\,m}$]{\includegraphics[width=.45\textwidth]{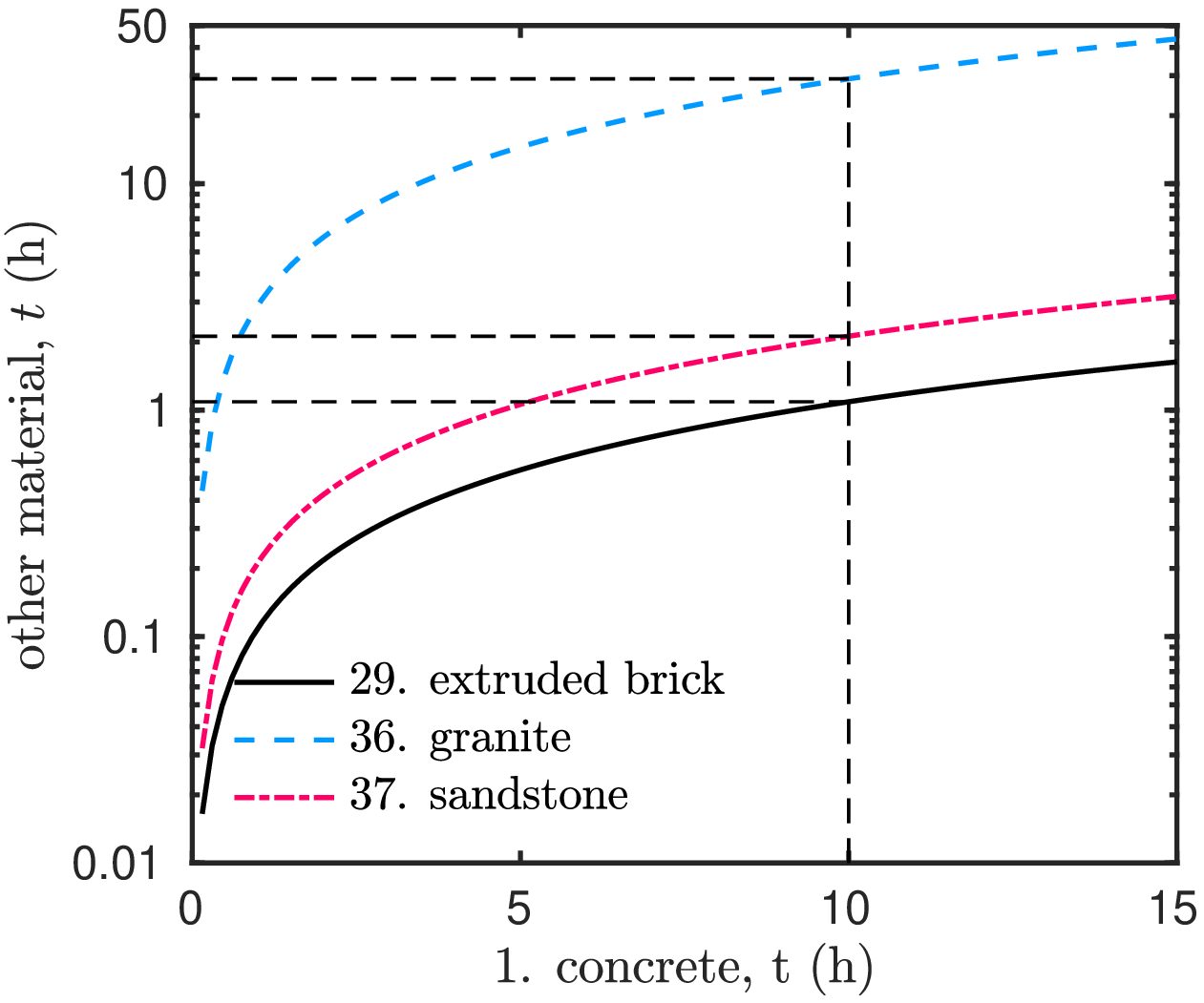}} \hspace{0.2cm}
\subfigure[\label{fig:gamma_FoQ_vt2} $\gamma \cdot \Fo_{\,q}$]{\includegraphics[width=.45\textwidth]{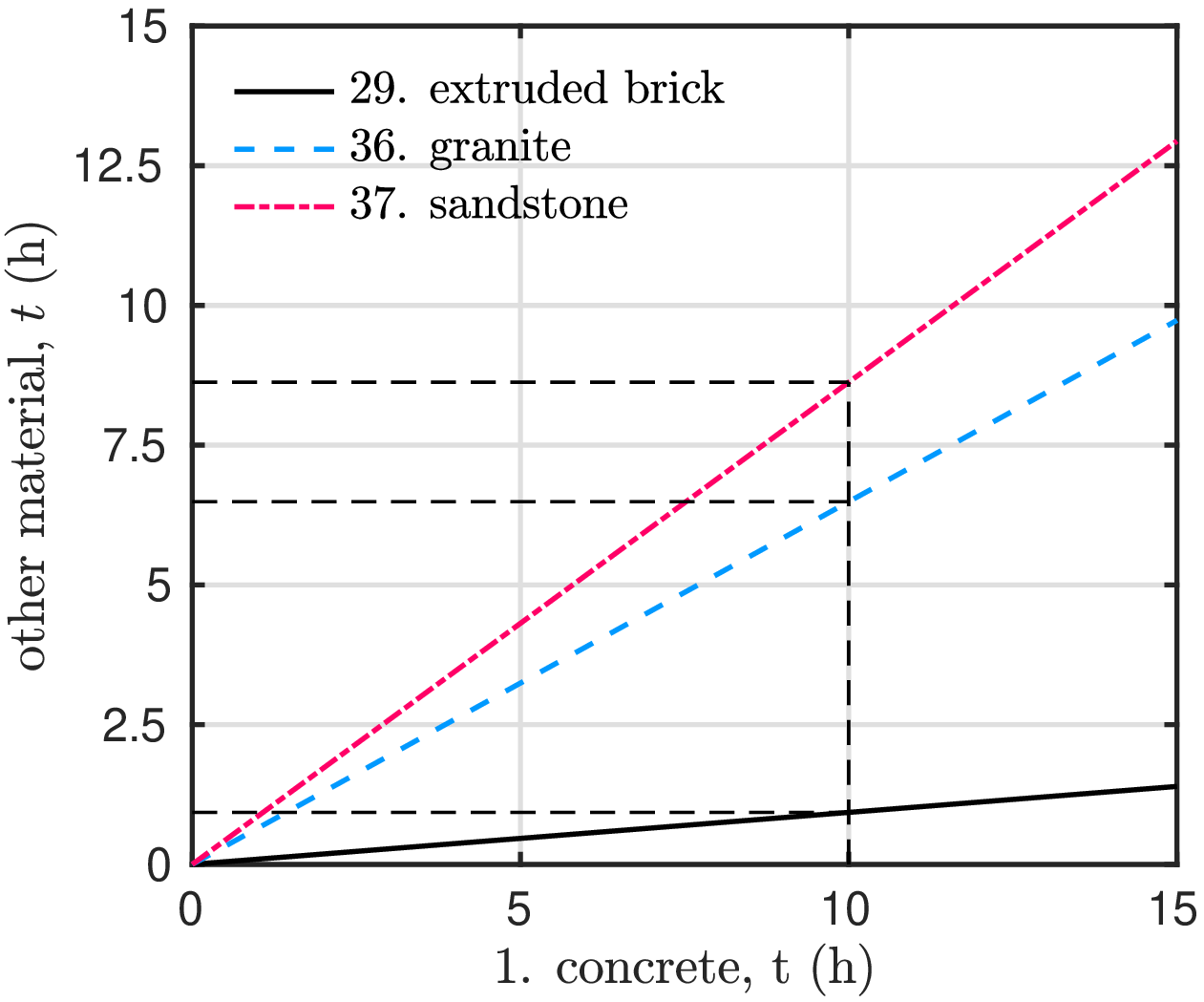}}
\caption{Geometric similarities, in terms of equivalent time, for the mass \emph{(a,c)} and heat \emph{(b,d)} \textsc{Fourier} numbers and coupling parameters for four materials.}
\label{fig:vt2}
\end{figure}

Another similarity analysis can be carried out. In the framework of planning experimental design, one can require to estimate the time of the diffusion processes in materials. This study compares the concrete with three other materials: extruded brick, granite and sandstone. The four materials have a reference length of $L^{\,\rf} \egal 10 \ \mathsf{cm}\,$. Figures~\ref{fig:FoM_vt2} to ~\ref{fig:gamma_FoQ_vt2} show the similarities in terms of equivalent time of transfer for the four dimensionless numbers. As can be remarked in Figure~\ref{fig:FoM_vt2}, similar velocity of mass transfer should be observed for a time observation of $10 \ \mathsf{h}$ in concrete, $0.6 \ \mathsf{h}$ in extruded brick, $3.8 \ \mathsf{h}$ in granite and $1.1 \ \mathsf{h}$ in sandstone. In terms of heat transfer, the functions for sandstone and granite are very closed to the unity one. Thus, it can be deduced that sandstone, granite and concrete have a very similar characteristic time of heat transfer. However, the heat transfer in extruded brick is around two times slower than in concrete. Important differences in the behavior of the materials can be remarked in Figure~\ref{fig:delta_FoM_vt2}. The mass transfer under temperature gradient is slower in granite than in concrete. Lastly, for the latent heat transfer, the all functions in Figure~\ref{fig:gamma_FoQ_vt2} are under the unity function. It indicates that the mechanism is faster for the three materials than for concrete. A similar experiment in concrete and in brick would take $10 \ \mathsf{h}$ and $54 \ \mathsf{min}\,$, respectively.

\subsubsection{Dynamic similarity analysis between two experimental configurations}

The issue is to investigate the aging of hygroscopic material properties under cyclic conditions of heat and mass transfer. An experimental design can be defined for the analysis. One considers a material with a characteristic length $L^{\,\rf}\,$, as given in Figure~\ref{fig:Lref} for several materials. The characteristic frequency of daily hygrothermal variations in a building is $24 \ \mathsf{h}\,$, with a given amplitude of temperature and vapor pressure. To analyze the aging of the properties, the experiment requires to be carried out over a long time, such as one year. This is an important constraint in the achievement of the investigations. The dimensionless numbers can be used to find similitude relations and define an equivalent experimental design with a reduced duration. The equivalent configuration is denoted with a upper bar $\bar{  }$ and the initial one without. The similitude relations between equivalent and initial configurations yield to a system of equations among the dimensionless numbers: 
\begin{align}
\label{eq:similitude_system}
\begin{cases}
\Fo^{\,q} & \egal \overline{\Fo^{\,q}} \,, \\
\Fo^{\,m}  & \egal \overline{\Fo^{\,m}} \,, \\
\eta & \egal \overline{\eta} \,, \\
\gamma & \egal \overline{\gamma} \,,
\end{cases}
\qquad \& \qquad
\begin{cases}
\delta & \egal \overline{\delta} \,,  \\
\Bi^{\,q} & \egal \overline{\Bi^{\,q}} \,, \\
\Bi^{\,qm} & \egal \overline{\Bi^{\,qm}} \,, \\
\Bi^{\,m} & \egal \overline{\Bi^{\,m}} \,.
\end{cases}
\end{align}
Keeping the same initial condition and reference temperature and vapor pressure between the two configurations, the system~\eqref{eq:similitude_system} leads to the following results: 
\begin{align}
\label{eq:similitude_law}
t^{\,\rf} \egal \Pi^{\,2} \cdot \overline{t^{\,\rf}} \,, \quad
L^{\,\rf} \egal \Pi \cdot \overline{L^{\,\rf}} \,, \quad
h_{\,q} \egal \frac{1}{\Pi} \cdot \overline{h_{\,q}} \,, \quad
h_{\,m} \egal \frac{1}{\Pi} \cdot \overline{h_{\,m}} \,
\end{align}
where $\Pi$ is the proportionality factor between the initial and equivalent configurations. Respecting the similitude laws~\eqref{eq:similitude_law} ensures that the dimensionless problems between the two configurations are completely equal. Note that the system~\eqref{eq:similitude_system} induces also the equality between the nonlinear distorsion coefficients since the same material is considered. 

\begin{figure}
\centering
\includegraphics[width=.95\textwidth]{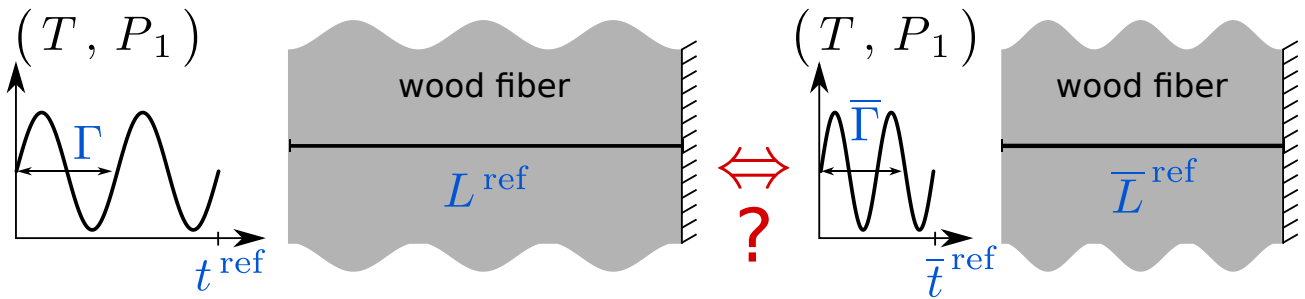}
\caption{Illustration of the dynamic similitude problem to plan experimental design.}
\label{fig:exp_similarity}
\end{figure}

\begin{figure}
\centering
\subfigure[\label{fig:BCPT_ft}]{\includegraphics[width=.45\textwidth]{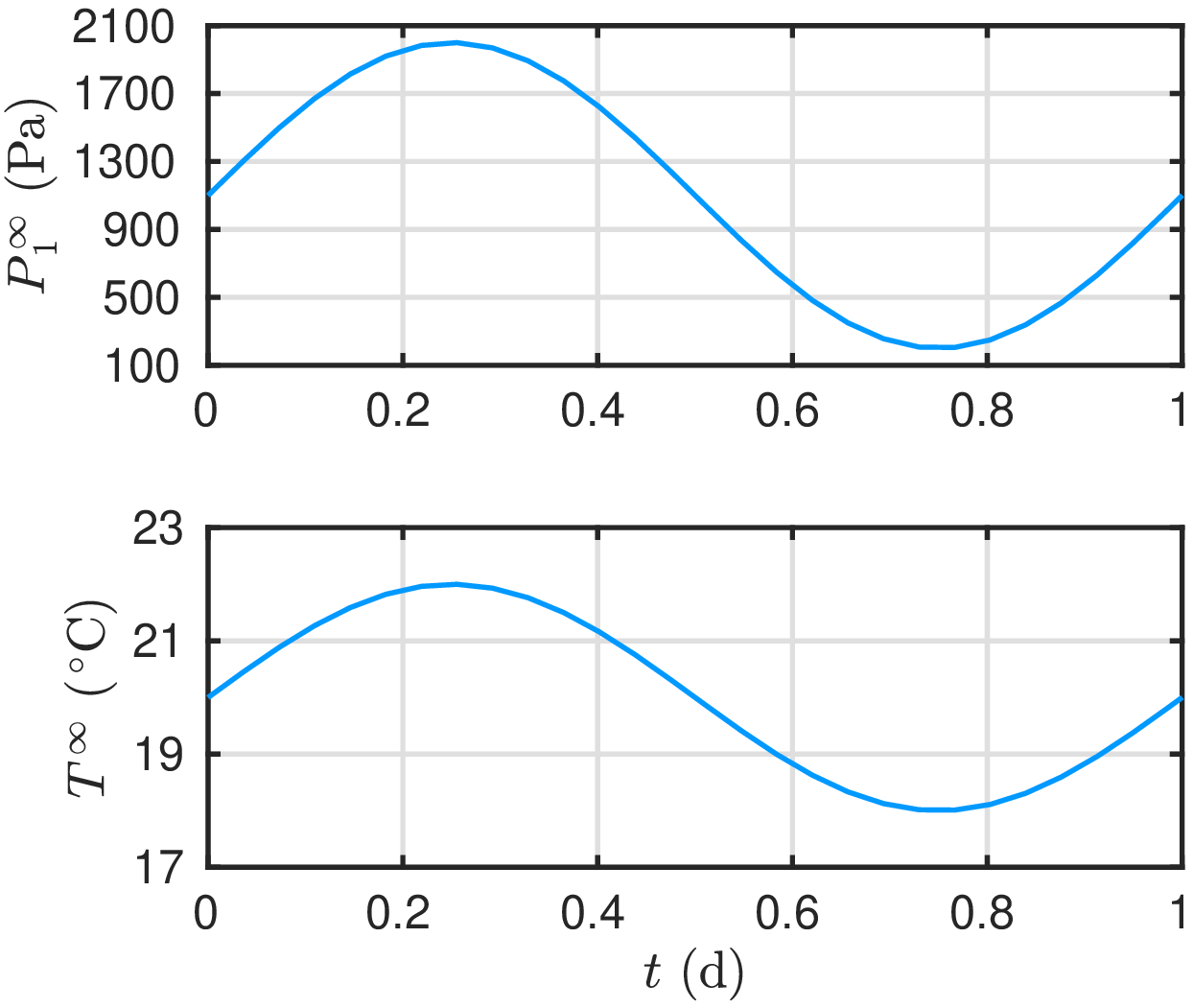}} \hspace{0.2cm}
\subfigure[\label{fig:BCPTeq_ft}]{\includegraphics[width=.45\textwidth]{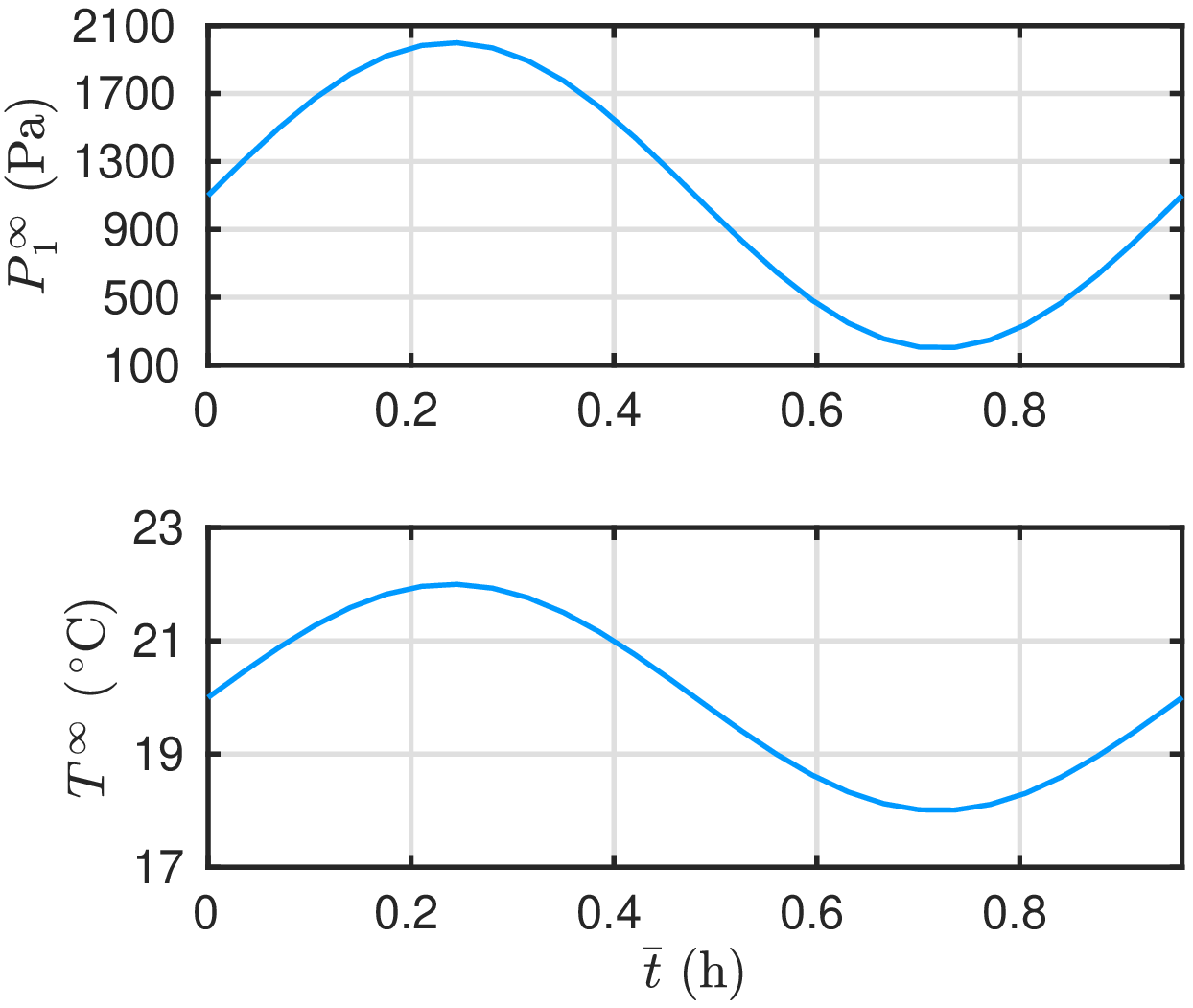}} 
\caption{Time evolution of the boundary conditions over one period cycle for the initial \emph{(a)} and equivalent \emph{(b)} configurations.}
\end{figure}

It is proposed to evaluate the reliability of this approach using simulations for a realistic case study as illustrated in Figure~\ref{fig:exp_similarity}. The initial configuration considers a wood fiber material of $L^{\,\rf} \egal 20 \ \mathsf{cm}\,$. To evaluate the aging of its material properties, the experimental design submits one side of the material to daily variations of temperature and vapor pressure. The boundary conditions of time period $\Gamma \egal 24 \ \mathsf{h}$ are illustrated in Figure~\ref{fig:BCPT_ft}. The surface coefficients of this side are  $h_{\,m}^{\,\rf} \egal 5 \e{-9} \ \mathsf{s\,.m^{\,-1}}$ and $h_{\,q}^{\,\rf} \egal 5 \ \mathsf{W\,.\,m^{\,-2}\,.\,K^{\,-1}}\,$. Another side of the material is set as adiabatic. The total time of the experiment is $t^{\,\rf} \egal 365 \ \mathsf{d}\,$. 

The hygrothermal similitude laws~\eqref{eq:similitude_law} with a proportionality factor $\Pi \egal 0.2$ give the following equivalent configuration. The length of the material is reduced to $\overline{L^{\,\rf}} \egal 4 \ \mathsf{cm}\,$. The boundary conditions vary with a shorter time period of $\overline{\Gamma} \egal 0.96 \ \mathsf{h}\,$. The surface coefficients are increased to $\overline{h_{\,m}^{\,\rf}} \egal 2.8 \e{-8} \ \mathsf{s\,.m^{\,-1}}$ and $\overline{h_{\,q}^{\,\rf}} \egal 25 \ \mathsf{W\,.\,m^{\,-2}\,.\,K^{\,-1}}\,$. The boundary conditions are given in Figure~\ref{fig:BCPTeq_ft}, highlighting a decrease of the period cycle from $1 \ \mathsf{d}$ for the initial configuration to less than $1 \ \mathsf{h}$ ($57.6 \ \mathsf{min}$). The experiment requires also a significantly reduced horizon of $\overline{t^{\,\rf}} \egal 14.6 \ \mathsf{d}\,$. 

\begin{figure}
\centering
\subfigure[\label{fig:TP_ft}]{\includegraphics[width=.45\textwidth]{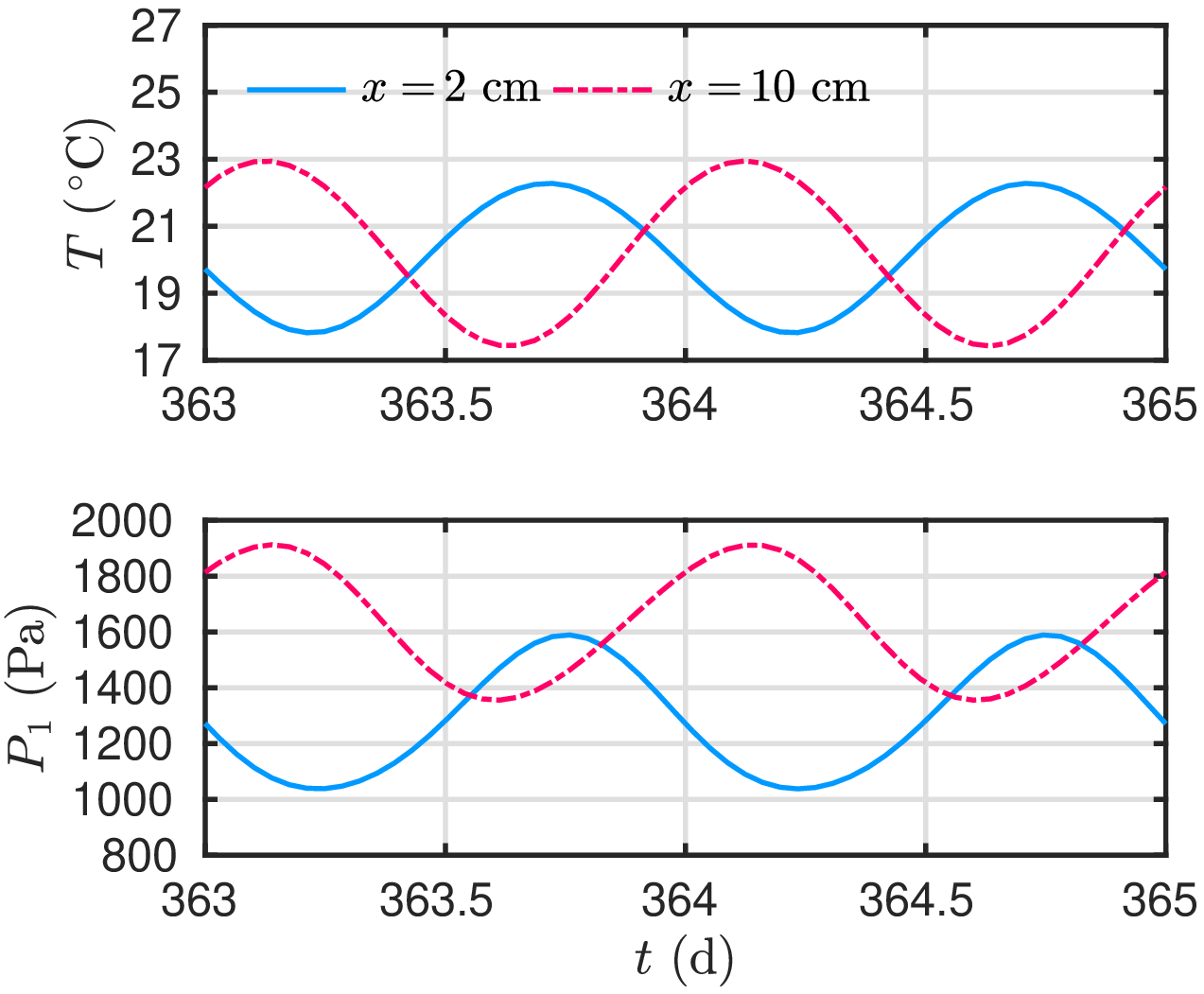}} \hspace{0.2cm}
\subfigure[\label{fig:TPeq_ft}]{\includegraphics[width=.45\textwidth]{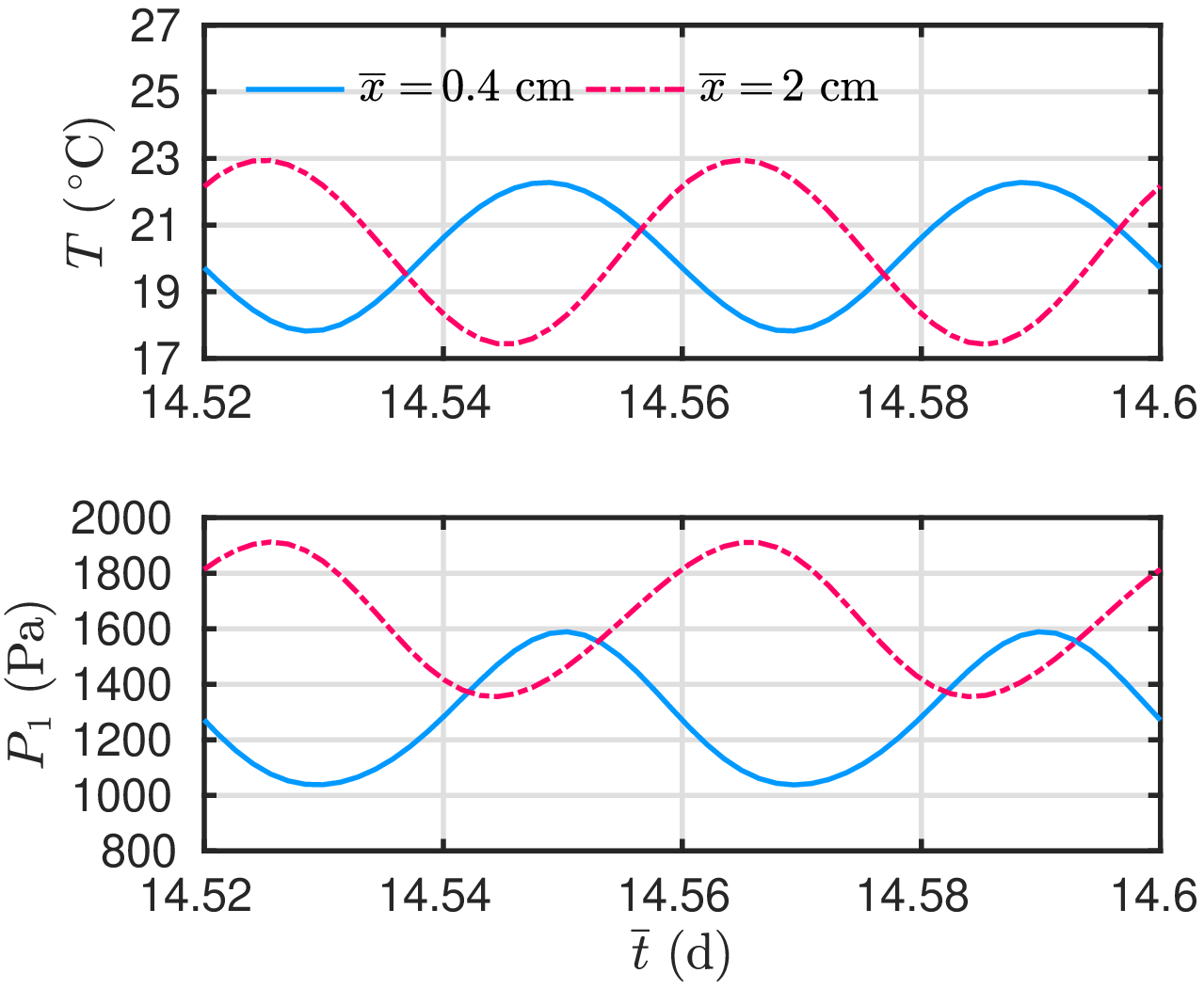}}\\
\subfigure[\label{fig:TP_fx}]{\includegraphics[width=.45\textwidth]{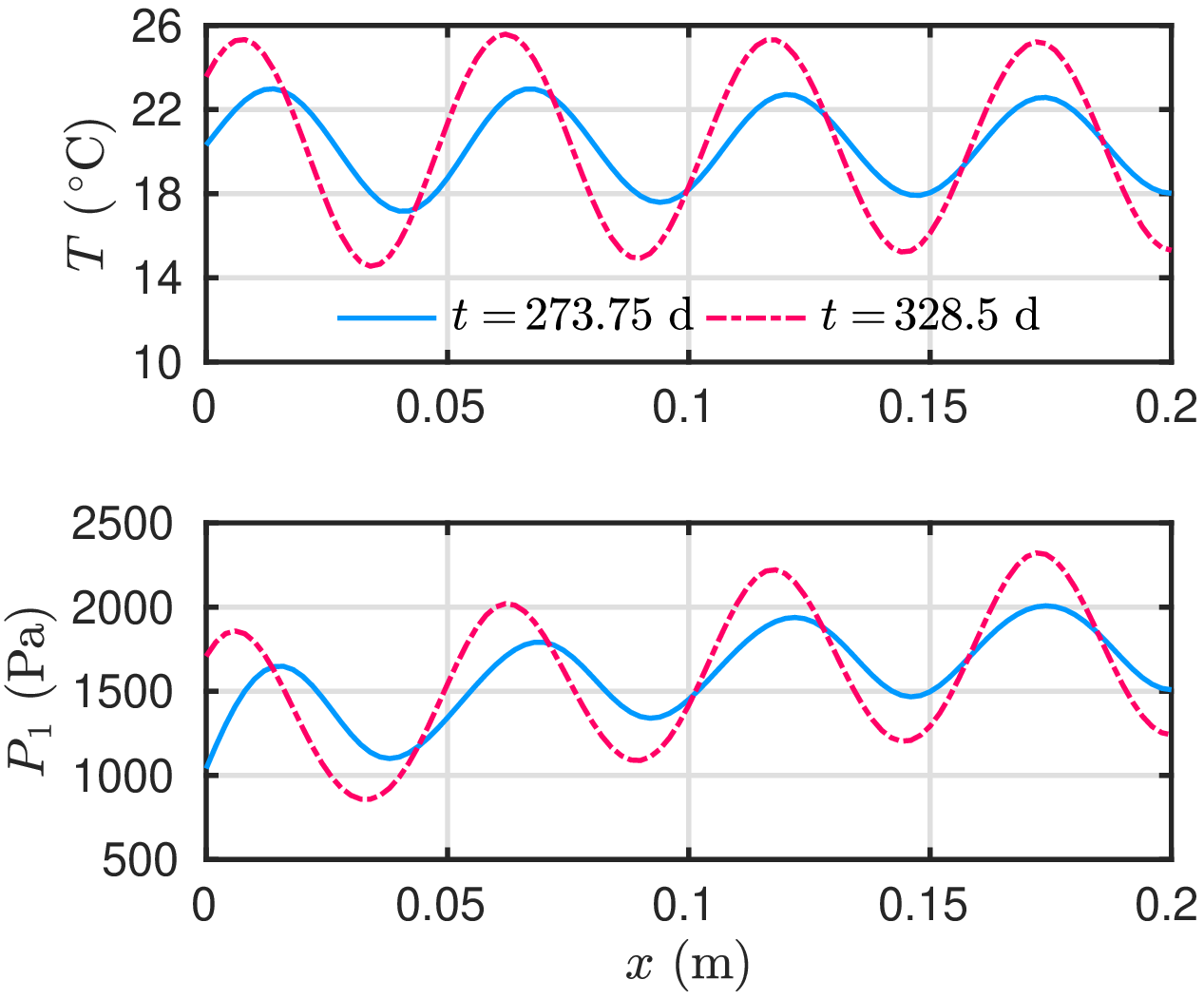}} \hspace{0.2cm}
\subfigure[\label{fig:TPeq_fx}]{\includegraphics[width=.45\textwidth]{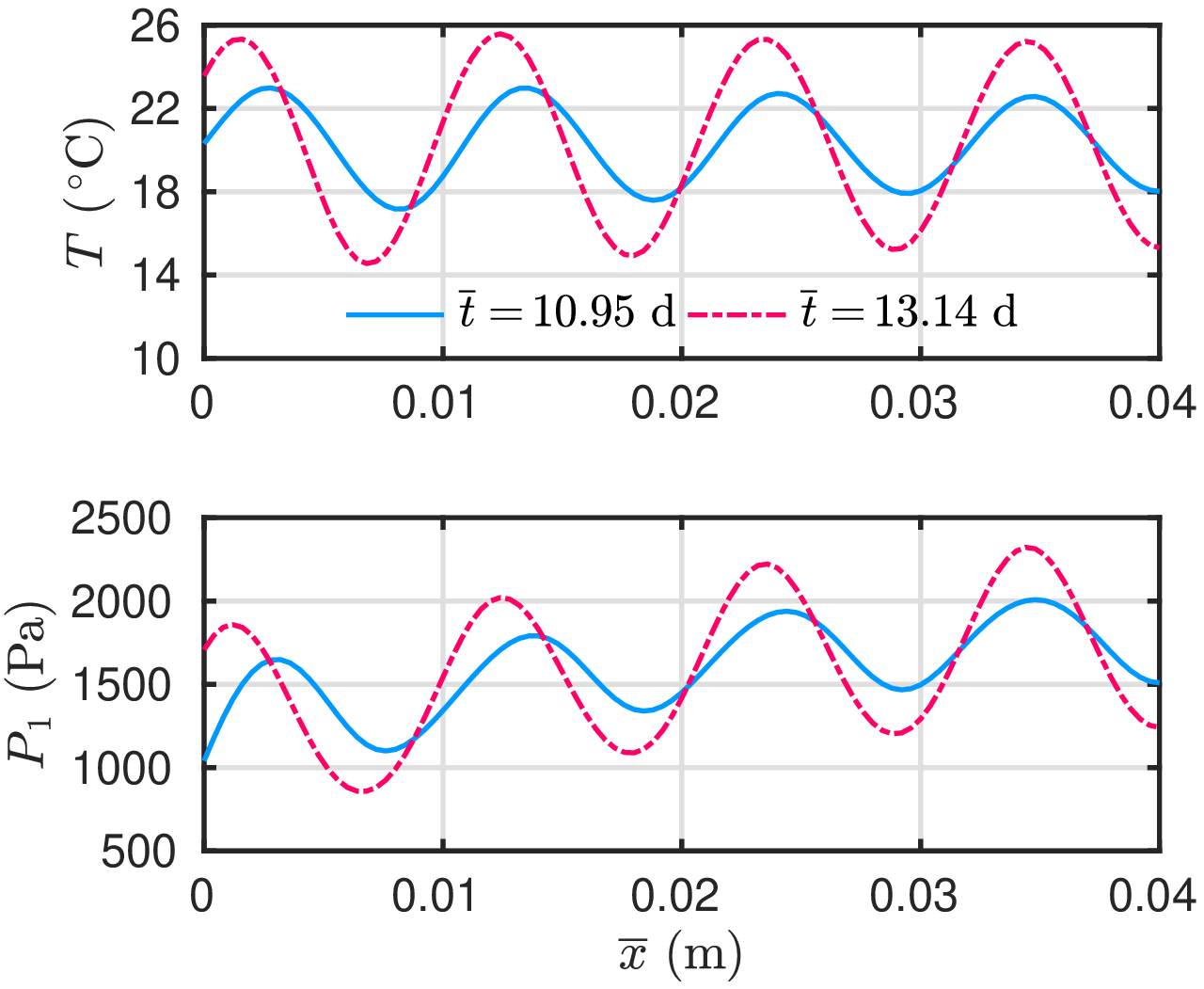}} 
\caption{Evolution of the temperature and vapor pressure according to space and time for the initial \emph{(a,c)} and equivalent \emph{(b,d)} configurations.}
\end{figure}

\begin{figure}
\centering
\subfigure[\label{fig:EW_ft}]{\includegraphics[width=.45\textwidth]{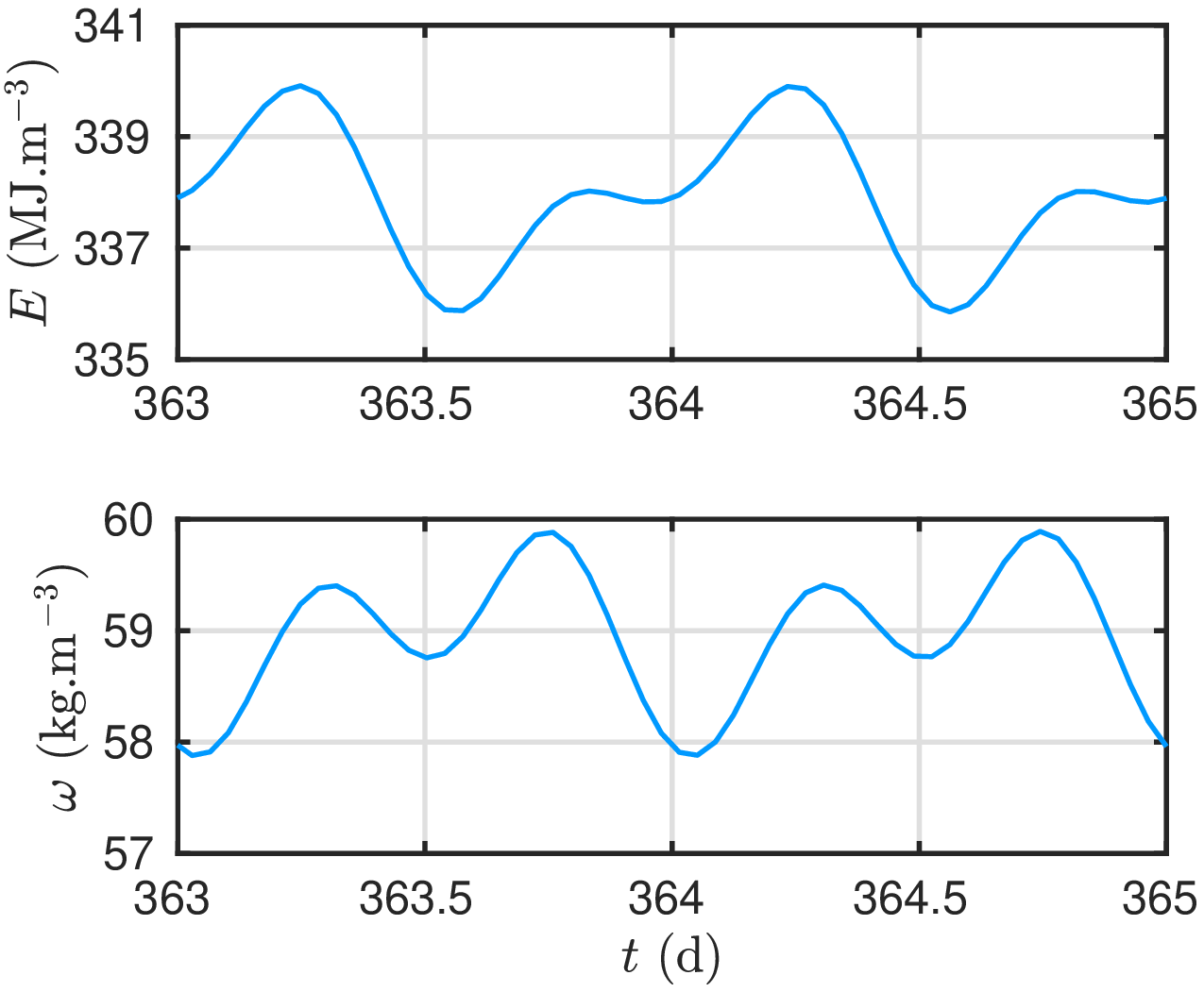}} \hspace{0.2cm}
\subfigure[\label{fig:EWeq_ft}]{\includegraphics[width=.45\textwidth]{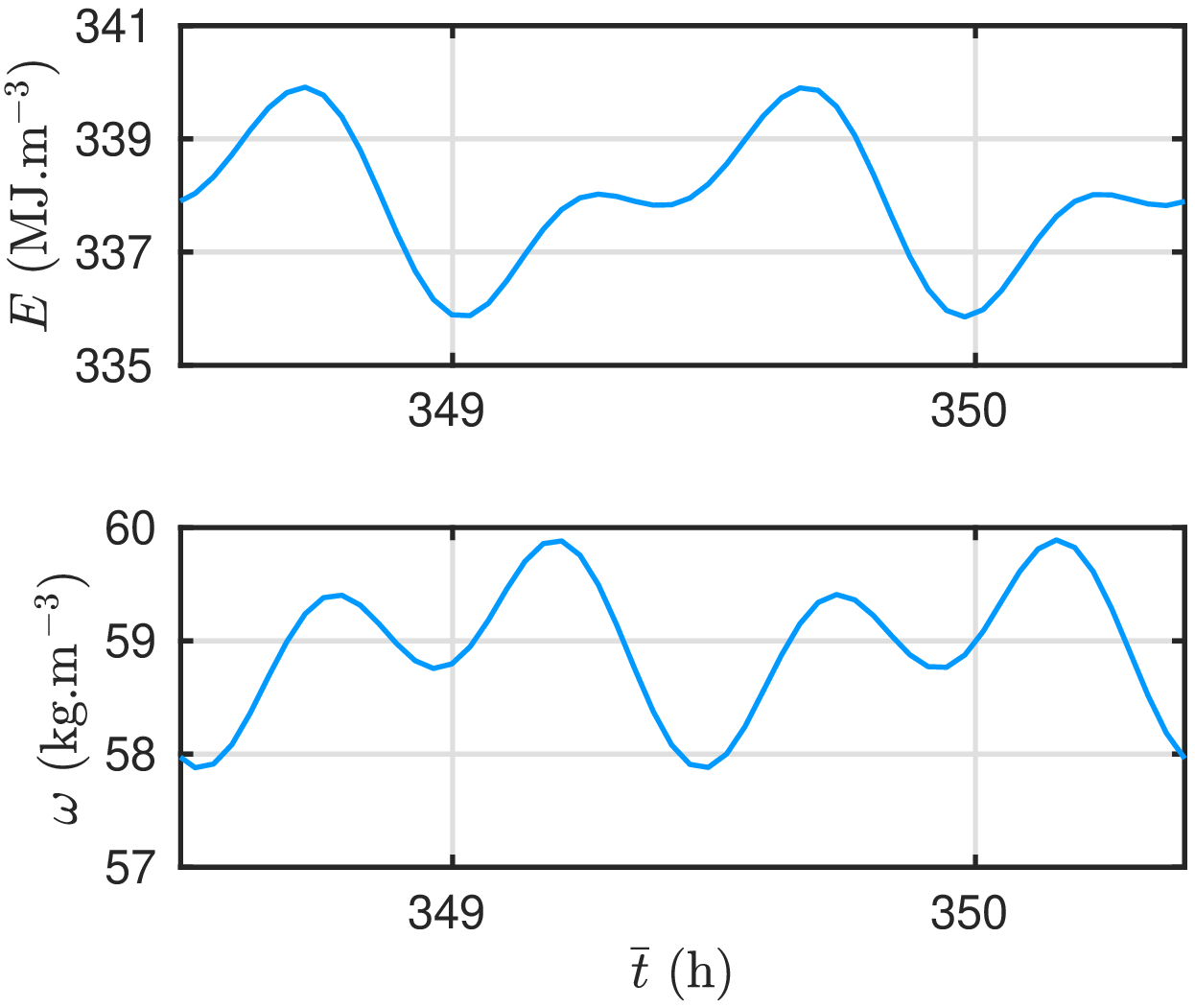}}
\caption{Evolution of the volumetric  internal energy and water content according time for the initial \emph{(a)} and equivalent \emph{(b)} configurations.}
\end{figure}

It is possible to verify that both configurations lead to the same dimensionless problem with the following coefficients: 
\begin{align*}
\Fo^{\,m}  & \egal 6.89 \,,
&& \Fo^{\,q}  \egal 94.76 \,,
&& \delta  \egal 0.97 \,,
&& \gamma  \egal 0.64 \,, \\ 
\eta & \egal 0.43 \,,
&& \Bi^{\,q} \egal 25.49 \,,
&& \Bi^{\,m}  \egal 16.56 \,,
&& \Bi^{\,qm} \egal 16.56 \,.
\end{align*}
Therefore the phenomena occurring in the material from the two configurations are similar. Measurements are equivalent in both configurations. Those results can be proven by carrying simulations, using the one dimensional numerical model established in \cite{Berger_2019c}. Figures~\ref{fig:TP_ft} and \ref{fig:TPeq_ft} compare the time evolution of temperature and vapor pressure in the material for both configurations. The dimensionless points of observations, \emph{i.e.} measurements, are $ \chi \egal \bigl\{\, 0.1 \,,\, 0.5 \,\bigr\}\,$, as reported in Table~\ref{tab:point_of_observation}. From a physical point of view, the observations correspond to sensors placed at $x  \egal \bigl\{\, 2 \,,\, 10 \,\bigr\} \ \mathsf{cm}$ and $\overline{x}  \egal \bigl\{\, 0.4 \,,\, 2 \,\bigr\} \ \mathsf{cm} \,$ in the initial and equivalent configurations, respectively. Looking at Figures~\ref{fig:TP_ft} and \ref{fig:TPeq_ft}, the time evolution during the last two days of the experiment is exactly the same for both configurations. The last two astronomical days correspond to $t^{\,\star} \, \in \, \bigl[\,0.9945 \,,\, 1 \,\bigr]\,$, $t \, \in \, \bigl[\,362.99 \,,\,  365 \,\bigr] \ \mathsf{d}$ and  $\overline{t} \, \in \, \bigl[\,348.48 \,,\, 350.4 \,\bigr] \ \mathsf{h}$ for the dimensionless, initial and equivalent configurations, respectively. A similar analysis can be carried for the profiles of the fields in Figures~\ref{fig:TP_fx} and  \ref{fig:TPeq_fx}. The equivalence chosen time of observation between the configurations is reported in Table~\ref{tab:point_of_observation}. If one could observe with a specific experimental device those profiles at $\tau \egal \bigl\{\, 0.75 \,,\, 0.9 \,\bigr\}\,$, there would be a perfect agreement between the initial and equivalent configurations. As illustrated in Figures~\ref{fig:EW_ft} and \ref{fig:EWeq_ft}, there is also a perfect concordance between the two experimental configurations for thermodynamic variables such as the volumetric internal energy and moisture content of the material. Those results highlight that the dimensionless representations can be used to define transfer similitude laws between two experimental designs. The laws can be used to define experimental design with relaxed constraint by reducing, for instance, the time of experiments. Note that the dynamic similarity could be validated by carrying out an experimental campaign for the two equivalent configurations. Then, a measurement of temperature and relative humidity inside the material together with mass content can be compared in order to validate the dynamic similarity between the two configurations.

\begin{table}[h!]
\centering
\caption{Point of observation of the field according to the similitude laws.}
\label{tab:point_of_observation}
\begin{tabular}{ccc}
\hline
\hline
\textit{Dimensionless representation}
& \textit{Initial configuration}
& \textit{Equivalent configuration} \\
\hline
\multicolumn{3}{c}{Space coordinate} \\
$\chi \ \unit{-}$
& $x \ \unit{cm}$
& $\overline{x} \ \unit{cm}$ \\
$0.1$
& $2$
& $0.4$ \\
$0.5$
& $10$
& $2$ \\
\hline
\multicolumn{3}{c}{Time coordinate} \\
$\tau \ \unit{-}$
& $t \ \unit{d}$
& $\overline{t} \ \unit{d}$ \\
$0.75$
& $273.75$
& $10.95$ \\
$0.9$
& $328.5$
& $13.14$ \\
\hline
\hline
\end{tabular}
\end{table}

\section{Conclusion}

Within the environmental context, designers and engineers require tools to understand the dominant phenomena of heat and mass transfer without using computation tools. To tackle this issue, the dimensionless analysis of heat and mass transfer in building porous material is proposed. It is based on scaling the governing equations in order to introduce the so-called dimensionless numbers with their nonlinear distortion. The analysis of these quantities has several advantages as illustrated in this article. 

First advantage is that the dimensionless analysis permits a clear physical understanding of the phenomena of heat and mass transfer in building porous materials as shown in Section~\ref{sec:mapping_phenomena}. A first analysis gives the range of variations of the dimensionless numbers at a fixed reference time and length for seven categories of materials. Some categories of materials can be highlighted regarding their sensitivity to the processes. It may help in the modeling of the physical process by simplifying certain phenomena for some materials, and as a consequence, to reduce the complexity of the numerical model. 

In Section~\ref{sec:comparison_materials}, a mapping of the dimensionless numbers for $49$ materials is proposed for the given characteristic lengths. The latter is defined according to a conventional use of a material in buildings. Some regions of dominant physical processes of the materials are highlighted. The aim is to observe the competition between heat and mass transfer for each material. The importance of the latent and sensible heat transfer can also be seen with this mapping.

The dimensionless number gives an overview of the dominant processes in the defined reference conditions. Since the coefficients of heat and mass transfer are nonlinear, \emph{i.e.} depending on the fields, the dimensionless analysis is extended in Section~\ref{sec:non_linear_behavior}. The so-called distortion of the dimensionless numbers is analyzed for two materials. The behavior is compared according to vapor pressure and temperature, represented by the dimensionless variables $u$ and $v\,$, respectively. 

In Section~\ref{sec:multi_layer} the investigations are presented for two wall configurations composed of several layers. Here, the dimensionless numbers are evaluated for each layer of the wall. The two configurations are then compared depending on the direction of the heat or mass flows. It is possible to identify the layers where the transfer is the fastest. \jules{Note that general conclusions over the whole configurations is not possible when the dimensionless numbers are not completely ordered. In such case, further investigations are required using building simulation programs.}

The second advantage is highlighted in Section~\ref{sec:similarity_analysis}. The similarity analysis can be used in the context of planning experiments or designing the efficiency of walls. The kinetic, geometric and dynamic similarities are presented. Firstly, the similarity in terms of a material length is presented for a given reference time. The equivalent length of the wood wool with three other insulation materials is computed according to the different heat and mass transfer processes. The second case establishes the equivalent time of heat and mass transfer in the concrete with three other materials. Lastly, the equivalency in terms of heat and mass transfer is stated between two experimental configurations. 

Other advantages of dimensionless formulation should be underlined, particularly when developing numerical models. The number of variables is reduced in the dimensionless equations compared to the physical ones. This may be very relevant within the context of determining materials properties using experimental observations of the field. In \cite{Berger_2019}, the unknown parameters to estimate are reduced from four to two between the dimensionless and dimensional models. Then, from a numerical point of view, solving one dimensionless problem is equivalent to solving a whole class of dimensionless problems sharing the same scaling parameters. It also promotes the share and development of numerical models for efficient, accurate and fast computing of the equations. Last, the floating point arithmetic is designed to minimize the rounding errors when computer manipulates numbers of the same magnitude \citep{Kahan_1979}. Since the floating point numbers have the highest density within the interval $(\, 0,\,1 \,)\,$, it is wise to carry numerical analysis with dimensionless equations. This lack of accuracy has been illustrated in \cite{Berger_2019b} where the computation of the equation of heat or mass transfer in physical and dimensionless forms are compared. When solving the equation in its physical dimension, the significant digits accuracy of the solution can lose one order.

In the end, the dimensionless analysis is an efficient tool for evaluating the dominant processes in the framework of modeling  or performing experiments of heat and mass transfer. It may also be appropriate for designers and engineers to operate a first order analysis of wall configurations. It is important to mention that the proposed dimensionless analysis strongly depends on the nine material properties presented in Section~\ref{sec:porous_mat_properties}. Thus, future works should focus on developing a consolidated database of building materials properties, integrating also more complex phenomena such as the effect of temperature on isotherm sorption curve \cite{Rahim_2016}. The analysis also rely on the mathematical model. Here, the model of heat and mass transfer is written with the temperature and vapor pressure potentials. The scaling of equations can be extended to any other model, considering other potentials such as moisture content or capillary pressure. Though it should be noted that it is required to have the material properties in accordance to the mathematical model. An important link can be done between the macroscopic formulation used here and the microscopic description of transfer in porous media \cite{ArticleA,ArticleB,ArticleF}. In this way, the dimensionless number could be linked to microscopic properties of the material. It would enable to extend the determination of the dimensionless numbers and their nonlinear distortion.

\section*{Acknowledgments}

The authors acknowledge the Junior Chair Research program ``Building performance assessment, evaluation and enhancement'' from the University of Savoie Mont Blanc in collaboration with The French Atomic and Alternative Energy Center (CEA) and Scientific and Technical Center for Buildings (CSTB). The authors thanks the grants from the French Environment and Energy Management Agency (ADEME). The authors also acklowledge the French and Brazilian agencies for their financial supports through the project CAPES--COFECUB, as well as the CNPQ of the Brazilian  Ministry of Education and of the Ministry of Science, Technology and Innovation, respectively, for co-funding.

\section*{Nomenclature and symbols}

\begin{tabular}{|cll|}
\hline
\multicolumn{3}{|c|}{\emph{Latin letters, physical parameters}} \\
$A$ & water adsorption coefficient & $\unit{kg\,.\,m^{\,-2}\,.\,s^{\,-0.5}}$ \\
$c_{\,m}$ & mass storage coefficient & $\unit{kg\,.\,m^{\,-3}}$ \\
$c_{\,mq}$ & mass storage coefficient due to temperature & $\unit{kg \,.\,Pa\,.\,m^{\,-3}\,.\,K^{\,-1}}$ \\
$c_{\,q}\,,\,c_{\,0}\,,\,c_{\,1}\,,\,c_{\,2}$ & heat storage coefficient & $\unit{ J \,.\, m^{\,-3}\,.\,K^{\,-1}}$ \\
$D_{\,2}$ & liquid diffusion coefficient & $\unit{m^{\,2}\,.\,s^{\,-1}}$ \\
$h_{\,m}$ & surface mass transfer coefficient  & $\unit{s\,.m^{\,-1}}$ \\
$h_{\,q}$ & surface heat transfer coefficient  & $\unit{W\,.\,m^{\,-2}\,.\,K^{\,-1}}$ \\
$k_{\,1}\,,\,k_{\,2}\,,\,k_{\,m}\,,\,k_{\,qm}$ & mass permeability coefficient & $\unit{s}$ \\
$k_{\,mq}$ & mass transfer coefficient under temperature gradient & $\unit{kg\,.\,m^{\,-1}\,.\,s^{\,-1}\,.\,K^{\,-1}}$ \\
$k_{\,q}$ & heat transfer coefficient & $\unit{W\,.\,m^{\,-1}\,.\,K^{\,-1}}$ \\
$L$ & length & $\unit{m}$ \\
$P\,,\,P_{\,1}\,,\,P_{\,2}\,,\,P_{\,3}\,,\,\Psat\,,\,\deltaP$ & pressure & $\unit{Pa}$ \\
$R$ & gas constant & $\unit{ J \,.\, kg^{\,-1} \,.\,K^{\,-1}}$ \\
$r_{\,12}\,,\,r_{\,12\,,\,0}$ & latent heat of evaporation & $\unit{ J \,.\, kg^{\,-1}}$ \\
$T\,,\,\deltaT$ & temperature & $\unit{K}$ \\
$t$ & time & $\unit{s}$ \\
$x$ & space coordinate & $\unit{m}$ \\
\multicolumn{3}{|c|}{\emph{Greek letters, physical parameters}} \\
$\alpha_{\,1}$ & \textsc{Oswin} model coefficient & $\unit{-}$ \\
$\beta$ & thermal conductivity variation with water content & $\unit{W\,.\,m^{\,2}\,.\,kg^{\,-1}\,.\,K^{\,-1}}$ \\
$\mu$ & vapor resistant factor & $\unit{-}$ \\
$\omega$ & mass content & $\unit{kg\,.\,m^{\,-3}}$ \\
$\phi$ & relative humidity & $\unit{-}$ \\
$\rho\,,\,\rho_{\,0}\,,\,\rho_{\,1}\,,\,\rho_{\,2}$ & specific mass & $\unit{kg\,.\,m^{\,-3}}$ \\
\hline
\end{tabular}

\hspace{2cm}
\bigskip

\begin{tabular}{|cll|}
\hline
\multicolumn{3}{|c|}{\emph{Dimensionless parameters}} \\
$\Bi^{\,q}\,,\,\Bi^{\,qm}$ & surface heat transfer \textsc{Biot} number & \\
$\Bi^{\,m}$ & surface mass transfer \textsc{Biot} number & \\
$\Fo^{\,q}$ & heat transfer \textsc{Fourier} number &  \\
$\Fo^{\,m}$ & mass transfer \textsc{Fourier} number &  \\
$c_{\,m}^{\,\star} \,,\,c_{\,q}^{\,\star}$ & storage distortion coefficient & \\
$k_{\,q}^{\,\star} \,,\,k_{\,m}^{\,\star}\,,\, k_{\,qm}^{\,\star}\,,\,k_{\,mq}^{\,\star}$ & permeability distortion coefficient & \\
$r_{\,12}^{\,\star}$ & latent heat vapor distortion coefficient & \\
$u$ & vapor pressure field &  \\
$v$ & temperature field & \\
$\gamma\,,\,\delta \,,\, \eta$ & numbers translating influence of the heat or mass transfer on each other & \\
$\chi$ & space domain & \\
$\tau$ & time domain & \\
\hline
\end{tabular}
\hspace{2cm}
\bigskip

\begin{tabular}{|cll|}
\hline
\multicolumn{3}{|c|}{\emph{Subscripts and superscripts}} \\
$i$ & inside &  \\
$m$ & mass transfer &  \\
$o$ & outside &  \\
$q$ & heat transfer &  \\
$mq$ & coupled mass coefficient under heat process &  \\
$qm$ & coupled heat coefficient under mass process &  \\
$\rf$ & reference value &  \\
$0$ & initial &  \\
$\infty$ & air ambient field &  \\
$\star$ & dimensionless parameter &  \\
\hline
\end{tabular}

\hspace{2cm}
\bigskip

The following symbols are used in the mathematical notation:
\begin{itemize}
\item $ \egal $  designates the equality between two scalar numbers. $a \egal b$ means that the scalars $a$ and $b$ are the same.
\item $\, \eqdef \,$ stands for a definition. $a \, \eqdef \, b$ means that $a$ is defined to be equal to $b\,$.
\end{itemize}

\bibliographystyle{unsrt}  
\bibliography{references}

\appendix

\section{Material properties}
\label{sec:mat_properties}

\begin{table}[h!]
\centering
\caption{Material properties, part 1.}
\label{tab:mat_properties_1}
\rotatebox{90}{
\begin{tabular}{c ccc ccc ccc}
\hline
\hline
\multirow{2}{*}{Material}
& $\rho_{\,0}$ & $c_{\,0}$ & $k_{\,q\,,\,0}$ 
& $\beta$ & $\mu$ & $\alpha_{\,1}$ 
& $\omega_{\,1}$ & $A$ & $\omega_{\,f}$ \\	
& $\unit{kg\,.\,m^{\,-3}}$ & $\unit{J\,.\,kg^{\,-1}\,.\,K^{\,-1}}$ & $\unit{W\,.\,m^{\,-2}\,.\,K^{\,-1}}$
& $\unit{W\,.\,m^{\,2}\,.\,kg^{\,-1}\,.\,K^{\,-1}}$ & $\unit{-} $ & $\unit{-} $ 
& $\unit{kg\,.\,m^{\,-3}}$ & $\unit{kg \,.\, m^{\,-2}\,.\,s^{\,-0.5}}$ & $\unit{kg\,.\,m^{\,-3}}$ \\
\hline \hline
1. Concrete
& $2104$ & $776$ & $1.37$
& $0.005$ & $76$ & $0.38$
& $66.47$ & $0.0125$ & $144$ \\ 
2. Pumice concrete 
& $672$ & $850$ & $0.14$ 
& $0$ & $4$ & $2.04$ 
& $1.65$ & $0.0467$ & $291$ \\ 
3. Cement Flooring Mid. Lay. 
& $1940$ & $1000$ & $1.4$ 
& $0$ & $102$ & $0.31$ 
& $25.54$ & $0.0183$ & $147$ \\ 
4. Cement Flooring Up. Lay. 
& $1940$ & $1000$ & $1.4$ 
& $0$ & $102$ & $0.31$ 
& $25.54$ & $0.0183$ & $147$ \\ 
5. Cement Flooring Un. Lay. 
& $1940$ & $1000$ & $1.4$ 
& $0$ & $138$ & $0.32$ 
& $26.35$ & $0.012$ & $140$ \\ 
6. Mineral Cementing Material 
& 1436 & 1000 & 0.61 
& 0.003 & 25 & 1.88 
& 1.38 & 0.0007 & 264 \\ 
7. Mineral Finishing Coat 
& 1482 & 1000 & 0.95 
& 0.005 & 17 & 1.53 
& 1.07 & 0.002 & 321 \\ 
8. Silicon Resin Finishing Coat 
& 1475 & 1000 & 0.69 
& 0.004 & 74 & 1.12 
& 0.61 & 0.0002 & 303 \\ 
9. Anti Mold 
& 465 & 1173 & 0.11 
& 0 & 8.41 & 0.58 
& 27.18 & 0.0135 & 105 \\ 
10. Resin Finishing Coat 
& 1100 & 850 & 0.7 
& 0 & 1000 & 0.86 
& 3.03 & 0.0017 & 100 \\ 
11. Silicone Resin Final Coat 
& 1475 & 1000 & 0.7 
& 0 & 74 & 1.12 
& 0.61 & 0.0002 & 303 \\ 
12. Calcium Silicate 
& 270 & 1162 & 0.07 
& 0 & 3.85 & 0.09 
& 17.56 & 1.115 & 793 \\ 
13. Insulating Plaster
& 1128 & 1006 & 0.37 
& 0 & 10.9 & 0.44 
& 47.62 & 0.0317 & 298 \\ 
14. Wood Wool 
& 469 & 1470 & 0.07 
& 0 & 9.4 & 2.03 
& 1.63 & 0.0365 & 329 \\ 
15. Clay 
& 368 & 885 & 0.07 
& 0 & 7.7 & 0.49 
& 5.69 & 0.055 & 284 \\ 
16. Cellulose CPH 
& 92.8 & 2005 & 0.05 
& 0 & 2.42 & 0.56 
& 3.87 & 0.0583 & 735 \\ 
17. Cellulose Blowin 
& 55 & 2544 & 0.04 
& 0.001 & 2 & 0.45 
& 3.46 & 0.56 & 494 \\ 
18. Cellulose 
& 50 & 1600 & 0.04 
& 0 & 1.8 & 0.51 
& 2.7 & 0.157 & 426 \\ 
19. Flat Pannel 
& 39 & 850 & 0.04 
& 0 & 1.5 & 0.59 
& 2.2 & 0.016 & 348 \\ 
20. Wood Fiber 1 
& 168 & 1700 & 0.04 
& 0 & 3.3 & 0.41 
& 9.76 & 0.0033 & 526 \\ 
21. Wood Fiber 3 
& 165 & 1700 & 0.04 
& 0 & 2.9 & 0.29 
& 17.78 & 0.0015 & 240 \\ 
22. Wood Fiber 4 
& 159 & 1700 & 0.05 
& 0 & 3.3 & 0.41 
& 14.78 & 0.0018 & 833 \\ 
23. Wood Chips 
& 65 & 1500 & 0.04 
& 0 & 2.5 & 0.34 
& 5.63 & 0.0533 & 188 \\ 
24. Aerated Concrete 
& 392.2 & 1081 & 0.09 
& 0 & 7.38 & 0.88 
& 5.63 & 0.043 & 259 \\ 
\hline
\hline
\end{tabular}}
\end{table}

\begin{table}[h!]
	\centering
	\caption{Material properties, part 2.}
	\label{tab:mat_properties_2}
	\rotatebox{90}{
		\begin{tabular}{c ccc ccc ccc}
			\hline
			\hline
			\multirow{2}{*}{Material}
			& $\rho_{\,0}$ & $c_{\,0}$ & $k_{\,q\,,\,0}$ 
			& $\beta$ & $\mu$ & $\alpha_{\,1}$ 
			& $\omega_{\,1}$ & $A$ & $\omega_{\,f}$ \\	
			& $\unit{kg\,.\,m^{\,-3}}$ & $\unit{J\,.\,kg^{\,-1}\,.\,K^{\,-1}}$ & $\unit{W\,.\,m^{\,-2}\,.\,K^{\,-1}}$
			& $\unit{W\,.\,m^{\,2}\,.\,kg^{\,-1}\,.\,K^{\,-1}}$ & $\unit{-} $ & $\unit{-} $ 
			& $\unit{kg\,.\,m^{\,-3}}$ & $\unit{kg \,.\, m^{\,-2}\,.\,s^{\,-0.5}}$ & $\unit{kg\,.\,m^{\,-3}}$ \\
			\hline \hline
25. Straw Clay Brick 
& 1036 & 1000 & 0.4 
& 0 & 7.1 & 0.68 
& 16.05 & 0.1117 & 450 \\ 
26. Insulation Brick 
& 600 & 850 & 0.1 
& 0.001 & 16 & 0.96
& 2.9 & 0.0983 & 188 \\ 
27. Solid Brick 
& 2060 & 839 & 0.83 
& 0.003 & 159 & 0.19 
& 0.74 & 0.106 & 148 \\ 
28. Old Brick 
& 1800 & 850 & 0.6 
& 0 & 15 & 0.29 
& 3 & 0.29 & 230 \\ 
29. Extruded Brick 
& 1650 & 850 & 0.6 
& 0 & 9.5 & 0.53 
& 4.4 & 0.46 & 370 \\ 
30. Gypsum Board 
& 732 & 1384 & 0.19 
& 0.002 & 6.8 & 0.77 
& 2.55 & 0.13 & 353 \\ 
31. Cement Plaster 
& 2000 & 850 & 1.2 
& 0 & 25 & 0.87 
& 10.4 & 0.0083 & 280 \\ 
32. Plaster 
& 850 & 1000 & 0.3
& 0 & 8.3 & 0.52
& 3.46 & 0.29 & 400 \\ 
33. Clay Mortar 
& 1568 & 488 & 0.48 
& 0.002 & 11.4 & 0.41 
& 21.81 & 0.175 & 375 \\ 
34. Clay Plaster 
& 1514 & 1000 & 0.65 
& 0 & 18.8 & 0.48 
& 9.65 & 0.0467 & 294 \\ 
35. Calcareous Sandstone 
& 1900 & 1000 & 1 
& 0 & 28 & 0.45 
& 14.44 & 0.415 & 250 \\ 
36. Granite 
& 2453 & 702 & 1.66 
& 0.005 & 54 & 0.23 
& 5.46 & 0.0086 & 50 \\ 
37. Sandstone 
& 2224 & 771 & 1.68 
& 0.006 & 73 & 1.44 
& 0.31 & 0.0816 & 89 \\ 
38. Tuff
& 1450 & 925 & 0.34 
& 0.002 & 10.4 & 0.13 
& 63.07 & 0.0983 & 259 \\ 
39. Longitudinal Oak 
& 685 & 1500 & 0.3 
& 0 & 8 & 0.36 
& 70.43 & 0.0108 & 500 \\ 
40. Radial Oak 
& 685 & 1500 & 0.13 
& 0 & 140 & 0.36 
& 70.43 & 0.0017 & 500 \\ 
41. Old Oak 
& 740 & 1600 & 0.15 
& 0 & 223 & 0.29 
& 69.92 & 0.0017 & 349 \\ 
42. OSB 
& 595 & 1700 & 0.1 
& 0 & 165 & 0.23 
& 69.93 & 0.002 & 814 \\ 
43. Plywood 
& 427 & 1600 & 0.09 
& 0 & 188 & 0.87 
& 20.37 & 0.0022 & 573 \\ 
44. Chipboard 
& 620 & 1700 & 0.11 
& 0 & 44 & 0.79 
& 35.25 & 0.033 & 520 \\ 
45. Timbered Mortar 
& 960 & 1000 & 0.17 
& 0 & 17.3 & 1.3 
& 1.71 & 0.032 & 360 \\ 
46. Longitudinal Spruce 
& 455 & 1500 & 0.23 
& 0 & 4.3 & 0.38 
& 46.57 & 0.0108 & 600 \\ 
47. Radial Spruce 
& 455 & 1500 & 0.09 
& 0 & 130 & 0.38 
& 46.57 & 0.0015 & 600 \\ 
48. Microstrandboard 
& 664 & 1700 & 0.13 
& 0 & 92.4 & 0.37 
& 54.41 & 0.0018 & 587 \\ 
49. DWD 
& 528 & 1700 & 0.14 
& 0 & 10 & 0.25 
& 49.59 & 0.0017 & 667 \\ 
\hline
\hline
\end{tabular}}
\end{table}

\end{document}